\documentclass[apj,twocolumn]{openjournal}

\usepackage{xcolor}
\usepackage{textgreek}
\usepackage[utf8]{inputenc}
\usepackage[english]{babel}

\usepackage{hyperref}
\hypersetup{
    unicode, 
    colorlinks=true,
    linkcolor=linkcolor,
    citecolor=linkcolor,
    filecolor=linkcolor,
    urlcolor=linkcolor,
}
\usepackage{color,colortbl}
\definecolor{linkcolor}{rgb}{0.0,0.3,0.5}
\usepackage{tensind}
\tensordelimiter{?}
\DeclareGraphicsExtensions{.bmp,.png,.jpg,.pdf}
\usepackage{verbatim}
\usepackage[normalem]{ulem}
\usepackage{orcidlink}
\usepackage{soul}

\usepackage{amsmath}
\usepackage{enumitem}
\setlist[enumerate]{itemsep=2pt, parsep=0pt, topsep=4pt}

\begin{document}
\title{Cataclysmic variables from Gaia XP spectra: The Gaia emission-line magnitude (GEM) diagram}

\author{Ilkham Galiullin$^{1}$\, \orcidlink{0000-0001-5778-2355}}
\author{Vladislav Dodon\,\orcidlink{0009-0003-7483-0090}$^{1}$}
\author{Antonio C. Rodriguez\,\orcidlink{0000-0003-4189-9668}$^{2}$}
\author{Askar Sibgatullin\,\orcidlink{0009-0007-8224-316X}$^{1}$}

\email{Corresponding author: IlhIGaliullin@kpfu.ru}
\affiliation{$^1$Kazan Federal University, Kremlevskaya Str. 18, 420008, Kazan, Russia}
\affiliation{$^2$Center for Astrophysics $|$ Harvard \& Smithsonian, 60 Garden Street, Cambridge, MA 02138, USA}

\begin{abstract}
{\it Gaia} Data Release 3 (DR3) opens a new window for the census of Galactic stellar populations in all-sky surveys, specifically for identifying cataclysmic variables (CVs). While the majority of CVs are distributed below the main sequence on the {\it Gaia} Hertzsprung-Russell (HR) diagram, identifying them among dominant background stellar populations remains a significant challenge. Here, we present the Gaia Emission-line Magnitude (GEM) diagram --- a tool designed to isolate CV candidates from background stars located below the main sequence. Using {\it Gaia} photometric data and low-resolution BP/RP (XP) spectra, we construct the GEM diagram by plotting the extinction-corrected absolute magnitude ($M_{G0}$) against the pseudo-equivalent width of the  H$\alpha$ line ($pgEW_{\mathrm{H}\alpha}$), which is computed via a Gaussian line-profile approximation. We find that known CVs and major stellar background populations, including white dwarfs, hot subdwarfs, and white dwarf-main sequence binaries, occupy distinct regions on the GEM diagram, reflecting both their physical properties and the low-resolution nature of the {\it Gaia} XP spectra. We define an empirical selection line on the GEM diagram to separate true H$\alpha$-emitting CVs from background contaminants. As a proof of concept, we apply the GEM diagram to a {\it Gaia} volume-limited sample within 250 pc ($G < 17.5~\mathrm{mag}$) distributed below the main sequence, recovering known CVs. Additionally, we identify three new CV candidates whose accreting nature is supported by archival X-ray data and optical variability. Compared to traditional optical color-based selection techniques, the GEM diagram uses {\it Gaia}'s low-resolution spectra for initial target identification in all-sky surveys. This provides a more reliable framework for the statistical selection of CV candidates prior to ground-based spectroscopic follow-up and can be scaled to future {\it Gaia} data releases  to map Galactic emission-line star populations.
\end{abstract}

\maketitle

\section{Introduction}
\label{sec:intro}

Cataclysmic variables (CVs) are semi-detached binaries in which a primary white dwarf (WD) accretes matter from a companion star that fills its Roche lobe \citep{2003cvs..book.....W}.  Being the most numerous accreting binaries in the Milky Way, CVs serve as good laboratories for studying binary evolution, accretion physics, and high-energy phenomena \citep[see for reviews,][]{2017PASP..129f2001M,2026SSRv..222...32S}. New CVs are commonly discovered as by-products of wide-field photometric, spectroscopic, and X-ray surveys rather than through dedicated systematic searches in all-sky surveys \citep[e.g.,][]{1996A&A...307..459M,2002AJ....123..430S,2009MNRAS.397.2170G,2020AJ....159..198S}. A recent example of a systematic approach is the X-ray main sequence \citep{2024PASP..136e4201R}, which combines {\it Gaia} colors with the X-ray-to-optical flux ratio to efficiently identify accreting compact objects \citep[e.g.,][]{2024A&A...690A.374G, 2025PASP..137a4201R, 2025ApJ...991..125L}.

The {\it Gaia} mission has transformed stellar astronomy by providing homogeneous all-sky astrometry and broadband photometry for more than a billion sources \citep{2016A&A...595A...1G, 2018A&A...616A...1G, 2021A&A...649A...1G, 2023A&A...674A...1G}. {\it Gaia} Data Release 3 (DR3) introduced low-resolution mean spectra from the Blue and Red Photometers (BP/RP), commonly referred to as XP spectra, for approximately 220 million sources \citep{2023A&A...674A...1G, 2023A&A...674A...2D}.  These spectra cover a wavelength range of $\lambda \simeq 330$--$1050$~nm. While the resolving power of {\it Gaia} XP spectra is insufficient for detailed line analyses, sufficiently strong lines (e.g., the H$\alpha$ line) can be detected at that resolution. Recent studies have used {\it Gaia} XP spectra to estimate stellar atmospheric parameters \citep{2023A&A...674A..27A,2023ApJS..267....8A,2023MNRAS.524.1855Z}, as well as to systematically search for H$\alpha$ emitters \citep{2026A&A...711A.215C, 2026arXiv260621420M, rodriguez2026emissiongaiaxp}, polluted WDs \citep{2024ApJ...977...31P, 2024ApJ...970..181K}, metal-poor stars \citep{2024MNRAS.52710937Y}, binary stars \citep{2025ApJS..279...47L, 2025A&A...704A.126L}, carbon stars \citep{2025ApJ...982..184R}, hot subdwarfs \citep{2026A&A...708A..23A}, and young stellar objects \citep{2025A&A...699A.145D}. These diverse applications demonstrate that the vast number of available {\it Gaia} XP spectra, combined with precise parallaxes and photometry, provides a powerful statistical foundation for the systematic classification and candidate selection of Galactic stellar populations.

On the {\it Gaia} Hertzsprung–Russell (HR) diagram, the region below the main sequence is primarily dominated by non-accreting background stellar populations, including isolated WDs, hot subdwarfs, and inactive binaries. \citet{2020MNRAS.492L..40A} demonstrated that the majority of known CVs are also distributed below the main sequence, placing them co-spatial with these stellar classes. This alignment makes CVs indistinguishable from the background stellar populations through broadband photometry alone. In the optical spectrum, CVs exhibit prominent emission lines, most notably the H$\alpha$ line, while most background populations remain spectrally featureless or display absorption lines. The primary exception arises from white dwarf-main sequence (WD+MS) binaries. In these systems, real chromospheric H$\alpha$ emission from the active M-dwarf companion, combined with steep titanium oxide (TiO) molecular absorption bands, can create pseudo-lines in low-resolution {\it Gaia} XP spectra. Analyzing the specific spectral features in the {\it Gaia} XP spectra of both CVs and background stellar populations allows us to define empirical selection boundaries between these groups. To date, no systematic search for CV candidates has been performed using {\it Gaia} XP spectra.

In this paper, we present the {\it Gaia} Emission-line Magnitude (GEM) diagram, a diagnostic tool designed to isolate CV candidates from the dominant background stellar populations distributed below the main sequence on the {\it Gaia} HR diagram. The GEM diagram improves upon traditional color-based selection methods by using the spectroscopic information available in {\it Gaia} XP data. This approach provides a practical framework to statistically select CV candidates prior to ground-based spectroscopic follow-up. As a proof of concept, we apply the GEM diagram to a {\it Gaia} volume-limited sample within 250~pc ($G < 17.5~\mathrm{mag}$). Our approach recovers the known CVs within this volume while simultaneously identifying three promising new CV candidates. 

This paper is structured as follows. In Section~\ref{sec:data}, we describe the compilation of the sample of known CVs and background stellar populations used to construct the GEM diagram. In Section~\ref{sec:methods}, we describe the H$\alpha$ line identification pipeline and the calculation of pseudo-equivalent widths from {\it Gaia} XP spectra. In Section~\ref{sec:results}, we present the GEM diagram, define our empirical selection criterion, and demonstrate its application to a {\it Gaia} volume-limited sample within 250~pc. Finally, we summarize our results in Section~\ref{sec:conclusion}.

\section{Data}
\label{sec:data}

\begin{figure}
    \centering
    \includegraphics[width = \linewidth]{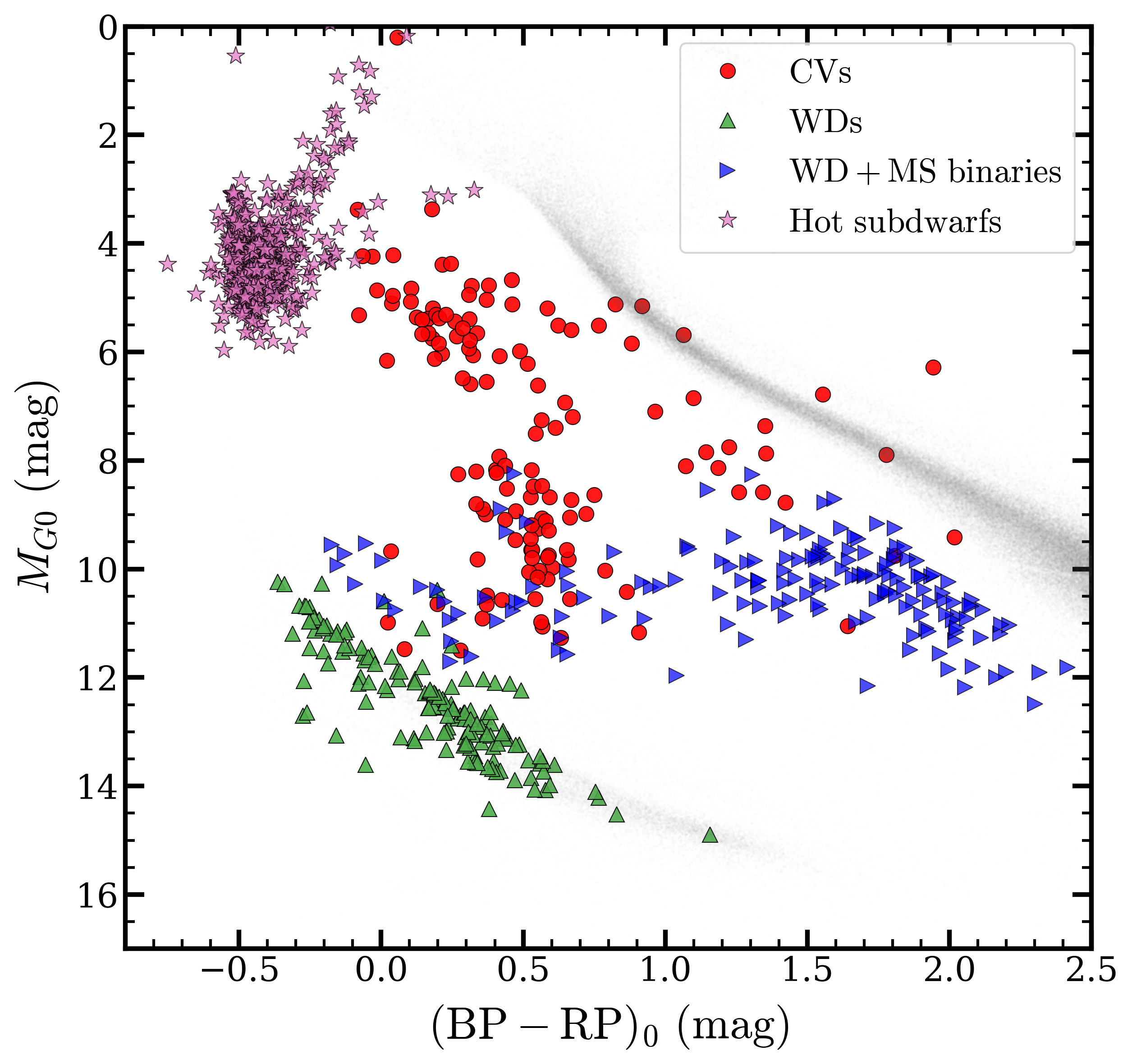}
    \caption{Major stellar populations distributed below the main sequence on the {\it Gaia} HR diagram. For illustrative purposes, {\it Gaia} objects within a 100~pc sample are shown in gray. Symbols correspond to different populations: CVs (red dots); hot subdwarfs (pink stars); WDs (green triangles); and WD+MS binaries (blue right triangles). These stellar populations were compiled from different catalogs (see Section~\ref{sec:data}).Only sources with prominent H$\alpha$ features (emission or absorption) in their {\it Gaia} XP spectra  are plotted (significance threshold of $|\text{pgEW}_{\rm H\alpha}/\text{pgEW}_{\rm err}| \ge 3$; see Section~\ref{sec:methods} for more details).}
    \label{fig:classes}
\end{figure}

The region below the main sequence on the {\it Gaia} HR diagram is primarily occupied by isolated hot subdwarfs, WDs, and WD+MS binaries (see Figure~\ref{fig:classes}).  None of these stars are expected to show prominent emission lines in the optical spectra, with the exception of active M dwarfs within WD+MS systems. Therefore, any source located below the main sequence on the {\it Gaia} HR diagram that demonstrates strong emission lines serves as a reliable CV candidate.

To characterize the {\it Gaia} XP spectral signatures of both the CVs and background stellar populations below the main sequence on the {\it Gaia} HR diagram, we downloaded catalogs of known systems and then cross-matched them with {\it Gaia} DR3 \citep{2023A&A...674A...1G}. We applied the following quality cuts to the {\it Gaia} data across all catalogs:
\begin{enumerate}
    \item $G<17.5$ mag
    \item $\tt parallax\_over\_error > 5$
    \item $\tt RUWE < 1.4$\footnote{We required a Renormalised Unit Weight Error ($\tt RUWE < 1.4$) to ensure good astrometric solutions \citep[e.g.,][]{2023A&A...674A...1G}.}
    \item $\tt has\_xp\_continuous=True$
\end{enumerate}
We describe each of these stellar populations in more detail below.

-- {\bf Cataclysmic variables.} We compiled a sample of known CVs from multiple catalogs: SDSS I--IV \citep{2023MNRAS.524.4867I}; LAMOST Data Release 11 (DR11) \citep{2012RAA....12.1197C,2026yCat.5162....0L}; the final version (December 31, 2015) of the Ritter and Kolb (R\&K) catalog \citep{2003A&A...404..301R}; and the catalog of polars, low-accretion rate polars (LARP), and candidate objects \citep[PolarCat;][]{2025A&A...698A.106S}. For the R\&K catalog, we kept only sources with a positional accuracy better than 2$\arcsec$, resulting in 1399 objects. We cross-matched this catalog with {\it Gaia} DR3 using a 2$\arcsec$ search radius, leaving 1313 objects. We then applied our {\it Gaia} quality cuts to all the CV catalogs. This reduced the sample size to 48 objects for SDSS I--IV, 124 for LAMOST DR11, 335 for the R\&K catalog, 35 for PolarCat polars, and 11 for PolarCat LARPs. We merged these catalogs to create a sample of unique CVs based on their {\it Gaia} source identification number, prioritizing the R\&K catalog if a source was duplicated in SDSS I--IV or LAMOST DR11. This merger resulted in 372 unique CVs. We defined five general CV classes by grouping specific catalog sub-types into broader categories. Specifically, dwarf novae were grouped into ``Dwarf Nova'' (139 objects), nova-like variables into ``Nova-like'' (133 objects), polars and intermediate polars into ``Magnetic CVs'' (49 objects), novae into ``Nova'' (11 objects), and LARPs into ``LARPs'' (11 objects). The remaining 29 objects without a specific classification were grouped as ``CVs''.

-- {\bf Hot subdwarfs.} We used the catalog of known hot subdwarfs from \citet{2022A&A...662A..40C}, which was compiled from {\it Gaia} DR3 and contains 6616 unique sources. First, we applied {\it Gaia} quality cuts, which reduced our sample size to 3614 objects. We then filtered this subset based on the provided SIMBAD spectral classifications by grouping related subtypes into broader categories. Specifically, the ``sdO'', ``sdOC'', ``sdOB'', ``He-sdO'', and ``sdBO'' classes were combined into a broad ``sdO'' class, while the ``sdB'', ``sdBC'', and ``sdBN'' labels were combined into a broad ``sdB'' class. Additionally, any composite spectral label containing a `+' symbol was grouped into the ``sd binary'' class. This classification yielded a final subset of 2412 hot subdwarfs, including 2198 isolated systems and 214 subdwarf binaries.

-- {\bf White dwarfs.} We used the 40 pc WD sample of \citet{2024MNRAS.527.8687O}, constructed from {\it Gaia} DR3 data. The sample contains 1076 spectroscopically confirmed WDs spanning various spectral classes. We applied the {\it Gaia} quality cuts, which reduced our sample to 876 sources. 

-- {\bf WD+MS binaries.} We used the catalog of WD + MS binaries from \cite{2025A&A...699A.153R}, which contains 1312 objects. We applied the {\it Gaia} quality cuts to this dataset, which reduced our sample to 829 sources.

For each of these catalogs, we downloaded the {\it Gaia} XP spectral coefficients \citep{2023A&A...674A...3M, 2023A&A...674A...2D} using the Python module {\tt astroquery}. For all objects, we computed their geometric distances based on parallaxes.  We then computed the extinction-corrected absolute $M_{G0}$ magnitudes and $(BP-RP)_0$ colors 
based on  the extinction value $E(B-V)$ from  a combination of three dust maps \citep[{\tt Combined19map};][]{2003A&A...409..205D,2006A&A...453..635M,2019ApJ...887...93G}, implemented in the {\tt mwdust}\footnote{\url{https://github.com/jobovy/mwdust}} package \citep{2016ApJ...818..130B}.

\newpage

\section{Methods}
\label{sec:methods}

\begin{figure}
    \centering
    \includegraphics[width=1\linewidth]{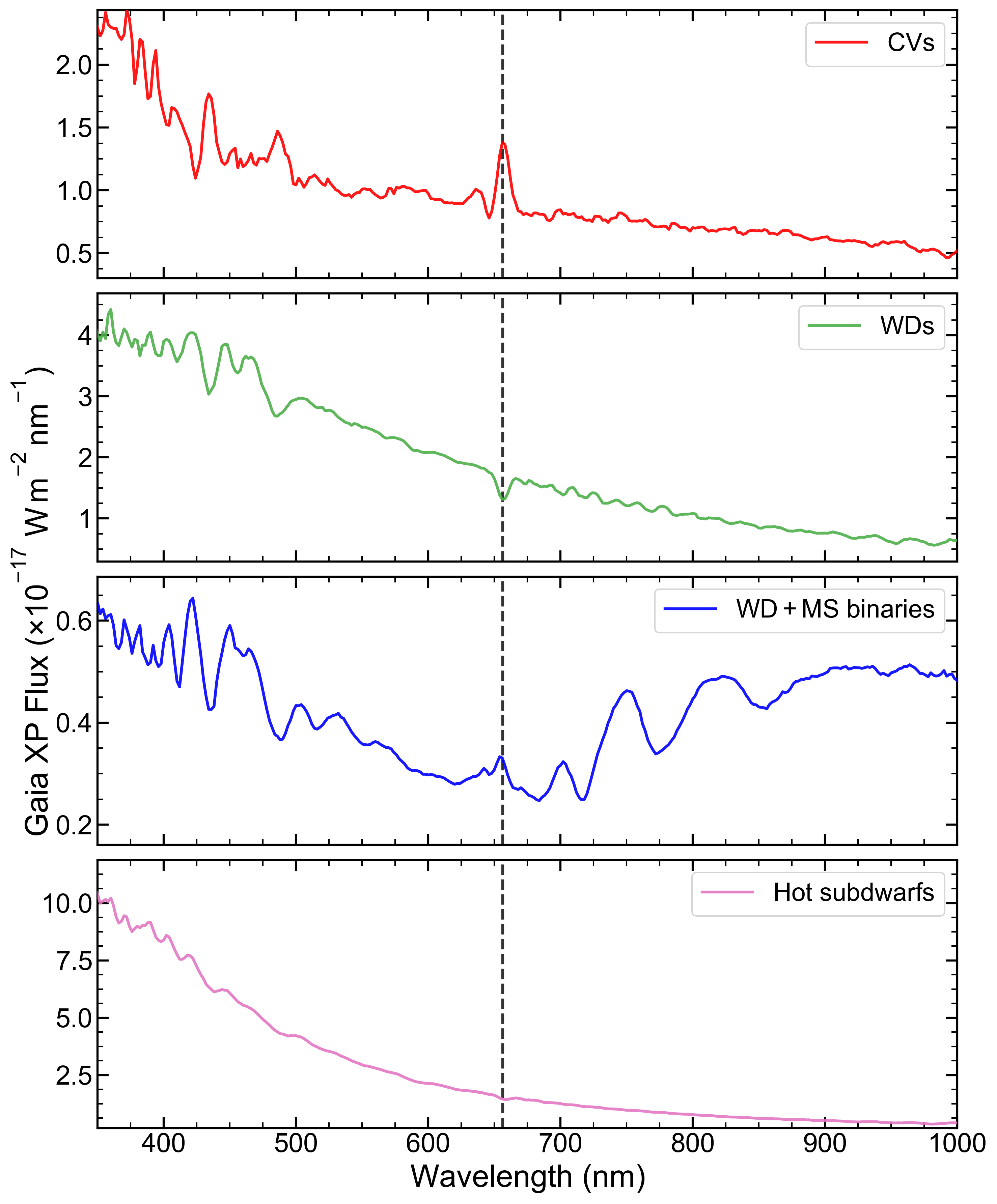}
    \caption{Median {\it Gaia} XP spectra for different stellar populations distributed below the main sequence on the {\it Gaia} HR diagram. From top to bottom, panels display: CVs; WDs; WD+MS binaries; and hot subdwarfs. The position of H$\alpha$ is indicated by a vertical dashed line. Only objects where the H$\alpha$ line was identified were used to compute the median spectra. A prominent H$\alpha$ emission line is visible for CVs, whereas absorption lines are present in the WDs and hot subdwarfs. The H$\alpha$  emission feature in WD+MS binaries likely results from chromospheric emission, pseudo-emission artifacts of TiO molecular bands from the M-dwarf companion in the low-resolution {\it Gaia}~XP spectra, or a combination of both.
 }
    \label{fig:spectra_classes}
\end{figure}

\subsection{Line search in Gaia XP spectra}

We used the Python package {\tt GaiaXPy}\footnote{\url{https://gaia-dpci.github.io/GaiaXPy-website/}} v2.1.4 \citep{2023A&A...674A..33G} to sample the {\it Gaia} XP spectra from their continuous representation onto a discrete wavelength grid. To search for emission or absorption lines, we used the {\tt find\_extrema} tool\footnote{\url{https://gaia-dpci.github.io/GaiaXPy-website/tutorials/Linefinder\%20tutorial.html}}
\citep{2023A&A...671A..52W}. This tool provides a list of all detected extrema and their properties within the internally calibrated mean {\it Gaia} XP spectra. The extrema are identified where the first derivative of the spectrum is equal to zero.  For each detected extremum, the tool provides a line identifier ({\tt line\_name} parameter) based on the spectral band (BP or RP) and the pseudo-wavelength. The wavelength at the extremum ({\tt wavelength} parameter in units of nm) is calculated using the {\tt GaiaXPy} dispersion function, and the flux value ({\tt line\_flux} parameter in units of $\text{W~nm}^{-1}~\text{m}^{-2}$) is measured at that position. The inflection points are found as the closest roots of the second derivative of the spectrum. These inflection points are then used to find the local line width ({\tt width} parameter in units of nm) and to estimate the local continuum. Finally, the tool gives the line depth ({\tt depth} parameter in units of $\text{W~nm}^{-1}~\text{m}^{-2}$), which is the difference between the line flux and the estimated local continuum. The significance in the internally calibrated {\it Gaia} XP spectrum ({\tt sig\_pwl} parameter) and in the externally calibrated spectrum ({\tt significance} parameter)  are calculated as the ratio of the line depth to the flux error at the extremum.

Figure~\ref{fig:spectra_classes} shows representative median {\it Gaia}~XP spectra for the main classes of objects considered in this work. The optical spectra of CVs are typically characterized by a blue continuum along with strong Balmer, He\,I and He\,II emission lines arising from the accretion flow. In non-magnetic CVs, these emission lines often exhibit double-peaked profiles caused by Keplerian motion of matter in the accretion disk \citep[e.g.,][]{1981AcA....31..395S, 1986MNRAS.218..761H}. These double-peaked profiles are typically observed in high-inclination systems viewed near edge-on. While these structures are clearly visible in high-resolution spectra with peak separations of a few nanometers, the low-resolution {\it Gaia}~XP spectra \citep[see, e.g., $\lambda/\Delta\lambda \sim 25\text{--}100$;][]{2021A&A...652A..86C} are insufficient to resolve them. The {\it Gaia} XP spectra of WDs and hot subdwarfs generally exhibit blue continua similar to those observed in CVs, but show an absence of emission lines. The median {\it Gaia}~XP spectrum of WD+MS binaries shows prominent TiO molecular absorption bands arising from the M-dwarf companion. This creates pseudo-emission lines, specifically at wavelengths $\lambda \gtrsim 600$~nm due to {\it Gaia}'s low spectral resolution. Chromospherically active M~dwarfs can also exhibit true H$\alpha$ emission, which complicates the {\it Gaia}~XP spectra of WD+MS binaries. An M-dwarf companion can also be present in CVs, where the donor star contributes to the red part of the spectrum, showing pseudo-emission lines caused by molecular absorption bands. However, the relative strength of the M-dwarf companion and the corresponding spectral features depend on the accretion rate, inclination angle of the system, magnetic field strength, and the relative contributions of the donor star and WD to the observed spectrum. Both the median {\it Gaia}~XP spectrum of known CVs and that of the background stellar population show a smooth oscillatory structure. This structure arises from the low-resolution reconstruction and calibration procedure used to generate the {\it Gaia} XP spectra \citep{2023A&A...674A...2D,2023A&A...674A...3M}. These features can complicate the identification of weak spectral lines and may introduce spurious line-like signatures. For this reason, we focus our analysis on the H$\alpha$ line, where this feature in CVs is sufficiently strong to be detected with high significance.

To find the H$\alpha$ line among all extrema identified by the {\tt find\_extrema} tool, we cross-matched the measured {\tt wavelength} of each detected extremum with the expected wavelength of the H$\alpha$ line (656.3 nm). We selected the extrema that fell within a search window of 5 nm ($|\lambda_{\rm line} - {\rm 656.3}| \le 5$~nm) as the true H$\alpha$ line. For multiple matches in the search window caused by spurious, line-like oscillatory structures from {\it Gaia} XP spectra, we selected the extremum with the highest {\tt sig\_pwl} parameter, representing the highest line significance in the pseudo-wavelength space. We chose this relatively wide 5~nm search window to account for the high velocities and broad line profiles typical of CV accretion disks (up to $\sim$2200~$\text{km~s}^{-1}$). This window includes the extrema of most known CVs and the background stellar population used in this work (see Appendix~\ref{app:lines}). We additionally filtered the extrema, keeping only those with a width of less than 15~nm (${\tt width} \le 15$~nm). This excludes broad features that are not narrow H$\alpha$ emission or absorption lines. By applying the 5~nm search window and width cut to identify the H$\alpha$ line in the {\it Gaia}~XP spectra, we obtained a final sample consisting of 245 CVs, 1029 hot subdwarfs, 264 WDs, and 361 WD+MS binaries.

\subsection{Pseudo-Gaussian equivalent width}

The equivalent width (EW) of a spectral line relative to the local continuum is defined as 

\begin{gather}
EW = \int \left(\frac{F(\lambda) - F_c(\lambda)}{F_c(\lambda)}\right) d\lambda \approx \frac{1}{F_c} \int  \Delta F(\lambda) d\lambda, 
\label{eq:ew}
\end{gather}

where $F(\lambda)$ is the total flux at wavelength $\lambda$, $F_c(\lambda)$ is the wavelength-dependent continuum, $F_c$ represents the constant mean local continuum flux, and $\Delta F(\lambda) = F(\lambda) - F_c$ is the continuum-subtracted line flux. Equation~\ref{eq:ew} shows that for an absorption line, the equivalent width is negative ($EW < 0$), while for an emission line, it is positive ($EW > 0$).

We introduce the pseudo-Gaussian equivalent width ($pgEW$) as a proxy to characterize the line properties in low-resolution {\it Gaia} XP spectra. Assuming a Gaussian line profile, the total area of the spectral line in this approach is given by 
\begin{gather}
 \int  \Delta F(\lambda) d\lambda = A \sigma \sqrt{2\pi},
\label{eq:gaussian}
\end{gather}
where $A$ is the maximum amplitude (depth or height) of the line, and $\sigma$ is the standard deviation (Gaussian width). Under this Gaussian line profile approximation, the pseudo-Gaussian equivalent width is defined as
\begin{gather}
pgEW = \frac{A \times \sigma \times \sqrt{2\pi}}{F_c}.
\label{eq:gew}
\end{gather}
While standard equivalent width measurements typically require high-resolution spectroscopy, the $pgEW$ framework provides a fast, automated, and self-consistent proxy for line strength in low-resolution {\it Gaia} XP spectra. It is calculated directly from the native outputs of the {\tt find\_extrema} pipeline as:
\begin{gather}
pgEW= \frac{1.25 \times {\tt depth} \times {\tt width}}{{\tt line\_flux} - {\tt depth}}, 
\label{eq:pgew}
\end{gather}
where the distance between the inflection points for a Gaussian line profile is $2\sigma$, corresponding to  ${\tt width} = 2\sigma$. The factor of $1.25$ derives from the numerical value of $\sqrt{2\pi}/2 \approx 1.25$. The error of the pseudo-Gaussian equivalent width ($pgEW_{\rm err}$) is calculated using the line significance given by the {\tt find\_extrema} tool, and is given by:
\begin{gather}
pgEW_{\rm err} = | pgEW| \times \left( \frac{1}{{\tt significance}} \right).
\label{eq:pgew_err}
\end{gather}

We used the pseudo-Gaussian equivalent width as a proxy for the H$\alpha$ line properties  ($\mathrm{pgEW}_{\mathrm{H}\alpha}$). We note that the $\mathrm{pgEW}_{\mathrm{H}\alpha}$ value will differ significantly from a standard\footnote{While the standard equivalent width of an emission line is negative, $\text{pgEW}$ is defined to be positive for emission features for simplicity.} $\mathrm{EW}$ measured through high-resolution spectroscopy, primarily due to the low resolution of the {\it Gaia} XP spectra and our Gaussian line profile with constant continuum assumption. However, since  $\mathrm{pgEW}_{\mathrm{H}\alpha}$ is calculated uniformly across all objects, it serves as a consistent baseline for our target selection methodology. We applied a quality cut to the $\mathrm{pgEW}_{\mathrm{H}\alpha}$ parameter, requiring a significance threshold of $|\mathrm{pgEW}_{\mathrm{H}\alpha}/\mathrm{pgEW}_{\mathrm{err}}| \ge 3$. This resulted in a final sample of 137 CVs, 637 hot subdwarfs, 142 WDs, and 159 WD+MS binaries (see Figure ~\ref{fig:classes}). 

\section{Results and Discussion}
\label{sec:results}

\begin{figure*}
    \centering
    \resizebox{\linewidth}{!}{
        \includegraphics[height=10cm]{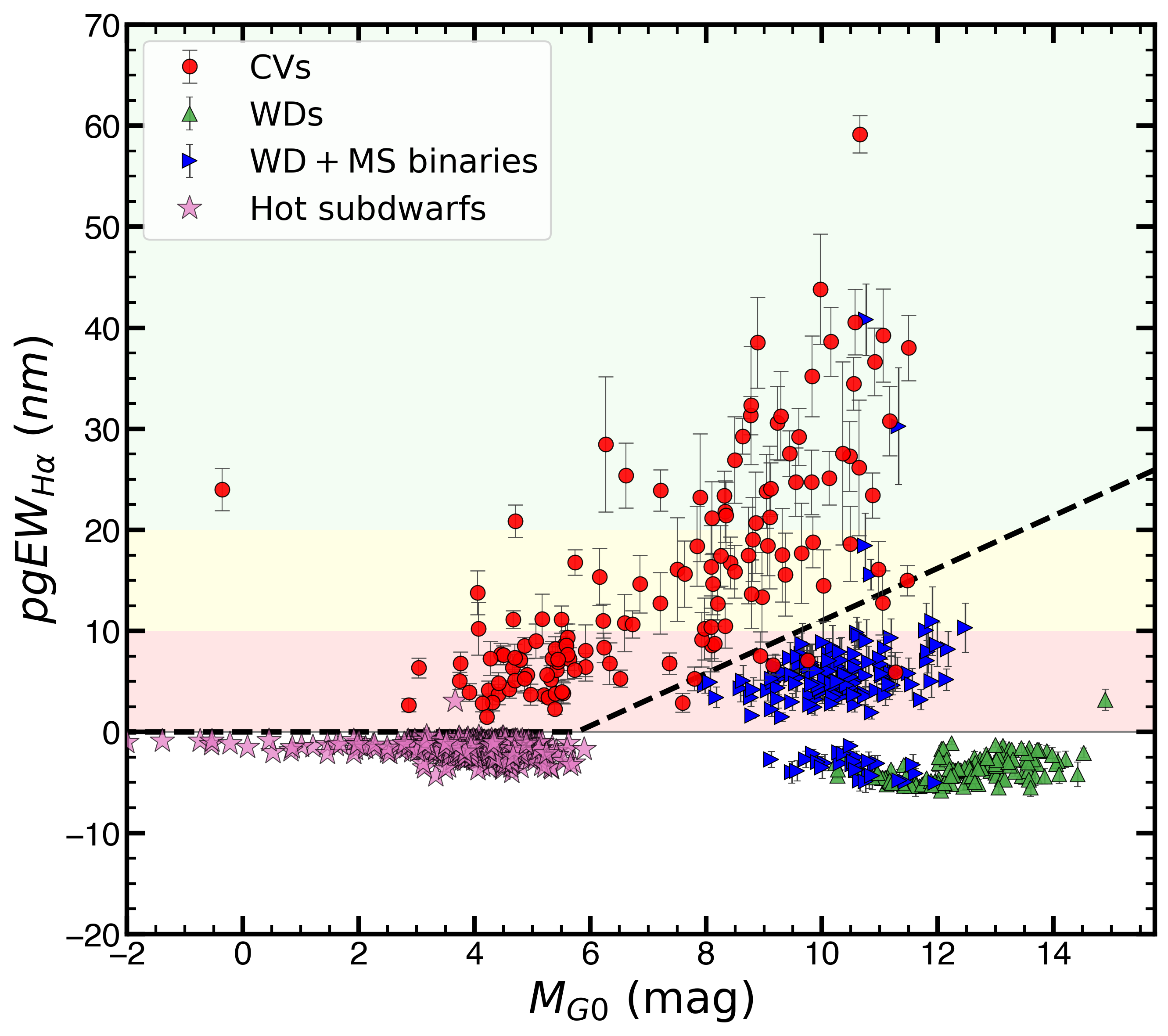}
        \includegraphics[height=10cm]{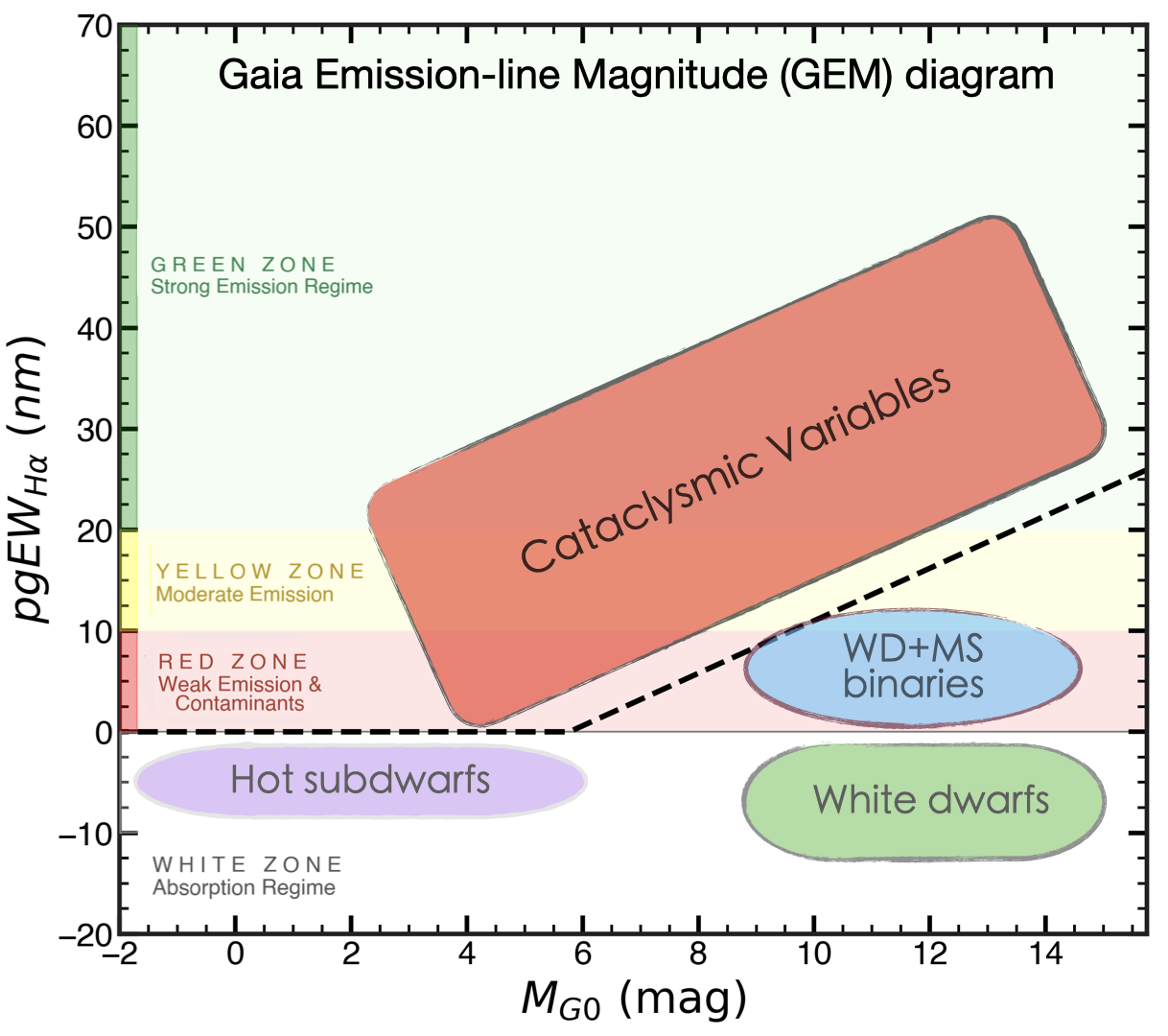}
        }
    \caption{GEM diagram for objects distributed below the main sequence on the {\it Gaia} HR diagram. The horizontal and vertical axes represent the absolute magnitude ($M_{G0}$) and the pseudo-Gaussian equivalent width ($pgEW_{\rm H\alpha}$), respectively. Left: Distribution of different stellar populations on the GEM diagram. The symbols for different populations are the same as in Figure~\ref{fig:classes}. Right: Schematic view of the GEM diagram. The dashed line shows the empirical threshold used to select CV candidates (see Equation~\ref{eq:selection}).}
    \label{fig:gem}
\end{figure*}

\begin{figure}
    \centering
    \includegraphics[width=1\linewidth]{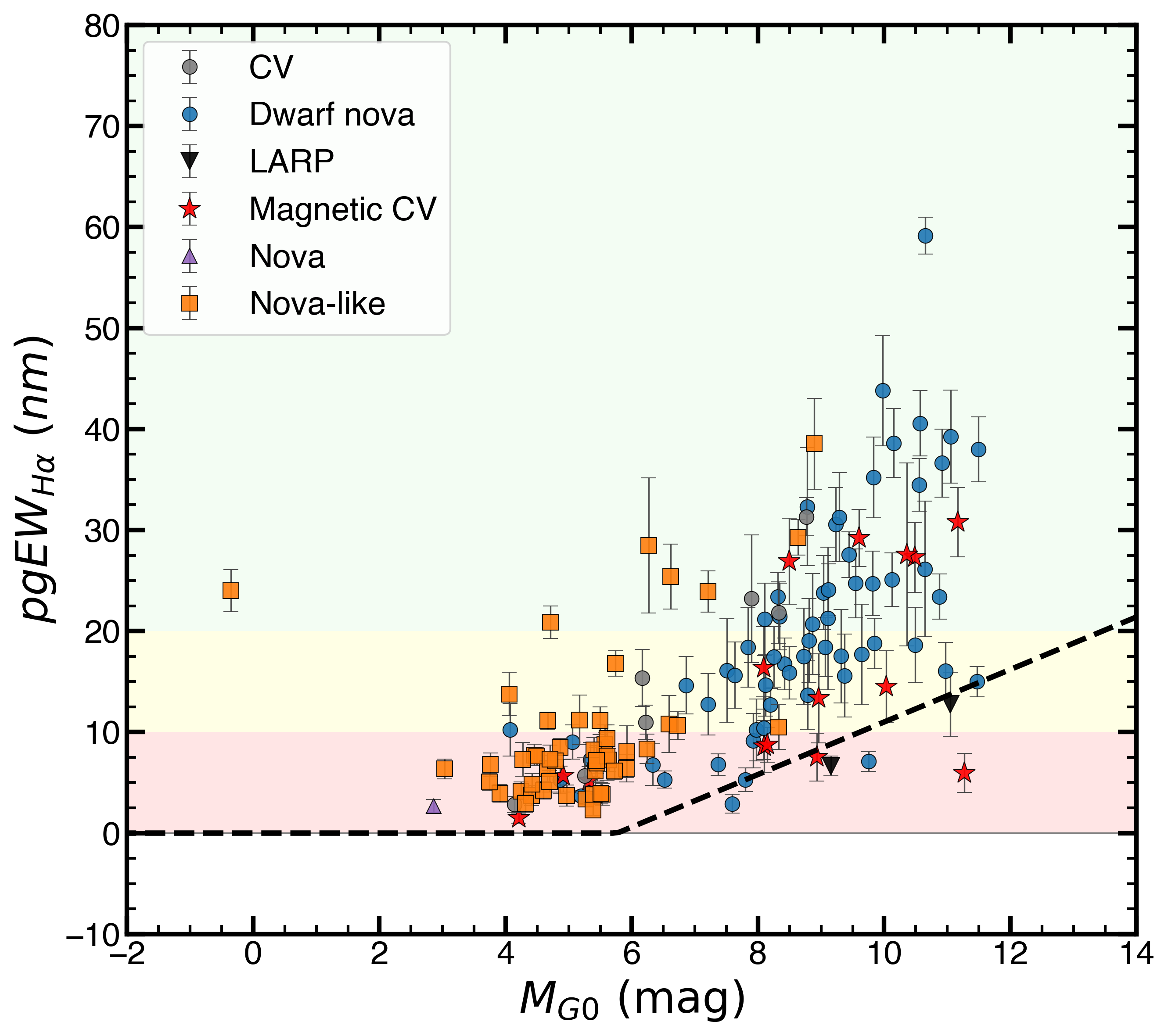}
    \caption{GEM diagram for known CVs, categorized by their respective classes.  Symbols correspond to: dwarf novae (blue dots); LARPs (black downward-pointing triangles); magnetic CVs (red stars); novae (purple upward-pointing triangles); and nova-like variables (orange squares).}
    \label{fig:pgews_cvs}
\end{figure}

\subsection{Gaia emission-line magnitude (GEM) diagram}

We introduce the Gaia Emission-line Magnitude (GEM) diagram as a diagnostic tool that combines {\it Gaia} colors with spectroscopic information from {\it Gaia}~XP spectra to isolate CV candidates. This diagram plots the H$\alpha$ line proxy, $\mathrm{pgEW}_{\mathrm{H}\alpha}$, against the absolute {\it Gaia} $G$-band magnitude ($M_{G0}$). Figure~\ref{fig:gem} shows the GEM diagram for the known CVs along with the background stellar populations, including hot subdwarfs, WDs, and WD+MS binaries. These sources occupy distinct regions on the GEM diagram based on their physical properties and {\it Gaia}~XP spectral features:

-- \textbf{Cataclysmic variables.} CVs are distributed across the GEM diagram in the region with $pgEW_{\rm H\alpha} \gtrsim 0$~nm and $M_{G0} \in [-1, 12]$~mag. Figure~\ref{fig:pgews_cvs} shows the GEM diagram specifically for the different types of CVs (137 objects in total): 60 dwarf novae, 50 nova-like variables, 16 magnetic CVs, 2 LARPs, 1 nova, and 8 unclassified CVs. Two distinct CV populations (nova-like variables and dwarf novae) are visible on the GEM diagram, reflecting fundamental differences in their underlying physics. The majority of nova-like variables clump within $0 \lesssim pgEW_{\rm H\alpha} \lesssim 10$~nm and $M_{G0} \in [4, 6]$~mag. Because the optical spectra of nova-like variables are dominated by intense continuum emission from a hot, steady-state accretion disk, their H$\alpha$ emission lines appear characteristically weak or diluted compared to dwarf novae. In contrast, the dwarf novae occupy a much broader, fainter regime where $pgEW_{\rm H\alpha} \gtrsim 10$~nm and $M_{G0} \in [6, 12]$~mag. Within this population, the $pgEW_{\rm H\alpha}$ increases systematically as the system brightness drops ($M_{G0}$ increases). This trend continues toward $M_{G0} \approx 10$~mag, where the {\it Gaia}~XP spectra of dwarf novae show strong, prominent H$\alpha$ emission lines. The distribution of nova-like variables and dwarf novae on the GEM diagram traces the physical transition from disk-dominated light to the line-dominated emission typical of cooler, low-mass transfer systems. Furthermore, CVs are highly variable objects, meaning individual sources can change their positions on the diagram. Specifically, dwarf novae exhibit state transitions that can cause a system to temporarily show H$\alpha$ absorption lines ($pgEW_{\rm H\alpha} < 0$), shifting its position down into the hot subdwarf or WD regions. While the majority of known CVs cluster in their unique region, several known CVs are distributed across the WD+MS binary region on the GEM diagram. Their \textit{Gaia}~XP spectra show M-dwarf donor signatures along with a weak H$\alpha$ emission line. Notably, the two LARPs in our sample are distributed close to the WD+MS binary region. The magnetic CVs in our sample do not cluster tightly on the GEM diagram, being instead distributed across the nova-like variable, dwarf nova, and WD+MS binary regions.

-- \textbf{Hot subdwarfs.} Hot subdwarfs show strong H$\alpha$ absorption lines, primarily falling in the region $pgEW_{\rm H\alpha} \lesssim 0$~nm with absolute magnitudes in the range of $M_{G0} \in [-2, 6]$~mag. A few hot subdwarfs show weak positive values ($pgEW_{\rm H\alpha} \sim 2$~nm). This artifact is not caused by real emission lines, but rather by systematic continuum fluctuations in the {\it Gaia}~XP spectra.

-- \textbf{White dwarfs.} WDs show strong H$\alpha$ absorption lines, primarily falling in the region $pgEW_{\rm H\alpha} \lesssim 0$~nm with absolute magnitudes in the range of $M_{G0} \in [10, 15]$~mag. One object shows a weak positive value ($pgEW_{\rm H\alpha} \sim 2$~nm). This artifact is not caused by real emission lines, but rather by systematic continuum fluctuations in the {\it Gaia}~XP spectra.

-- \textbf{WD+MS binaries.} This population splits into two distinct groups on the GEM diagram. The first group has $pgEW_{\rm H\alpha} \lesssim 0$~nm and $M_{G0} \in [9, 13]$~mag, where the WD flux dominates the emission and H$\alpha$ absorption lines are seen in the {\it Gaia}~XP spectra. The second group forms a separate population with $0 \lesssim pgEW_{\rm H\alpha} \lesssim 10$~nm and $M_{G0} \in [9, 13]$~mag. These positive pseudo-emission lines are primarily caused by TiO molecular absorption features in the M-dwarf companion, which mimic an H$\alpha$ emission line in the low-resolution {\it Gaia}~XP spectra. However, a subset of WD+MS binaries within this region can still exhibit true H$\alpha$ emission lines due to the chromospheric activity of the companion. A few objects in the WD+MS binary catalog have $pgEW_{\rm H\alpha} \gtrsim 15$~nm and are located in the CV region\footnote{Two of these sources, Gaia~DR3~6097584212806297984 ($pgEW_{\rm H\alpha} \approx 18$~nm) and Gaia~DR3~5890007786261623680 ($pgEW_{\rm H\alpha} \approx 16$~nm), show M-dwarf signatures in their {\it Gaia}~XP spectra and are classified as emission-line stars in the SIMBAD database, though a CV nature cannot be definitively ruled out.}. Notably, one source, Gaia~DR3~1951666884868218112 ($pgEW_{\rm H\alpha} \approx 41$~nm), is a known CV \citep[TCP J21290156+3631056;][]{2023AJ....165..163C}, while the other source, Gaia~DR3~3021820276571880064 ($pgEW_{\rm H\alpha} \approx 30$~nm), was independently identified by our GEM diagram as a new CV candidate in the {\it Gaia} 250~pc sample (see Appendix~\ref{app:identification}).

Based on the distribution of sources in the GEM diagram, we define a simple, phenomenological zonation using the $pgEW_{\rm H\alpha}$ parameter space. This approach provides a practical framework for identifying the dominant source types across different emission regimes:

\begin{itemize}
    \item \textbf{White Zone} ($pgEW_{\rm H\alpha} \lesssim 0$~nm) -- The Absorption Regime: This region is dominated by sources showing H$\alpha$ absorption lines, such as WDs, hot subdwarfs, and a fraction of the WD+MS binaries.
    
    \item \textbf{Red Zone} ($0 \lesssim pgEW_{\rm H\alpha} \lesssim 10$~nm) -- The Weak Emission/Contaminant Regime: This intermediate zone hosts a complex mixture of sources. It contains the majority of bright nova-like variables, some of dwarf novae and magnetic CVs. However, this zone is also heavily contaminated at fainter magnitudes by WD+MS binaries.
    
    \item \textbf{Yellow Zone} ($10 \lesssim pgEW_{\rm H\alpha} \lesssim 20$~nm) -- The Moderate Emission Regime: This transitional zone contains CVs alongside a small number of WD+MS binaries. 
    
    \item \textbf{Green Zone} ($pgEW_{\rm H\alpha} \gtrsim 20$~nm) -- The Strong Emission Regime: This high-purity region contains objects showing prominent H$\alpha$ emission lines. This zone is dominated by CVs, specifically dwarf novae. Contamination here is minimal, making any source in this zone a strong CV candidate.
\end{itemize}

While the flat color zones provide a useful descriptive overview of the GEM diagram, using them for target selection can lead to either missing true CVs or including a significant number of WD+MS binaries. We empirically define a boundary (a broken line) in the GEM diagram to capture the majority of true CV candidates while filtering out contaminants. This empirical selection line is defined as:
\begin{equation}
pgEW_{\rm H\alpha}^{\rm cut} = 
\begin{cases} 
2.6 \times M_{G0} - 15, & \text{for } M_{G0} \ge 5.8 \\
0, & \text{for } M_{G0} < 5.8 
\end{cases}
\label{eq:selection}
\end{equation}
The first condition in Equation~\ref{eq:selection} for $M_{G0} \ge 5.8$ isolates the majority of CVs from WD+MS binaries, while the second condition for $M_{G0} < 5.8$ includes nova-like variables and separates them from the majority of hot subdwarfs. A source is subsequently classified as a CV candidate if it lies above this boundary. This empirical boundary is shown as a dashed line in Figures~\ref{fig:gem} and~\ref{fig:pgews_cvs}.

We note that while the current implementation of the GEM diagram focuses explicitly on H$\alpha$ features (in either emission or absorption), the framework itself can be extended to other prominent lines. Due to this hydrogen-focused selection, our pipeline is generally insensitive to helium-dominated, ultra-compact AM~CVn binaries, which typically show strong helium features without detectable hydrogen lines. However, some short-period AM~CVn systems or systems in high mass-transfer states exhibit a prominent He~II Pickering line ($\rm 656.0\,nm$ in emission), which closely matches the wavelength of the H$\alpha$ feature \citep[e.g.,][]{2019MNRAS.485.1947G,  2021A&A...645A.114B, 2026A&A...713A.326G}. Consequently, the GEM diagram framework remains capable of detecting such helium-rich systems via this spectral overlap.

\subsection{Proof of concept: Application to the Gaia 250~pc sample}
\label{sec:250pc}

\begin{figure*}
    \centering
    \resizebox{\linewidth}{!}{%
        \includegraphics[height=10cm]{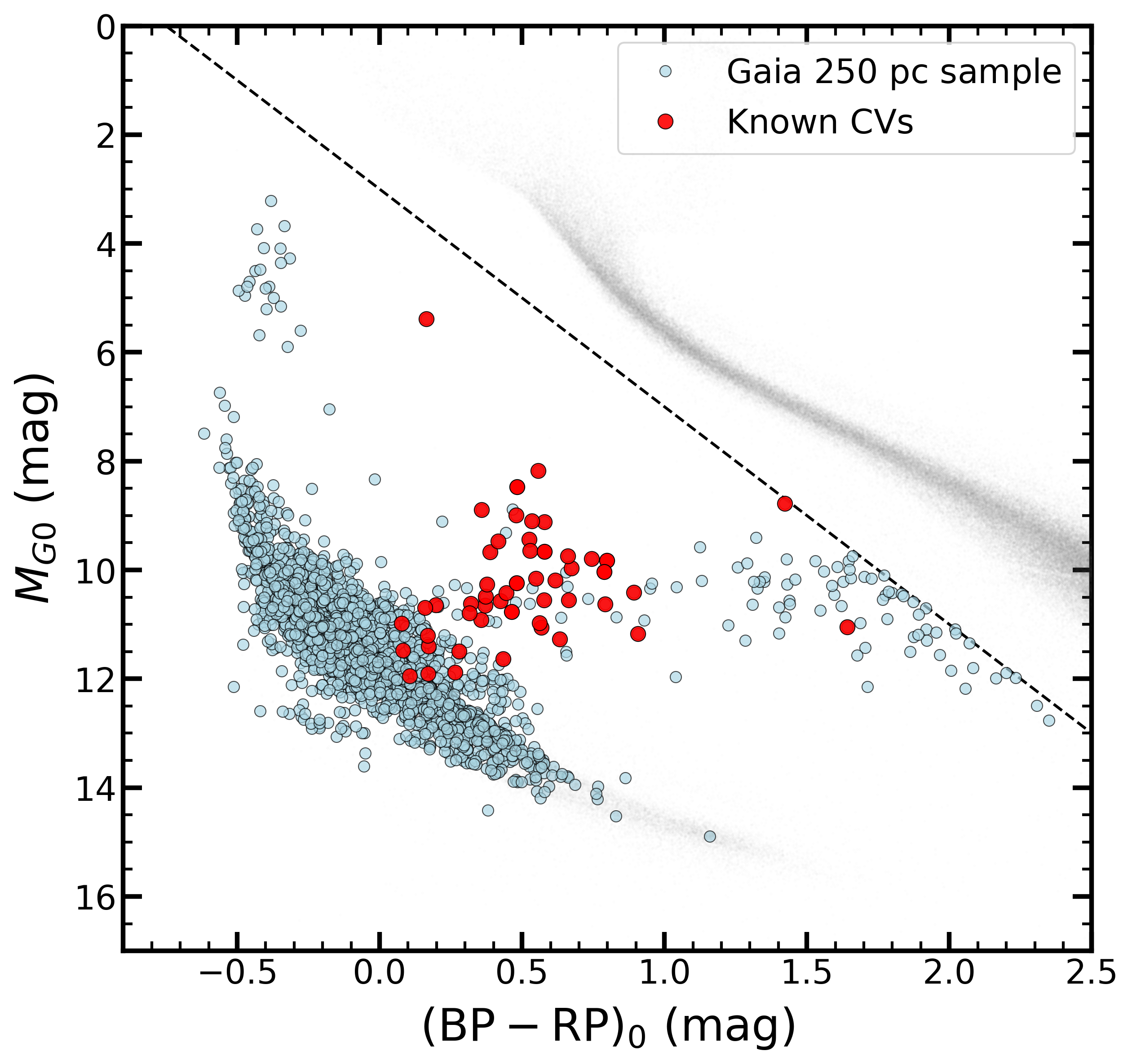}%
        \includegraphics[height=10cm]{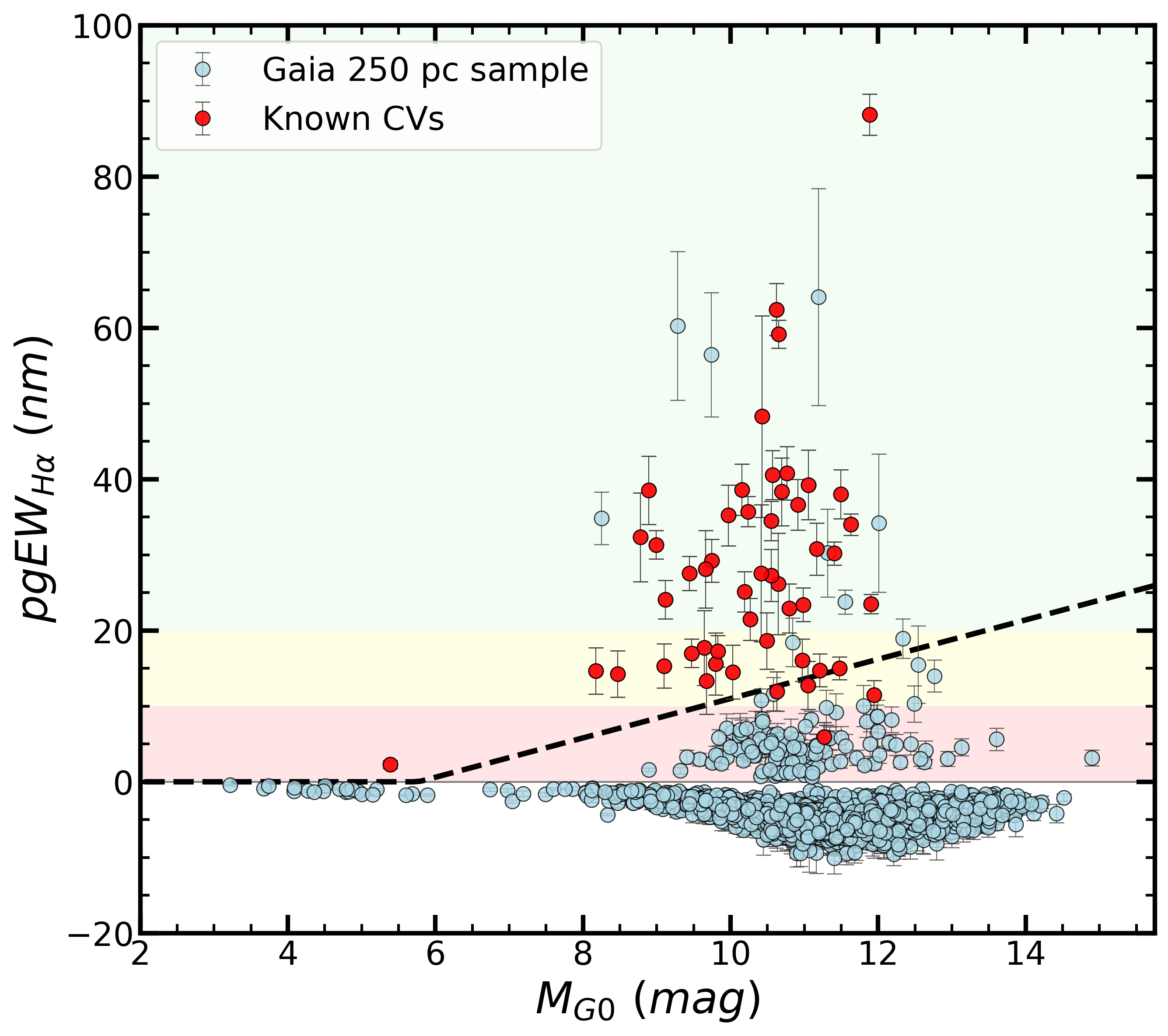}%
        }
    \caption{A proof of concept using {\it Gaia} objects within a 250~pc sample distributed below the main sequence (4207 objects with detected H$\alpha$ lines). Left: {\it Gaia} HR diagram for objects within the 250~pc sample. The dashed line shows the cut used to query {\it Gaia} data below the main sequence. For illustrative purposes, the {\it Gaia} 100~pc sample is shown in gray. Right: GEM diagram. Known CVs within the {\it Gaia} 250~pc sample are shown in red. Two known CVs that fall below the empirical selection cut are included (see Appendix~\ref{app:identification} for more details). The majority of these known CVs are distributed above the empirical selection line of the GEM diagram, demonstrating a clear separation between the CV population and the background stellar population. }
    \label{fig:250pc}
\end{figure*}

\begin{figure}
    \centering
    \includegraphics[width=1\linewidth]{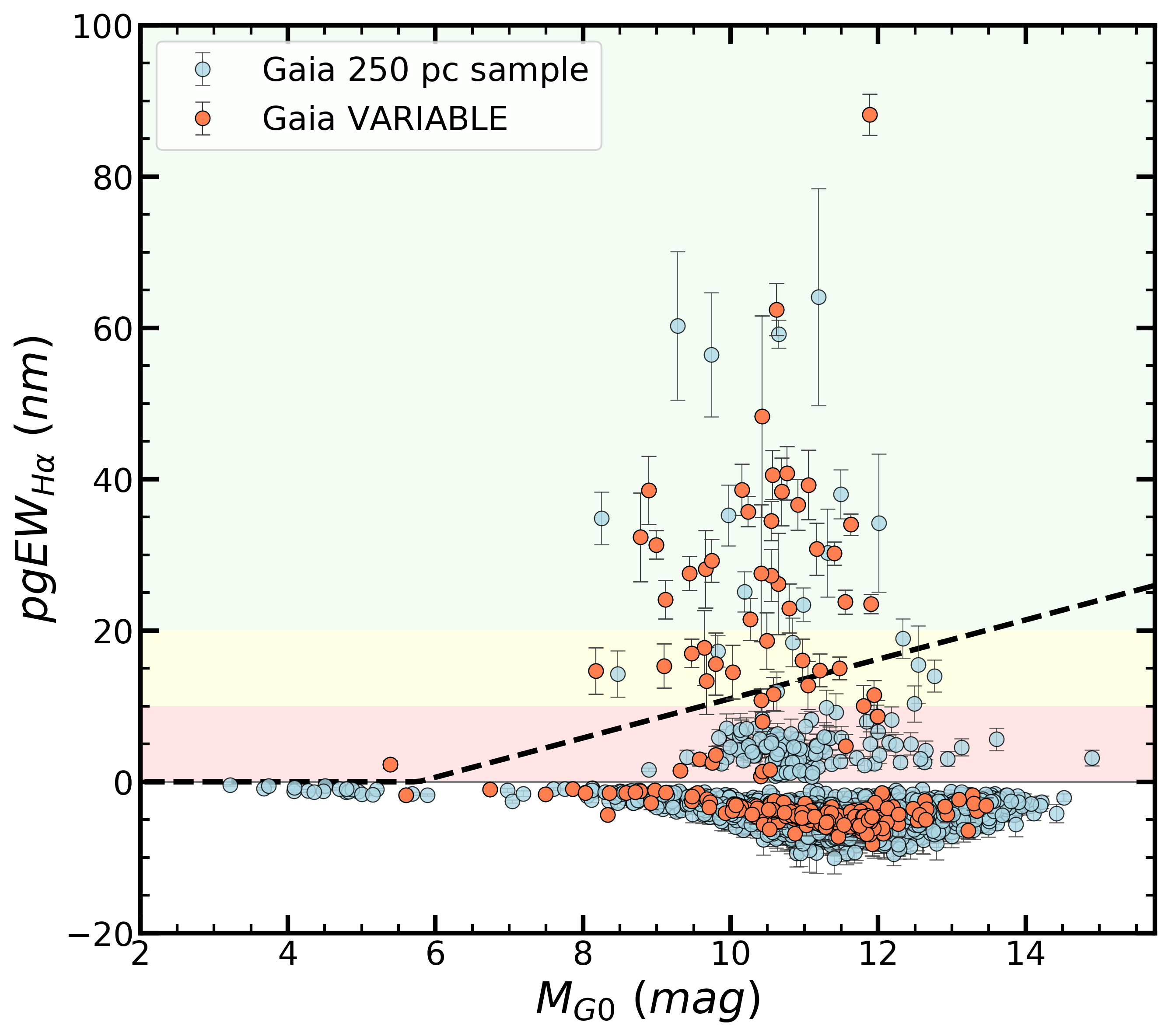}
    \caption{GEM diagram for {\it Gaia} objects within the 250~pc sample. Orange dots show variable objects identified by {\it Gaia}. Objects located above the empirical selection cut correspond to both known CVs and new CV candidates.}
    \label{fig:250pcvariable}
\end{figure}

To probe how the GEM diagram isolates CV candidates from stellar contaminants during a blind search, we constructed a {\it Gaia} volume-limited sample within 250~pc. The following criteria were applied to define this sample:
\begin{enumerate}
    \item $\tt parallax > 4.0$ mas
    \item $\tt parallax\_over\_error > 3$
    \item $\tt phot\_g\_mean\_flux\_over\_error  > 3$
    \item $\tt phot\_bp\_mean\_flux\_over\_error  > 3$
    \item $\tt phot\_rp\_mean\_flux\_over\_error  > 3$
    \item $ G<17.5$ mag
    \item $\tt RUWE < 1.4$
    \item $\tt has\_xp\_continuous=True$
\end{enumerate}
To keep only sources located below the main sequence, we applied the following color-magnitude cut:
\begin{enumerate}
    \item $\tt  -2\ <\ (BP-RP)\ < 3  $
    \item $\tt M_G > 3 + 4 \times (BP-RP)$ 
\end{enumerate}
This query  results in 12247 sources, from which we detected and identified the H$\alpha$ line in the {\it Gaia}~XP spectra of 6728 objects. Filtering these targets with a significance threshold of $|\text{pgEW}_{\rm H\alpha}/\text{pgEW}_{\rm err}| \ge 3$ yields a final sample of 4207 sources. Figure~\ref{fig:250pc} shows the {\it Gaia}~HR diagram within the 250~pc volume for the 4207 objects with detected H$\alpha$ lines (left panel) along with their corresponding positions on the GEM diagram (right panel). The distribution of sources on the GEM diagram follows a pattern similar to that in Figure~\ref{fig:gem}, clearly mapping out the expected populations of CVs, hot subdwarfs, WD+MS binaries, and WDs.

To select CV candidates out of the 4207 sources with detected H$\alpha$ features, we applied our empirical selection cut from Equation~\ref{eq:selection} using the criterion $(\text{pgEW}_{\rm H\alpha} + \text{pgEW}_{\rm err}) \ge \text{pgEW}_{\rm H\alpha}^{\rm cut}$, which resulted in a sample of 62 objects. We cross-matched these 62 candidates with the catalogs of known CVs described in Section~\ref{sec:data}. Additionally, we queried the SIMBAD database using a $2\arcsec$ search radius to verify their classifications. We found 43 known CVs and 19 sources with different SIMBAD classifications. A manual inspection of these 19 objects using the VizieR database revealed five additional known CVs. In total, out of our 62 CV candidates, we identified 48 known CVs, 11 contaminants with prominent H$\alpha$ emission lines (including young stellar objects, emission-line stars, and WD+MS binaries), and three new CV candidates: Gaia~DR3 227766037015779200, Gaia~DR3 3021820276571880064, and Gaia~DR3 3446531278731142912. The detailed source identification for the 62 CV candidates and the individual properties of the three new CV candidates are presented in Appendix~\ref{app:identification}. We note that emission-line stars or young stellar objects are generally not expected to contaminate the GEM diagram, as the majority of these populations reside on or above the main sequence on the {\it Gaia}~HR diagram. However, due to peculiar colors or extreme properties (for example viewing through high inclination edge-on disks), a small number of these objects can scatter below the main sequence and fall into the CV zone on the GEM diagram. 

Figure~\ref{fig:250pcvariable} shows the GEM diagram for the {\it Gaia} 250~pc sample along with variable objects identified by {\it Gaia} via the {\tt phot\_variable\_flag = VARIABLE} criterion. Out of the 62 CV candidates, 44 exhibit variability. Some of the objects distributed within the WD+MS binary region also show variability, which is likely driven by ellipsoidal modulations or reflection effects. Objects within the WD and hot subdwarf regions are variable as well, primarily due to stellar pulsations. Combining this variability information with the GEM diagram classification provides a powerful tool for identifying true variable CV candidates.

Out of the entire {\it Gaia} 250~pc sample (12247 objects that have passed our selection criteria above and are distributed below the main sequence), we identified 74 known CVs based on a cross-match with the SIMBAD database and our reference CV catalogs (see Appendix~\ref{app:identification}). This indicates an overall completeness of $\sim$65\% for {\it Gaia} 250~pc sample, driven primarily by the sensitivity of the H$\alpha$  line-detection step. Once an H$\alpha$ line is detected ($|\text{pgEW}_{\rm H\alpha}/\text{pgEW}_{\rm err}| \ge 3$), our empirical cut on the GEM diagram isolates the known CVs with a high efficiency of $\sim$96\%. We note that a detailed selection function and completeness analysis of the GEM diagram is beyond the scope of this work, as this {\it Gaia} 250~pc sample serves as an initial application of our diagnostic framework.

\section{Conclusion}
\label{sec:conclusion}

We present a diagnostic tool for stellar astronomy that uses {\it Gaia} spectroscopic and photometric data to isolate CV candidates from dominant background stellar populations distributed below the main sequence on the {\it Gaia}~HR diagram. By combining extinction-corrected absolute magnitudes ($M_{G0}$) with an H$\alpha$ line proxy, the pseudo-Gaussian equivalent width ($pgEW_{\mathrm{H}\alpha}$), we introduced the Gaia Emission-line Magnitude (GEM) diagram (see Figure~\ref{fig:gem}). The GEM diagram maps distinct stellar populations, including isolated WDs, hot subdwarfs, and WD+MS binaries, into unique regions of the parameter space based on their intrinsic spectrophotometric properties and {\it Gaia}~XP low-resolution line profiles. We isolate CVs from the dominant background stellar population by using an empirical target selection boundary on the GEM diagram (see Figure~\ref{fig:gem} and Equation~\ref{eq:selection}). This selection threshold filters out stellar contaminants, most notably WD+MS binaries where the low-resolution TiO molecular absorption bands from the M-dwarf companion create prominent absorption edges that mimic pseudo H$\alpha$ emission features in {\it Gaia} XP spectra.

As a proof of concept, we applied the GEM diagram to a volume-limited {\it Gaia} sample within $250~\mathrm{pc}$ ($G < 17.5~\mathrm{mag}$), targeting objects distributed below the main sequence on the {\it Gaia}~HR diagram (see Figure~\ref{fig:250pc}). We identified $62$ CV candidates above our empirical boundary. Based on cross-matches with the SIMBAD database, we recovered $48$ known CVs while missing only two previously confirmed systems. We identified three new CV candidates (Gaia DR3 227766037015779200, Gaia DR3 3021820276571880064, and Gaia DR3 3446531278731142912), whose accreting natures are independently supported by archival X-ray data (ROSAT, Swift, and SRG/eROSITA) and  optical variability in ZTF light curves. The remaining eleven objects are stellar contaminants, including emission-line stars, young stellar objects, and WD+MS binaries (see Appendix~\ref{app:identification} and Table~\ref{tab:objects}). Further spectroscopic follow-up observations of these sources are required to clarify their nature.

In conclusion, the GEM diagram is a promising tool for the initial classification of CV candidates using {\it Gaia} spectrophotometric data before performing ground-based spectroscopic observations. This approach provides a reliable framework to accelerate discoveries and expand our census of Galactic compact binaries using upcoming new data from {\it Gaia} in combination with the Rubin Observatory Legacy Survey of Space and Time (LSST) and other multi-wavelength missions. Furthermore, the GEM diagram can be applied to the entire {\it Gaia} dataset to help identify emission-line stars across the Milky Way.

\section*{Acknowledgments}
This work has made use of data from the European Space Agency (ESA) mission Gaia (https://www.cosmos.esa.int/gaia), processed by the Gaia Data Processing and Analysis Consortium (DPAC, https://www.cosmos.esa.int/web/gaia/dpac/consortium). Funding for the DPAC has been provided by national institutions, in particular the institutions participating in the Gaia Multilateral Agreement. This work has made use of the Python package GaiaXPy, developed and maintained by members of the Gaia Data Processing and Analysis Consortium (DPAC), and in particular, Coordination Unit 5 (CU5), and the Data Processing Centre located at the Institute of Astronomy, Cambridge, UK (DPCI). This research has made use of the VizieR catalogue access tool, CDS, Strasbourg, France \citep{10.26093/cds/vizier}. The original description of the VizieR service was published in \citet{vizier2000}. This research has made use of the SIMBAD database, operated at CDS, Strasbourg, France \citep{2000A&AS..143....9W}. We thank the anonymous referee for the insightful feedback and comments, which helped improve this manuscript.

\newpage
\bibliographystyle{apsrev4-1}
\bibliography{oja_template}

@ARTICLE{2019MNRAS.485.1947G,
       author = {{Green}, M.~J. and {Marsh}, T.~R. and {Steeghs}, D. and {Breedt}, E. and {Kupfer}, T. and {Rodr{\'\i}guez-Gil}, P. and {van Roestel}, J. and {Ashley}, R.~P. and {Wang}, L. and {Cukanovaite}, E. and {Outmani}, S.},
        title = "{Phase-resolved spectroscopy of Gaia14aae: line emission from near the white dwarf surface}",
      journal = {\mnras},
         year = 2019,
        month = may,
       volume = {485},
       number = {2},
        pages = {1947-1960},
          doi = {10.1093/mnras/stz469},
archivePrefix = {arXiv},
       eprint = {1902.01213},
 primaryClass = {astro-ph.SR},
       adsurl = {https://ui.adsabs.harvard.edu/abs/2019MNRAS.485.1947G}
}

@ARTICLE{2021A&A...645A.114B,
       author = {{Bakowska}, K. and {Marsh}, T.~R. and {Steeghs}, D. and {Nelemans}, G. and {Groot}, P.~J.},
        title = "{Spectroscopy of the helium-rich binary ES Ceti reveals accretion via a disc and evidence of eclipses}",
      journal = {\aap},
         year = 2021,
        month = jan,
       volume = {645},
          eid = {A114},
        pages = {A114},
          doi = {10.1051/0004-6361/202039266},
archivePrefix = {arXiv},
       eprint = {2011.09213},
 primaryClass = {astro-ph.SR},
       adsurl = {https://ui.adsabs.harvard.edu/abs/2021A&A...645A.114B}
}

@ARTICLE{2026A&A...713A.326G,
       author = {{Galiullin}, Ilkham and {Rodriguez}, Antonio C. and {El-Badry}, Kareem and {Suleimanov}, Valery and {Dodon}, Vladislav and {Sibgatullin}, Askar and {Brown}, Warren R. and {Burdge}, Kevin and {van Roestel}, Jan and {Berger}, Edo and et al.},
        title = "{A survey of ultra-compact high-state AM CVn binaries with ZTF and Gaia: New discoveries and observational constraints on the Galactic space density}",
      journal = {\aap},
         year = 2026,
        month = sep,
       volume = {713},
          eid = {A326},
        pages = {A326},
          doi = {10.1051/0004-6361/202660872},
archivePrefix = {arXiv},
       eprint = {2608.04179},
 primaryClass = {astro-ph.SR},
       adsurl = {https://scixplorer.org/abs/2026A&A...713A.326G}
}

@inproceedings{rodriguez2026emissiongaiaxp,
  title     = {From Active Stars to Black Holes II: Discovery of H-alpha Emission
               in Ultra-Low Resolution Gaia XP Spectra with Machine Learning},
  author    = {Rodriguez, Antonio C. and Villar, V. Ashley and Berger, Edo and Villa, Valeria},
  booktitle = {Proceedings of PAI 2026},
  year      = {2026}
}

@ARTICLE{2023A&A...674A..27A,
       author = {{Andrae}, R. and {Fouesneau}, M. and {Sordo}, R. and {Bailer-Jones}, C.~A.~L. and {Dharmawardena}, T.~E. and {Rybizki}, J. and {De Angeli}, F. and {Lindstr{\o}m}, H.~E.~P. and {Marshall}, D.~J. and {Drimmel}, R. and {Korn}, A.~J. and {Soubiran}, C. and {Brouillet}, N. and {Casamiquela}, L. and {Rix}, H.-W. and {Abreu Aramburu}, A. and {{\'A}lvarez}, M.~A. and {Bakker}, J. and {Bellas-Velidis}, I. and {Bijaoui}, A. and {Brugaletta}, E. and {Burlacu}, A. and {Carballo}, R. and {Chaoul}, L. and {Chiavassa}, A. and {Contursi}, G. and {Cooper}, W.~J. and {Creevey}, O.~L. and {Dafonte}, C. and {Dapergolas}, A. and {de Laverny}, P. and {Delchambre}, L. and {Demouchy}, C. and {Edvardsson}, B. and {Fr{\'e}mat}, Y. and {Garabato}, D. and {Garc{\'\i}a-Lario}, P. and {Garc{\'\i}a-Torres}, M. and {Gavel}, A. and {Gomez}, A. and {Gonz{\'a}lez-Santamar{\'\i}a}, I. and {Hatzidimitriou}, D. and {Heiter}, U. and {Jean-Antoine Piccolo}, A. and {Kontizas}, M. and {Kordopatis}, G. and {Lanzafame}, A.~C. and {Lebreton}, Y. and {Licata}, E.~L. and {Livanou}, E. and {Lobel}, A. and {Lorca}, A. and {Magdaleno Romeo}, A. and {Manteiga}, M. and {Marocco}, F. and {Mary}, N. and {Nicolas}, C. and {Ordenovic}, C. and {Pailler}, F. and {Palicio}, P.~A. and {Pallas-Quintela}, L. and {Panem}, C. and {Pichon}, B. and {Poggio}, E. and {Recio-Blanco}, A. and {Riclet}, F. and {Robin}, C. and {Santove{\~n}a}, R. and {Sarro}, L.~M. and {Schultheis}, M.~S. and {Segol}, M. and {Silvelo}, A. and {Slezak}, I. and {Smart}, R.~L. and {S{\"u}veges}, M. and {Th{\'e}venin}, F. and {Torralba Elipe}, G. and {Ulla}, A. and {Utrilla}, E. and {Vallenari}, A. and {van Dillen}, E. and {Zhao}, H. and {Zorec}, J.},
        title = "{Gaia Data Release 3. Analysis of the Gaia BP/RP spectra using the General Stellar Parameterizer from Photometry}",
      journal = {\aap},
         year = 2023,
        month = jun,
       volume = {674},
          eid = {A27},
        pages = {A27},
          doi = {10.1051/0004-6361/202243462},
archivePrefix = {arXiv},
       eprint = {2206.06138},
 primaryClass = {astro-ph.SR},
       adsurl = {https://ui.adsabs.harvard.edu/abs/2023A&A...674A..27A}
}

@ARTICLE{2025A&A...699A.145D,
       author = {{Delfini}, L. and {Vioque}, M. and {Ribas}, {\'A}. and {Hodgkin}, S.},
        title = "{Star formation and accretion rates within 500 pc as traced by Gaia DR3 XP spectra}",
      journal = {\aap},
         year = 2025,
        month = jul,
       volume = {699},
          eid = {A145},
        pages = {A145},
          doi = {10.1051/0004-6361/202453539},
archivePrefix = {arXiv},
       eprint = {2505.04699},
 primaryClass = {astro-ph.SR},
       adsurl = {https://ui.adsabs.harvard.edu/abs/2025A&A...699A.145D}
}

@ARTICLE{2026A&A...708A..23A,
       author = {{Ambrosch}, M. and {Viscasillas V{\'a}zquez}, C. and {Solano}, E. and {Ulla}, A. and {P{\'e}rez-Couto}, X. and {P{\'e}rez-Fern{\'a}ndez}, E. and {Med{\v{z}}i{\={u}}nas}, A. and {Manteiga}, M. and {Dafonte}, C. and {Drazdauskas}, A. and {Magrini}, L. and {Mikolaitis}, {\v{S}}. and {{\v{S}}atas}, V.},
        title = "{Detection of hot subdwarf binaries and He-poor hot subdwarf stars using machine learning methods and a large sample of Gaia XP spectra}",
      journal = {\aap},
         year = 2026,
        month = mar,
       volume = {708},
          eid = {A23},
        pages = {A23},
          doi = {10.1051/0004-6361/202558282},
archivePrefix = {arXiv},
       eprint = {2601.21727},
 primaryClass = {astro-ph.SR},
       adsurl = {https://ui.adsabs.harvard.edu/abs/2026A&A...708A..23A}
}

@ARTICLE{2025ApJ...982..184R,
       author = {{Roulston}, Benjamin R. and {Leonhardes-Barboza}, Naunet and {Green}, Paul J. and {Portnoi}, Evan},
        title = "{Carbon Stars from Gaia Data Release 3 and the Space Density of Dwarf Carbon Stars}",
      journal = {\apj},
         year = 2025,
        month = apr,
       volume = {982},
       number = {2},
          eid = {184},
        pages = {184},
          doi = {10.3847/1538-4357/adba53},
archivePrefix = {arXiv},
       eprint = {2501.18763},
 primaryClass = {astro-ph.SR},
       adsurl = {https://ui.adsabs.harvard.edu/abs/2025ApJ...982..184R}
}

@ARTICLE{2025A&A...704A.126L,
       author = {{Li}, Jiadong and {Rix}, Hans-Walter and {Ting}, Yuan-Sen and {M{\"u}ller-Horn}, Johanna and {El-Badry}, Kareem and {Liu}, Chao and {Seeburger}, Rhys and {Green}, Gregory M. and {Zhang}, Xiangyu},
        title = "{Millions of main-sequence binary stars from Gaia BP/RP spectra}",
      journal = {\aap},
         year = 2025,
        month = dec,
       volume = {704},
          eid = {A126},
        pages = {A126},
          doi = {10.1051/0004-6361/202556362},
archivePrefix = {arXiv},
       eprint = {2507.09622},
 primaryClass = {astro-ph.SR},
       adsurl = {https://ui.adsabs.harvard.edu/abs/2025A&A...704A.126L}
}

@ARTICLE{2025ApJS..279...47L,
       author = {{Li}, Jiadong and {Ting}, Yuan-Sen and {Rix}, Hans-Walter and {Green}, Gregory M. and {Hogg}, David W. and {Ren}, Juan-Juan and {M{\"u}ller-Horn}, Johanna and {Seeburger}, Rhys},
        title = "{Identification of 30,000 White Dwarf─Main-sequence Binary Candidates from Gaia DR3 BP/RP (XP) Low-resolution Spectra}",
      journal = {\apjs},
         year = 2025,
        month = aug,
       volume = {279},
       number = {2},
          eid = {47},
        pages = {47},
          doi = {10.3847/1538-4365/addf3a},
archivePrefix = {arXiv},
       eprint = {2501.14494},
 primaryClass = {astro-ph.SR},
       adsurl = {https://ui.adsabs.harvard.edu/abs/2025ApJS..279...47L}
}

@ARTICLE{2024MNRAS.52710937Y,
       author = {{Yao}, Yupeng and {Ji}, Alexander P. and {Koposov}, Sergey E. and {Limberg}, Guilherme},
        title = "{200 000 candidate very metal-poor stars in Gaia DR3 XP spectra}",
      journal = {\mnras},
         year = 2024,
        month = feb,
       volume = {527},
       number = {4},
        pages = {10937-10954},
          doi = {10.1093/mnras/stad3775},
archivePrefix = {arXiv},
       eprint = {2303.17676},
 primaryClass = {astro-ph.GA},
       adsurl = {https://ui.adsabs.harvard.edu/abs/2024MNRAS.52710937Y}
}

@ARTICLE{2023ApJS..267....8A,
       author = {{Andrae}, Ren{\'e} and {Rix}, Hans-Walter and {Chandra}, Vedant},
        title = "{Robust Data-driven Metallicities for 175 Million Stars from Gaia XP Spectra}",
      journal = {\apjs},
         year = 2023,
        month = jul,
       volume = {267},
       number = {1},
          eid = {8},
        pages = {8},
          doi = {10.3847/1538-4365/acd53e},
archivePrefix = {arXiv},
       eprint = {2302.02611},
 primaryClass = {astro-ph.SR},
       adsurl = {https://ui.adsabs.harvard.edu/abs/2023ApJS..267....8A}
}

@ARTICLE{2023MNRAS.524.1855Z,
       author = {{Zhang}, Xiangyu and {Green}, Gregory M. and {Rix}, Hans-Walter},
        title = "{Parameters of 220 million stars from Gaia BP/RP spectra}",
      journal = {\mnras},
         year = 2023,
        month = sep,
       volume = {524},
       number = {2},
        pages = {1855-1884},
          doi = {10.1093/mnras/stad1941},
archivePrefix = {arXiv},
       eprint = {2303.03420},
 primaryClass = {astro-ph.SR},
       adsurl = {https://ui.adsabs.harvard.edu/abs/2023MNRAS.524.1855Z}
}

@ARTICLE{2026arXiv260621420M,
       author = {{Malhotra}, S. and {Weiler}, M. and {Anders}, F. and {Carrasco}, J.~M. and {Garcia-Moreno}, G. and {Jordi}, C. and {Blagorodnova}, N. and {Castro-Ginard}, A. and {Casta{\~n}eda}, J. and {Boix}, F. and {Luri}, X.},
        title = "{XP-TEAL: Gaia XP Tool for Emission and Absorption Lines II. Gaia-HELIX catalogue of H$α$ emitters in Gaia BP/RP spectra}",
      journal = {arXiv e-prints},
         year = 2026,
        month = jun,
          eid = {arXiv:2606.21420},
        pages = {arXiv:2606.21420},
          doi = {10.48550/arXiv.2606.21420},
archivePrefix = {arXiv},
       eprint = {2606.21420},
 primaryClass = {astro-ph.SR},
       adsurl = {https://ui.adsabs.harvard.edu/abs/2026arXiv260621420M}
}

@ARTICLE{2026A&A...711A.215C,
       author = {{Carrasco}, J.~M. and {Weiler}, M. and {Malhotra}, S. and {Anders}, F. and {Jordi}, C. and {Masana}, E. and {Casta{\~n}eda}, J. and {Boix}, F.},
        title = "{XP-TEAL: Gaia XP tool for emission and absorption lines: I. Balmer lines in open clusters}",
      journal = {\aap},
         year = 2026,
        month = jul,
       volume = {711},
          eid = {A215},
        pages = {A215},
          doi = {10.1051/0004-6361/202659924},
archivePrefix = {arXiv},
       eprint = {2606.16520},
 primaryClass = {astro-ph.SR},
       adsurl = {https://ui.adsabs.harvard.edu/abs/2026A&A...711A.215C}
}

@ARTICLE{2024ApJ...970..181K,
       author = {{Kao}, Malia L. and {Hawkins}, Keith and {Rogers}, Laura K. and {Bonsor}, Amy and {Dunlap}, Bart H. and {Sanders}, Jason L. and {Montgomery}, M.~H. and {Winget}, D.~E.},
        title = "{Hunting for Polluted White Dwarfs and Other Treasures with Gaia XP Spectra and Unsupervised Machine Learning}",
      journal = {\apj},
         year = 2024,
        month = aug,
       volume = {970},
       number = {2},
          eid = {181},
        pages = {181},
          doi = {10.3847/1538-4357/ad5d6e},
archivePrefix = {arXiv},
       eprint = {2405.17667},
 primaryClass = {astro-ph.SR},
       adsurl = {https://ui.adsabs.harvard.edu/abs/2024ApJ...970..181K}
}

@ARTICLE{2024ApJ...977...31P,
       author = {{P{\'e}rez-Couto}, Xabier and {Pallas-Quintela}, Lara and {Manteiga}, Minia and {Villaver}, Eva and {Dafonte}, Carlos},
        title = "{Identifying New High-confidence Polluted White Dwarf Candidates Using Gaia XP Spectra and Self-organizing Maps}",
      journal = {\apj},
         year = 2024,
        month = dec,
       volume = {977},
       number = {1},
          eid = {31},
        pages = {31},
          doi = {10.3847/1538-4357/ad88f5},
archivePrefix = {arXiv},
       eprint = {2410.16015},
 primaryClass = {astro-ph.SR},
       adsurl = {https://ui.adsabs.harvard.edu/abs/2024ApJ...977...31P}
}

@ARTICLE{2012RAA....12.1197C,
       author = {{Cui}, Xiang-Qun and {Zhao}, Yong-Heng and {Chu}, Yao-Quan and {Li}, Guo-Ping and {Li}, Qi and {Zhang}, Li-Ping and {Su}, Hong-Jun and {Yao}, Zheng-Qiu and {Wang}, Ya-Nan and {Xing}, Xiao-Zheng and {Li}, Xin-Nan and {Zhu}, Yong-Tian and {Wang}, Gang and {Gu}, Bo-Zhong and {Luo}, A.-Li and {Xu}, Xin-Qi and {Zhang}, Zhen-Chao and {Liu}, Gen-Rong and {Zhang}, Hao-Tong and {Yang}, De-Hua and {Cao}, Shu-Yun and {Chen}, Hai-Yuan and {Chen}, Jian-Jun and {Chen}, Kun-Xin and {Chen}, Ying and {Chu}, Jia-Ru and {Feng}, Lei and {Gong}, Xue-Fei and {Hou}, Yong-Hui and {Hu}, Hong-Zhuan and {Hu}, Ning-Sheng and {Hu}, Zhong-Wen and {Jia}, Lei and {Jiang}, Fang-Hua and {Jiang}, Xiang and {Jiang}, Zi-Bo and {Jin}, Ge and {Li}, Ai-Hua and {Li}, Yan and {Li}, Ye-Ping and {Liu}, Guan-Qun and {Liu}, Zhi-Gang and {Lu}, Wen-Zhi and {Mao}, Yin-Dun and {Men}, Li and {Qi}, Yong-Jun and {Qi}, Zhao-Xiang and {Shi}, Huo-Ming and {Tang}, Zheng-Hong and {Tao}, Qing-Sheng and {Wang}, Da-Qi and {Wang}, Dan and {Wang}, Guo-Min and {Wang}, Hai and {Wang}, Jia-Ning and {Wang}, Jian and {Wang}, Jian-Ling and {Wang}, Jian-Ping and {Wang}, Lei and {Wang}, Shu-Qing and {Wang}, You and {Wang}, Yue-Fei and {Xu}, Ling-Zhe and {Xu}, Yan and {Yang}, Shi-Hai and {Yu}, Yong and {Yuan}, Hui and {Yuan}, Xiang-Yan and {Zhai}, Chao and {Zhang}, Jing and {Zhang}, Yan-Xia and {Zhang}, Yong and {Zhao}, Ming and {Zhou}, Fang and {Zhou}, Guo-Hua and {Zhu}, Jie and {Zou}, Si-Cheng},
        title = "{The Large Sky Area Multi-Object Fiber Spectroscopic Telescope (LAMOST)}",
      journal = {Research in Astronomy and Astrophysics},
         year = 2012,
        month = sep,
       volume = {12},
       number = {9},
        pages = {1197-1242},
          doi = {10.1088/1674-4527/12/9/003},
       adsurl = {https://ui.adsabs.harvard.edu/abs/2012RAA....12.1197C}
}

@ARTICLE{2017PASP..129f2001M,
       author = {{Mukai}, K.},
        title = "{X-Ray Emissions from Accreting White Dwarfs: A Review}",
      journal = {\pasp},
         year = 2017,
        month = jun,
       volume = {129},
       number = {976},
        pages = {062001},
          doi = {10.1088/1538-3873/aa6736},
archivePrefix = {arXiv},
       eprint = {1703.06171},
 primaryClass = {astro-ph.HE},
       adsurl = {https://ui.adsabs.harvard.edu/abs/2017PASP..129f2001M}
}

@ARTICLE{2000A&AS..143....9W,
       author = {{Wenger}, M. and {Ochsenbein}, F. and {Egret}, D. and {Dubois}, P. and {Bonnarel}, F. and {Borde}, S. and {Genova}, F. and {Jasniewicz}, G. and {Lalo{\"e}}, S. and {Lesteven}, S. and {Monier}, R.},
        title = "{The SIMBAD astronomical database. The CDS reference database for astronomical objects}",
      journal = {\aaps},
         year = 2000,
        month = apr,
       volume = {143},
        pages = {9-22},
          doi = {10.1051/aas:2000332},
archivePrefix = {arXiv},
       eprint = {astro-ph/0002110},
 primaryClass = {astro-ph},
       adsurl = {https://ui.adsabs.harvard.edu/abs/2000A&AS..143....9W}
}

@ARTICLE{2021A&A...652A..86C,
       author = {{Carrasco}, J.~M. and {Weiler}, M. and {Jordi}, C. and {Fabricius}, C. and {De Angeli}, F. and {Evans}, D.~W. and {van Leeuwen}, F. and {Riello}, M. and {Montegriffo}, P.},
        title = "{Internal calibration of Gaia BP/RP low-resolution spectra}",
      journal = {\aap},
         year = 2021,
        month = aug,
       volume = {652},
          eid = {A86},
        pages = {A86},
          doi = {10.1051/0004-6361/202141249},
archivePrefix = {arXiv},
       eprint = {2106.01752},
 primaryClass = {astro-ph.IM},
       adsurl = {https://ui.adsabs.harvard.edu/abs/2021A&A...652A..86C}
}

@ARTICLE{1981AcA....31..395S,
       author = {{Smak}, J.},
        title = "{On the emission lines from rotating gaseous disks.}",
      journal = {\actaa},
         year = 1981,
        month = jan,
       volume = {31},
        pages = {395-408},
       adsurl = {https://ui.adsabs.harvard.edu/abs/1981AcA....31..395S}
}

@ARTICLE{1986MNRAS.218..761H,
       author = {{Horne}, K. and {Marsh}, T.~R.},
        title = "{Emission line formation in accretion discs}",
      journal = {\mnras},
         year = 1986,
        month = feb,
       volume = {218},
        pages = {761-773},
          doi = {10.1093/mnras/218.4.761},
       adsurl = {https://ui.adsabs.harvard.edu/abs/1986MNRAS.218..761H}
}

@misc{10.26093/cds/vizier,
  doi = {10.26093/CDS/VIZIER},
  url = {https://vizier.cds.unistra.fr},
  author = {Ochsenbein,  Francois},
  title = {The VizieR database of astronomical catalogues},
  publisher = {CDS,  Centre de DonnÃ©es astronomiques de Strasbourg},
  year = {1996},
  copyright = {Refer to CDS usage}
}

@ARTICLE{vizier2000,
       author = {{Ochsenbein}, F. and {Bauer}, P. and {Marcout}, J.},
        title = "{The VizieR database of astronomical catalogues}",
      journal = {\aaps},
         year = 2000,
        month = apr,
       volume = {143},
        pages = {23-32},
          doi = {10.1051/aas:2000169},
archivePrefix = {arXiv},
       eprint = {astro-ph/0002122},
 primaryClass = {astro-ph},
       adsurl = {https://ui.adsabs.harvard.edu/abs/2000A&AS..143...23O}
}

@ARTICLE{2025MNRAS.536.1057I,
       author = {{Inight}, Keith and {G{\"a}nsicke}, Boris T. and {Schwope}, Axel and {Anderson}, Scott F. and {Breedt}, Elm{\'e} and {Brownstein}, Joel R. and {Demasi}, Sebastian and {Friedrich}, Susanne and {Hermes}, J.~J. and {Long}, Knox S. and {Mulvany}, Timothy and {Adamane Pallathadka}, Gautham and {Salvato}, Mara and {Scaringi}, Simone and {Schreiber}, Matthias R. and {Stringfellow}, Guy S. and {Thorstensen}, John R. and {Tovmassian}, Gagik and {Zakamska}, Nadia L.},
        title = "{Cataclysmic variables from Sloan Digital Sky Survey - V (2020-2023) identified using machine learning}",
      journal = {\mnras},
         year = 2025,
        month = jan,
       volume = {536},
       number = {2},
        pages = {1057-1076},
          doi = {10.1093/mnras/stae2524},
archivePrefix = {arXiv},
       eprint = {2406.19459},
 primaryClass = {astro-ph.SR},
       adsurl = {https://ui.adsabs.harvard.edu/abs/2025MNRAS.536.1057I}
}

@ARTICLE{2023AJ....165..163C,
       author = {{Canbay}, Remziye and {Bilir}, Sel{\c{c}}uk and {{\"O}zd{\"o}nmez}, Aykut and {Ak}, Tansel},
        title = "{Galactic Model Parameters and Spatial Density of Cataclysmic Variables in the Gaia Era: New Constraints on Population Models}",
      journal = {\aj},
         year = 2023,
        month = apr,
       volume = {165},
       number = {4},
          eid = {163},
        pages = {163},
          doi = {10.3847/1538-3881/acbead},
archivePrefix = {arXiv},
       eprint = {2302.11568},
 primaryClass = {astro-ph.SR},
       adsurl = {https://ui.adsabs.harvard.edu/abs/2023AJ....165..163C}
}

@ARTICLE{2020ApJS..249...18C,
       author = {{Chen}, Xiaodian and {Wang}, Shu and {Deng}, Licai and {de Grijs}, Richard and {Yang}, Ming and {Tian}, Hao},
        title = "{The Zwicky Transient Facility Catalog of Periodic Variable Stars}",
      journal = {\apjs},
         year = 2020,
        month = jul,
       volume = {249},
       number = {1},
          eid = {18},
        pages = {18},
          doi = {10.3847/1538-4365/ab9cae},
archivePrefix = {arXiv},
       eprint = {2005.08662},
 primaryClass = {astro-ph.SR},
       adsurl = {https://ui.adsabs.harvard.edu/abs/2020ApJS..249...18C}
}

@ARTICLE{2024A&A...684A.121F,
       author = {{Freund}, S. and {Czesla}, S. and {Predehl}, P. and {Robrade}, J. and {Salvato}, M. and {Schneider}, P.~C. and {Starck}, H. and {Wolf}, J. and {Schmitt}, J.~H.~M.~M.},
        title = "{The SRG/eROSITA all-sky survey. Identifying the coronal content with HamStar}",
      journal = {\aap},
         year = 2024,
        month = apr,
       volume = {684},
          eid = {A121},
        pages = {A121},
          doi = {10.1051/0004-6361/202348278},
archivePrefix = {arXiv},
       eprint = {2401.17282},
 primaryClass = {astro-ph.SR},
       adsurl = {https://ui.adsabs.harvard.edu/abs/2024A&A...684A.121F}
}

@ARTICLE{2022A&A...664A.105F,
       author = {{Freund}, S. and {Czesla}, S. and {Robrade}, J. and {Schneider}, P.~C. and {Schmitt}, J.~H.~M.~M.},
        title = "{The stellar content of the ROSAT all-sky survey}",
      journal = {\aap},
         year = 2022,
        month = aug,
       volume = {664},
          eid = {A105},
        pages = {A105},
          doi = {10.1051/0004-6361/202142573},
archivePrefix = {arXiv},
       eprint = {2205.12874},
 primaryClass = {astro-ph.SR},
       adsurl = {https://ui.adsabs.harvard.edu/abs/2022A&A...664A.105F}
}

@ARTICLE{2016A&A...588A.103B,
       author = {{Boller}, Th. and {Freyberg}, M.~J. and {Tr{\"u}mper}, J. and {Haberl}, F. and {Voges}, W. and {Nandra}, K.},
        title = "{Second ROSAT all-sky survey (2RXS) source catalogue}",
      journal = {\aap},
         year = 2016,
        month = apr,
       volume = {588},
          eid = {A103},
        pages = {A103},
          doi = {10.1051/0004-6361/201525648},
archivePrefix = {arXiv},
       eprint = {1609.09244},
 primaryClass = {astro-ph.HE},
       adsurl = {https://ui.adsabs.harvard.edu/abs/2016A&A...588A.103B}
}

@ARTICLE{2023MNRAS.518..174E,
       author = {{Evans}, P.~A. and {Page}, K.~L. and {Beardmore}, A.~P. and {Eyles-Ferris}, R.~A.~J. and {Osborne}, J.~P. and {Campana}, S. and {Kennea}, J.~A. and {Cenko}, S.~B.},
        title = "{A real-time transient detector and the living Swift-XRT point source catalogue}",
      journal = {\mnras},
         year = 2023,
        month = jan,
       volume = {518},
       number = {1},
        pages = {174-184},
          doi = {10.1093/mnras/stac2937},
archivePrefix = {arXiv},
       eprint = {2208.14478},
 primaryClass = {astro-ph.HE},
       adsurl = {https://ui.adsabs.harvard.edu/abs/2023MNRAS.518..174E}
}

@ARTICLE{2023A&A...674A..33G,
       author = {{Gaia Collaboration} and {Montegriffo}, P. and {Bellazzini}, M. and {De Angeli}, F. and {Andrae}, R. and {Barstow}, M.~A. and {Bossini}, D. and {Bragaglia}, A. and {Burgess}, P.~W. and {Cacciari}, C. and {Carrasco}, J.~M. and {Chornay}, N. and {Delchambre}, L. and {Evans}, D.~W. and {Fouesneau}, M. and {Fr{\'e}mat}, Y. and {Garabato}, D. and {Jordi}, C. and {Manteiga}, M. and {Massari}, D. and {Palaversa}, L. and {Pancino}, E. and {Riello}, M. and {Ruz Mieres}, D. and {Sanna}, N. and {Santove{\~n}a}, R. and {Sordo}, R. and {Vallenari}, A. and {Walton}, N.~A. and {Brown}, A.~G.~A. and {Prusti}, T. and {de Bruijne}, J.~H.~J. and {Arenou}, F. and {Babusiaux}, C. and {Biermann}, M. and {Creevey}, O.~L. and {Ducourant}, C. and {Eyer}, L. and {Guerra}, R. and {Hutton}, A. and {Klioner}, S.~A. and {Lammers}, U.~L. and {Lindegren}, L. and {Luri}, X. and {Mignard}, F. and {Panem}, C. and {Pourbaix}, D. and {Randich}, S. and {Sartoretti}, P. and {Soubiran}, C. and {Tanga}, P. and {Bailer-Jones}, C.~A.~L. and {Bastian}, U. and {Drimmel}, R. and {Jansen}, F. and {Katz}, D. and {Lattanzi}, M.~G. and {van Leeuwen}, F. and {Bakker}, J. and {Casta{\~n}eda}, J. and {Fabricius}, C. and {Galluccio}, L. and {Guerrier}, A. and {Heiter}, U. and {Masana}, E. and {Messineo}, R. and {Mowlavi}, N. and {Nicolas}, C. and {Nienartowicz}, K. and {Pailler}, F. and {Panuzzo}, P. and {Riclet}, F. and {Roux}, W. and {Seabroke}, G.~M. and {Th{\'e}venin}, F. and {Gracia-Abril}, G. and {Portell}, J. and {Teyssier}, D. and {Altmann}, M. and {Audard}, M. and {Bellas-Velidis}, I. and {Benson}, K. and {Berthier}, J. and {Blomme}, R. and {Busonero}, D. and {Busso}, G. and {C{\'a}novas}, H. and {Carry}, B. and {Cellino}, A. and {Cheek}, N. and {Clementini}, G. and {Damerdji}, Y. and {Davidson}, M. and {de Teodoro}, P. and {Nu{\~n}ez Campos}, M. and {Dell'Oro}, A. and {Esquej}, P. and {Fern{\'a}ndez-Hern{\'a}ndez}, J. and {Fraile}, E. and {Garc{\'\i}a-Lario}, P. and {Gosset}, E. and {Haigron}, R. and {Halbwachs}, J.-L. and {Hambly}, N.~C. and {Harrison}, D.~L. and {Hern{\'a}ndez}, J. and {Hestroffer}, D. and {Hodgkin}, S.~T. and {Holl}, B. and {Jan{\ss}en}, K. and {Jevardat de Fombelle}, G. and {Jordan}, S. and {Krone-Martins}, A. and {Lanzafame}, A.~C. and {L{\"o}ffler}, W. and {Marchal}, O. and {Marrese}, P.~M. and {Moitinho}, A. and {Muinonen}, K. and {Osborne}, P. and {Pauwels}, T. and {Recio-Blanco}, A. and {Reyl{\'e}}, C. and {Rimoldini}, L. and {Roegiers}, T. and {Rybizki}, J. and {Sarro}, L.~M. and {Siopis}, C. and {Smith}, M. and {Sozzetti}, A. and {Utrilla}, E. and {van Leeuwen}, M. and {Abbas}, U. and {{\'A}brah{\'a}m}, P. and {Abreu Aramburu}, A. and {Aerts}, C. and {Aguado}, J.~J. and {Ajaj}, M. and {Aldea-Montero}, F. and {Altavilla}, G. and {{\'A}lvarez}, M.~A. and {Alves}, J. and {Anderson}, R.~I. and {Anglada Varela}, E. and {Antoja}, T. and {et al.}},
        title = "{Gaia Data Release 3. The Galaxy in your preferred colours: Synthetic photometry from Gaia low-resolution spectra}",
      journal = {\aap},
         year = 2023,
        month = jun,
       volume = {674},
          eid = {A33},
        pages = {A33},
          doi = {10.1051/0004-6361/202243709},
archivePrefix = {arXiv},
       eprint = {2206.06215},
 primaryClass = {astro-ph.SR},
       adsurl = {https://ui.adsabs.harvard.edu/abs/2023A&A...674A..33G}
}

@ARTICLE{2019ApJ...887...93G,
       author = {{Green}, Gregory M. and {Schlafly}, Edward and {Zucker}, Catherine and {Speagle}, Joshua S. and {Finkbeiner}, Douglas},
        title = "{A 3D Dust Map Based on Gaia, Pan-STARRS 1, and 2MASS}",
      journal = {\apj},
         year = 2019,
        month = dec,
       volume = {887},
       number = {1},
          eid = {93},
        pages = {93},
          doi = {10.3847/1538-4357/ab5362},
archivePrefix = {arXiv},
       eprint = {1905.02734},
 primaryClass = {astro-ph.GA},
       adsurl = {https://ui.adsabs.harvard.edu/abs/2019ApJ...887...93G}
}

@ARTICLE{2003A&A...409..205D,
       author = {{Drimmel}, R. and {Cabrera-Lavers}, A. and {L{\'o}pez-Corredoira}, M.},
        title = "{A three-dimensional Galactic extinction model}",
      journal = {\aap},
         year = 2003,
        month = oct,
       volume = {409},
        pages = {205-215},
          doi = {10.1051/0004-6361:20031070},
archivePrefix = {arXiv},
       eprint = {astro-ph/0307273},
 primaryClass = {astro-ph},
       adsurl = {https://ui.adsabs.harvard.edu/abs/2003A&A...409..205D}
}

@ARTICLE{2006A&A...453..635M,
       author = {{Marshall}, D.~J. and {Robin}, A.~C. and {Reyl{\'e}}, C. and {Schultheis}, M. and {Picaud}, S.},
        title = "{Modelling the Galactic interstellar extinction distribution in three dimensions}",
      journal = {\aap},
         year = 2006,
        month = jul,
       volume = {453},
       number = {2},
        pages = {635-651},
          doi = {10.1051/0004-6361:20053842},
archivePrefix = {arXiv},
       eprint = {astro-ph/0604427},
 primaryClass = {astro-ph},
       adsurl = {https://ui.adsabs.harvard.edu/abs/2006A&A...453..635M}
}

@ARTICLE{2016ApJ...818..130B,
       author = {{Bovy}, Jo and {Rix}, Hans-Walter and {Green}, Gregory M. and {Schlafly}, Edward F. and {Finkbeiner}, Douglas P.},
        title = "{On Galactic Density Modeling in the Presence of Dust Extinction}",
      journal = {\apj},
         year = 2016,
        month = feb,
       volume = {818},
       number = {2},
          eid = {130},
        pages = {130},
          doi = {10.3847/0004-637X/818/2/130},
archivePrefix = {arXiv},
       eprint = {1509.06751},
 primaryClass = {astro-ph.GA},
       adsurl = {https://ui.adsabs.harvard.edu/abs/2016ApJ...818..130B}
}

@ARTICLE{2023A&A...674A...2D,
       author = {{De Angeli}, F. and {Weiler}, M. and {Montegriffo}, P. and {Evans}, D.~W. and {Riello}, M. and {Andrae}, R. and {Carrasco}, J.~M. and {Busso}, G. and {Burgess}, P.~W. and {Cacciari}, C. and {Davidson}, M. and {Harrison}, D.~L. and {Hodgkin}, S.~T. and {Jordi}, C. and {Osborne}, P.~J. and {Pancino}, E. and {Altavilla}, G. and {Barstow}, M.~A. and {Bailer-Jones}, C.~A.~L. and {Bellazzini}, M. and {Brown}, A.~G.~A. and {Castellani}, M. and {Cowell}, S. and {Delchambre}, L. and {De Luise}, F. and {Diener}, C. and {Fabricius}, C. and {Fouesneau}, M. and {Fr{\'e}mat}, Y. and {Gilmore}, G. and {Giuffrida}, G. and {Hambly}, N.~C. and {Hidalgo}, S. and {Holland}, G. and {Kostrzewa-Rutkowska}, Z. and {van Leeuwen}, F. and {Lobel}, A. and {Marinoni}, S. and {Miller}, N. and {Pagani}, C. and {Palaversa}, L. and {Piersimoni}, A.~M. and {Pulone}, L. and {Ragaini}, S. and {Rainer}, M. and {Richards}, P.~J. and {Rixon}, G.~T. and {Ruz-Mieres}, D. and {Sanna}, N. and {Sarro}, L.~M. and {Rowell}, N. and {Sordo}, R. and {Walton}, N.~A. and {Yoldas}, A.},
        title = "{Gaia Data Release 3. Processing and validation of BP/RP low-resolution spectral data}",
      journal = {\aap},
         year = 2023,
        month = jun,
       volume = {674},
          eid = {A2},
        pages = {A2},
          doi = {10.1051/0004-6361/202243680},
archivePrefix = {arXiv},
       eprint = {2206.06143},
 primaryClass = {astro-ph.IM},
       adsurl = {https://ui.adsabs.harvard.edu/abs/2023A&A...674A...2D}
}

@ARTICLE{2023A&A...674A...3M,
       author = {{Montegriffo}, P. and {De Angeli}, F. and {Andrae}, R. and {Riello}, M. and {Pancino}, E. and {Sanna}, N. and {Bellazzini}, M. and {Evans}, D.~W. and {Carrasco}, J.~M. and {Sordo}, R. and {Busso}, G. and {Cacciari}, C. and {Jordi}, C. and {van Leeuwen}, F. and {Vallenari}, A. and {Altavilla}, G. and {Barstow}, M.~A. and {Brown}, A.~G.~A. and {Burgess}, P.~W. and {Castellani}, M. and {Cowell}, S. and {Davidson}, M. and {De Luise}, F. and {Delchambre}, L. and {Diener}, C. and {Fabricius}, C. and {Fr{\'e}mat}, Y. and {Fouesneau}, M. and {Gilmore}, G. and {Giuffrida}, G. and {Hambly}, N.~C. and {Harrison}, D.~L. and {Hidalgo}, S. and {Hodgkin}, S.~T. and {Holland}, G. and {Marinoni}, S. and {Osborne}, P.~J. and {Pagani}, C. and {Palaversa}, L. and {Piersimoni}, A.~M. and {Pulone}, L. and {Ragaini}, S. and {Rainer}, M. and {Richards}, P.~J. and {Rowell}, N. and {Ruz-Mieres}, D. and {Sarro}, L.~M. and {Walton}, N.~A. and {Yoldas}, A.},
        title = "{Gaia Data Release 3. External calibration of BP/RP low-resolution spectroscopic data}",
      journal = {\aap},
         year = 2023,
        month = jun,
       volume = {674},
          eid = {A3},
        pages = {A3},
          doi = {10.1051/0004-6361/202243880},
archivePrefix = {arXiv},
       eprint = {2206.06205},
 primaryClass = {astro-ph.IM},
       adsurl = {https://ui.adsabs.harvard.edu/abs/2023A&A...674A...3M}
}

@ARTICLE{2020MNRAS.492L..40A,
       author = {{Abril}, Javier and {Schmidtobreick}, Linda and {Ederoclite}, Alessandro and {L{\'o}pez-Sanjuan}, Carlos},
        title = "{Disentangling cataclysmic variables in Gaia's HR diagram}",
      journal = {\mnras},
         year = 2020,
        month = feb,
       volume = {492},
       number = {1},
        pages = {L40-L44},
          doi = {10.1093/mnrasl/slz181},
archivePrefix = {arXiv},
       eprint = {1912.01531},
 primaryClass = {astro-ph.SR},
       adsurl = {https://ui.adsabs.harvard.edu/abs/2020MNRAS.492L..40A}
}

@ARTICLE{2022A&A...662A..40C,
       author = {{Culpan}, R. and {Geier}, S. and {Reindl}, N. and {Pelisoli}, I. and {Gentile Fusillo}, N. and {Vorontseva}, A.},
        title = "{The population of hot subdwarf stars studied with Gaia. IV. Catalogues of hot subluminous stars based on Gaia EDR3}",
      journal = {\aap},
         year = 2022,
        month = jun,
       volume = {662},
          eid = {A40},
        pages = {A40},
          doi = {10.1051/0004-6361/202243337},
archivePrefix = {arXiv},
       eprint = {2203.07938},
 primaryClass = {astro-ph.SR},
       adsurl = {https://ui.adsabs.harvard.edu/abs/2022A&A...662A..40C}
}

@ARTICLE{2024MNRAS.527.8687O,
       author = {{O'Brien}, Mairi W. and {Tremblay}, P.-E. and {Klein}, B.~L. and {Koester}, D. and {Melis}, C. and {B{\'e}dard}, A. and {Cukanovaite}, E. and {Cunningham}, T. and {Doyle}, A.~E. and {G{\"a}nsicke}, B.~T. and {Gentile Fusillo}, N.~P. and {Hollands}, M.~A. and {McCleery}, J. and {Pelisoli}, I. and {Toonen}, S. and {Weinberger}, A.~J. and {Zuckerman}, B.},
        title = "{The 40 pc sample of white dwarfs from Gaia}",
      journal = {\mnras},
         year = 2024,
        month = jan,
       volume = {527},
       number = {3},
        pages = {8687-8705},
          doi = {10.1093/mnras/stad3773},
archivePrefix = {arXiv},
       eprint = {2312.02735},
 primaryClass = {astro-ph.SR},
       adsurl = {https://ui.adsabs.harvard.edu/abs/2024MNRAS.527.8687O}
}

@ARTICLE{2025A&A...698A.106S,
       author = {{Schwope}, Axel D.},
        title = "{PolarCat: Catalog of polars, low-accretion rate polars, and candidate objects}",
      journal = {\aap},
         year = 2025,
        month = jun,
       volume = {698},
          eid = {A106},
        pages = {A106},
          doi = {10.1051/0004-6361/202554519},
archivePrefix = {arXiv},
       eprint = {2505.10337},
 primaryClass = {astro-ph.SR},
       adsurl = {https://ui.adsabs.harvard.edu/abs/2025A&A...698A.106S}
}

@ARTICLE{2003A&A...404..301R,
       author = {{Ritter}, H. and {Kolb}, U.},
        title = "{Catalogue of cataclysmic binaries, low-mass X-ray binaries   and related objects (Seventh edition)}",
      journal = {\aap},
         year = 2003,
        month = jun,
       volume = {404},
        pages = {301-303},
          doi = {10.1051/0004-6361:20030330},
archivePrefix = {arXiv},
       eprint = {astro-ph/0301444},
 primaryClass = {astro-ph},
       adsurl = {https://ui.adsabs.harvard.edu/abs/2003A&A...404..301R}
}

@ARTICLE{2023MNRAS.524.4867I,
       author = {{Inight}, Keith and {G{\"a}nsicke}, Boris T. and {Breedt}, Elm{\'e} and {Israel}, Henry T. and {Littlefair}, Stuart P. and {Manser}, Christopher J. and {Marsh}, Tom R. and {Mulvany}, Tim and {Pala}, Anna Francesca and {Thorstensen}, John R.},
        title = "{A catalogue of cataclysmic variables from 20 yr of the Sloan Digital Sky Survey with new classifications, periods, trends, and oddities}",
      journal = {\mnras},
         year = 2023,
        month = oct,
       volume = {524},
       number = {4},
        pages = {4867-4898},
          doi = {10.1093/mnras/stad2018},
archivePrefix = {arXiv},
       eprint = {2304.06749},
 primaryClass = {astro-ph.SR},
       adsurl = {https://ui.adsabs.harvard.edu/abs/2023MNRAS.524.4867I}
}

@ARTICLE{2025A&A...699A.153R,
       author = {{Rebassa-Mansergas}, Alberto and {Solano}, Enrique and {Brown}, Alex J. and {Parsons}, Steven G. and {Murillo-Ojeda}, Raquel and {Raddi}, Roberto and {Camisassa}, Maria and {Torres}, Santiago and {van Roestel}, Jan},
        title = "{Magnitude-limited catalogue of unresolved white dwarf-main sequence binaries from Gaia DR3}",
      journal = {\aap},
         year = 2025,
        month = jul,
       volume = {699},
          eid = {A153},
        pages = {A153},
          doi = {10.1051/0004-6361/202554700},
archivePrefix = {arXiv},
       eprint = {2505.15895},
 primaryClass = {astro-ph.SR},
       adsurl = {https://ui.adsabs.harvard.edu/abs/2025A&A...699A.153R}
}

@ARTICLE{2023A&A...671A..52W,
       author = {{Weiler}, M. and {Carrasco}, J.~M. and {Fabricius}, C. and {Jordi}, C.},
        title = "{Analysing spectral lines in Gaia low-resolution spectra}",
      journal = {\aap},
         year = 2023,
        month = mar,
       volume = {671},
          eid = {A52},
        pages = {A52},
          doi = {10.1051/0004-6361/202244764},
archivePrefix = {arXiv},
       eprint = {2211.06946},
 primaryClass = {astro-ph.IM},
       adsurl = {https://ui.adsabs.harvard.edu/abs/2023A&A...671A..52W}
}

@ARTICLE{2016A&A...595A...1G,
       author = {{Gaia Collaboration} and {Prusti}, T. and {de Bruijne}, J.~H.~J. and {Brown}, A.~G.~A. and {Vallenari}, A. and {Babusiaux}, C. and {Bailer-Jones}, C.~A.~L. and {Bastian}, U. and {Biermann}, M. and {Evans}, D.~W. and {Eyer}, L. and {Jansen}, F. and {Jordi}, C. and {Klioner}, S.~A. and {Lammers}, U. and {Lindegren}, L. and {Luri}, X. and {Mignard}, F. and {Milligan}, D.~J. and {Panem}, C. and {Poinsignon}, V. and {Pourbaix}, D. and {Randich}, S. and {Sarri}, G. and {Sartoretti}, P. and {Siddiqui}, H.~I. and {Soubiran}, C. and {Valette}, V. and {van Leeuwen}, F. and {Walton}, N.~A. and {Aerts}, C. and {Arenou}, F. and {Cropper}, M. and {Drimmel}, R. and {H{\o}g}, E. and {Katz}, D. and {Lattanzi}, M.~G. and {O'Mullane}, W. and {Grebel}, E.~K. and {Holland}, A.~D. and {Huc}, C. and {Passot}, X. and {Bramante}, L. and {Cacciari}, C. and {Casta{\~n}eda}, J. and {Chaoul}, L. and {Cheek}, N. and {De Angeli}, F. and {Fabricius}, C. and {Guerra}, R. and {Hern{\'a}ndez}, J. and {Jean-Antoine-Piccolo}, A. and {Masana}, E. and {Messineo}, R. and {Mowlavi}, N. and {Nienartowicz}, K. and {Ord{\'o}{\~n}ez-Blanco}, D. and {Panuzzo}, P. and {Portell}, J. and {Richards}, P.~J. and {Riello}, M. and {Seabroke}, G.~M. and {Tanga}, P. and {Th{\'e}venin}, F. and {Torra}, J. and {Els}, S.~G. and {Gracia-Abril}, G. and {Comoretto}, G. and {Garcia-Reinaldos}, M. and {Lock}, T. and {Mercier}, E. and {Altmann}, M. and {Andrae}, R. and {Astraatmadja}, T.~L. and {Bellas-Velidis}, I. and {Benson}, K. and {Berthier}, J. and {Blomme}, R. and {Busso}, G. and {Carry}, B. and {Cellino}, A. and {Clementini}, G. and {Cowell}, S. and {Creevey}, O. and {Cuypers}, J. and {Davidson}, M. and {De Ridder}, J. and {de Torres}, A. and {Delchambre}, L. and {Dell'Oro}, A. and {Ducourant}, C. and {Fr{\'e}mat}, Y. and {Garc{\'\i}a-Torres}, M. and {Gosset}, E. and {Halbwachs}, J.-L. and {Hambly}, N.~C. and {Harrison}, D.~L. and {Hauser}, M. and {Hestroffer}, D. and {Hodgkin}, S.~T. and {Huckle}, H.~E. and {Hutton}, A. and {Jasniewicz}, G. and {Jordan}, S. and {Kontizas}, M. and {Korn}, A.~J. and {Lanzafame}, A.~C. and {Manteiga}, M. and {Moitinho}, A. and {Muinonen}, K. and {Osinde}, J. and {Pancino}, E. and {Pauwels}, T. and {Petit}, J.-M. and {Recio-Blanco}, A. and {Robin}, A.~C. and {Sarro}, L.~M. and {Siopis}, C. and {Smith}, M. and {Smith}, K.~W. and {Sozzetti}, A. and {Thuillot}, W. and {van Reeven}, W. and {Viala}, Y. and {Abbas}, U. and {Abreu Aramburu}, A. and {Accart}, S. and {Aguado}, J.~J. and {Allan}, P.~M. and {Allasia}, W. and {Altavilla}, G. and {{\'A}lvarez}, M.~A. and {Alves}, J. and {Anderson}, R.~I. and {Andrei}, A.~H. and {Anglada Varela}, E. and {Antiche}, E. and {Antoja}, T. and {Ant{\'o}n}, S. and {Arcay}, B. and {Atzei}, A. and {Ayache}, L. and {Bach}, N. and {Baker}, S.~G. and {Balaguer-N{\'u}{\~n}ez}, L. and {Barache}, C. and {Barata}, C. and {Barbier}, A. and {Barblan}, F. and {Baroni}, M. and {Barrado y Navascu{\'e}s}, D. and {Barros}, M. and {Barstow}, M.~A. and {Becciani}, U. and {Bellazzini}, M. and {Bellei}, G. and {Bello Garc{\'\i}a}, A. and {Belokurov}, V. and {Bendjoya}, P. and {Berihuete}, A. and {Bianchi}, L. and {Bienaym{\'e}}, O. and {Billebaud}, F. and {Blagorodnova}, N. and {Blanco-Cuaresma}, S. and {Boch}, T. and {Bombrun}, A. and {Borrachero}, R. and {Bouquillon}, S. and {Bourda}, G. and {Bouy}, H. and {Bragaglia}, A. and {Breddels}, M.~A. and {Brouillet}, N. and {Br{\"u}semeister}, T. and {Bucciarelli}, B. and {Budnik}, F. and {Burgess}, P. and {Burgon}, R. and {Burlacu}, A. and {Busonero}, D. and {Buzzi}, R. and {Caffau}, E. and {Cambras}, J. and {Campbell}, H. and {Cancelliere}, R. and {Cantat-Gaudin}, T. and {Carlucci}, T. and {Carrasco}, J.~M. and {Castellani}, M. and {Charlot}, P. and {Charnas}, J. and {Charvet}, P. and {Chassat}, F. and {Chiavassa}, A. and {Clotet}, M. and {Cocozza}, G. and {Collins}, R.~S. and {Collins}, P. and {Costigan}, G.},
        title = "{The Gaia mission}",
      journal = {\aap},
         year = 2016,
        month = nov,
       volume = {595},
          eid = {A1},
        pages = {A1},
          doi = {10.1051/0004-6361/201629272},
archivePrefix = {arXiv},
       eprint = {1609.04153},
 primaryClass = {astro-ph.IM},
       adsurl = {https://ui.adsabs.harvard.edu/abs/2016A&A...595A...1G}
}

@ARTICLE{2018A&A...616A...1G,
       author = {{Gaia Collaboration} and {Brown}, A.~G.~A. and {Vallenari}, A. and {Prusti}, T. and {de Bruijne}, J.~H.~J. and {Babusiaux}, C. and {Bailer-Jones}, C.~A.~L. and {Biermann}, M. and {Evans}, D.~W. and {Eyer}, L. and {Jansen}, F. and {Jordi}, C. and {Klioner}, S.~A. and {Lammers}, U. and {Lindegren}, L. and {Luri}, X. and {Mignard}, F. and {Panem}, C. and {Pourbaix}, D. and {Randich}, S. and {Sartoretti}, P. and {Siddiqui}, H.~I. and {Soubiran}, C. and {van Leeuwen}, F. and {Walton}, N.~A. and {Arenou}, F. and {Bastian}, U. and {Cropper}, M. and {Drimmel}, R. and {Katz}, D. and {Lattanzi}, M.~G. and {Bakker}, J. and {Cacciari}, C. and {Casta{\~n}eda}, J. and {Chaoul}, L. and {Cheek}, N. and {De Angeli}, F. and {Fabricius}, C. and {Guerra}, R. and {Holl}, B. and {Masana}, E. and {Messineo}, R. and {Mowlavi}, N. and {Nienartowicz}, K. and {Panuzzo}, P. and {Portell}, J. and {Riello}, M. and {Seabroke}, G.~M. and {Tanga}, P. and {Th{\'e}venin}, F. and {Gracia-Abril}, G. and {Comoretto}, G. and {Garcia-Reinaldos}, M. and {Teyssier}, D. and {Altmann}, M. and {Andrae}, R. and {Audard}, M. and {Bellas-Velidis}, I. and {Benson}, K. and {Berthier}, J. and {Blomme}, R. and {Burgess}, P. and {Busso}, G. and {Carry}, B. and {Cellino}, A. and {Clementini}, G. and {Clotet}, M. and {Creevey}, O. and {Davidson}, M. and {De Ridder}, J. and {Delchambre}, L. and {Dell'Oro}, A. and {Ducourant}, C. and {Fern{\'a}ndez-Hern{\'a}ndez}, J. and {Fouesneau}, M. and {Fr{\'e}mat}, Y. and {Galluccio}, L. and {Garc{\'\i}a-Torres}, M. and {Gonz{\'a}lez-N{\'u}{\~n}ez}, J. and {Gonz{\'a}lez-Vidal}, J.~J. and {Gosset}, E. and {Guy}, L.~P. and {Halbwachs}, J.-L. and {Hambly}, N.~C. and {Harrison}, D.~L. and {Hern{\'a}ndez}, J. and {Hestroffer}, D. and {Hodgkin}, S.~T. and {Hutton}, A. and {Jasniewicz}, G. and {Jean-Antoine-Piccolo}, A. and {Jordan}, S. and {Korn}, A.~J. and {Krone-Martins}, A. and {Lanzafame}, A.~C. and {Lebzelter}, T. and {L{\"o}ffler}, W. and {Manteiga}, M. and {Marrese}, P.~M. and {Mart{\'\i}n-Fleitas}, J.~M. and {Moitinho}, A. and {Mora}, A. and {Muinonen}, K. and {Osinde}, J. and {Pancino}, E. and {Pauwels}, T. and {Petit}, J.-M. and {Recio-Blanco}, A. and {Richards}, P.~J. and {Rimoldini}, L. and {Robin}, A.~C. and {Sarro}, L.~M. and {Siopis}, C. and {Smith}, M. and {Sozzetti}, A. and {S{\"u}veges}, M. and {Torra}, J. and {van Reeven}, W. and {Abbas}, U. and {Abreu Aramburu}, A. and {Accart}, S. and {Aerts}, C. and {Altavilla}, G. and {{\'A}lvarez}, M.~A. and {Alvarez}, R. and {Alves}, J. and {Anderson}, R.~I. and {Andrei}, A.~H. and {Anglada Varela}, E. and {Antiche}, E. and {Antoja}, T. and {Arcay}, B. and {Astraatmadja}, T.~L. and {Bach}, N. and {Baker}, S.~G. and {Balaguer-N{\'u}{\~n}ez}, L. and {Balm}, P. and {Barache}, C. and {Barata}, C. and {Barbato}, D. and {Barblan}, F. and {Barklem}, P.~S. and {Barrado}, D. and {Barros}, M. and {Barstow}, M.~A. and {Bartholom{\'e} Mu{\~n}oz}, S. and {Bassilana}, J.-L. and {Becciani}, U. and {Bellazzini}, M. and {Berihuete}, A. and {Bertone}, S. and {Bianchi}, L. and {Bienaym{\'e}}, O. and {Blanco-Cuaresma}, S. and {Boch}, T. and {Boeche}, C. and {Bombrun}, A. and {Borrachero}, R. and {Bossini}, D. and {Bouquillon}, S. and {Bourda}, G. and {Bragaglia}, A. and {Bramante}, L. and {Breddels}, M.~A. and {Bressan}, A. and {Brouillet}, N. and {Br{\"u}semeister}, T. and {Brugaletta}, E. and {Bucciarelli}, B. and {Burlacu}, A. and {Busonero}, D. and {Butkevich}, A.~G. and {Buzzi}, R. and {Caffau}, E. and {Cancelliere}, R. and {Cannizzaro}, G. and {Cantat-Gaudin}, T. and {Carballo}, R. and {Carlucci}, T. and {Carrasco}, J.~M. and {Casamiquela}, L. and {Castellani}, M. and {Castro-Ginard}, A. and {Charlot}, P. and {Chemin}, L. and {Chiavassa}, A. and {Cocozza}, G. and {Costigan}, G. and {Cowell}, S. and {Crifo}, F. and {Crosta}, M. and {Crowley}, C. and {Cuypers}, J. and {Dafonte}, C. and {Damerdji}, Y. and {Dapergolas}, A. and {David}, P. and {David}, M. and {de Laverny}, P. and {De Luise}, F.},
        title = "{Gaia Data Release 2. Summary of the contents and survey properties}",
      journal = {\aap},
         year = 2018,
        month = aug,
       volume = {616},
          eid = {A1},
        pages = {A1},
          doi = {10.1051/0004-6361/201833051},
archivePrefix = {arXiv},
       eprint = {1804.09365},
 primaryClass = {astro-ph.GA},
       adsurl = {https://ui.adsabs.harvard.edu/abs/2018A&A...616A...1G}
}

@ARTICLE{2021A&A...649A...1G,
       author = {{Gaia Collaboration} and {Brown}, A.~G.~A. and {Vallenari}, A. and {Prusti}, T. and {de Bruijne}, J.~H.~J. and {Babusiaux}, C. and {Biermann}, M. and {Creevey}, O.~L. and {Evans}, D.~W. and {Eyer}, L. and {Hutton}, A. and {Jansen}, F. and {Jordi}, C. and {Klioner}, S.~A. and {Lammers}, U. and {Lindegren}, L. and {Luri}, X. and {Mignard}, F. and {Panem}, C. and {Pourbaix}, D. and {Randich}, S. and {Sartoretti}, P. and {Soubiran}, C. and {Walton}, N.~A. and {Arenou}, F. and {Bailer-Jones}, C.~A.~L. and {Bastian}, U. and {Cropper}, M. and {Drimmel}, R. and {Katz}, D. and {Lattanzi}, M.~G. and {van Leeuwen}, F. and {Bakker}, J. and {Cacciari}, C. and {Casta{\~n}eda}, J. and {De Angeli}, F. and {Ducourant}, C. and {Fabricius}, C. and {Fouesneau}, M. and {Fr{\'e}mat}, Y. and {Guerra}, R. and {Guerrier}, A. and {Guiraud}, J. and {Jean-Antoine Piccolo}, A. and {Masana}, E. and {Messineo}, R. and {Mowlavi}, N. and {Nicolas}, C. and {Nienartowicz}, K. and {Pailler}, F. and {Panuzzo}, P. and {Riclet}, F. and {Roux}, W. and {Seabroke}, G.~M. and {Sordo}, R. and {Tanga}, P. and {Th{\'e}venin}, F. and {Gracia-Abril}, G. and {Portell}, J. and {Teyssier}, D. and {Altmann}, M. and {Andrae}, R. and {Bellas-Velidis}, I. and {Benson}, K. and {Berthier}, J. and {Blomme}, R. and {Brugaletta}, E. and {Burgess}, P.~W. and {Busso}, G. and {Carry}, B. and {Cellino}, A. and {Cheek}, N. and {Clementini}, G. and {Damerdji}, Y. and {Davidson}, M. and {Delchambre}, L. and {Dell'Oro}, A. and {Fern{\'a}ndez-Hern{\'a}ndez}, J. and {Galluccio}, L. and {Garc{\'\i}a-Lario}, P. and {Garcia-Reinaldos}, M. and {Gonz{\'a}lez-N{\'u}{\~n}ez}, J. and {Gosset}, E. and {Haigron}, R. and {Halbwachs}, J.-L. and {Hambly}, N.~C. and {Harrison}, D.~L. and {Hatzidimitriou}, D. and {Heiter}, U. and {Hern{\'a}ndez}, J. and {Hestroffer}, D. and {Hodgkin}, S.~T. and {Holl}, B. and {Jan{\ss}en}, K. and {Jevardat de Fombelle}, G. and {Jordan}, S. and {Krone-Martins}, A. and {Lanzafame}, A.~C. and {L{\"o}ffler}, W. and {Lorca}, A. and {Manteiga}, M. and {Marchal}, O. and {Marrese}, P.~M. and {Moitinho}, A. and {Mora}, A. and {Muinonen}, K. and {Osborne}, P. and {Pancino}, E. and {Pauwels}, T. and {Petit}, J.-M. and {Recio-Blanco}, A. and {Richards}, P.~J. and {Riello}, M. and {Rimoldini}, L. and {Robin}, A.~C. and {Roegiers}, T. and {Rybizki}, J. and {Sarro}, L.~M. and {Siopis}, C. and {Smith}, M. and {Sozzetti}, A. and {Ulla}, A. and {Utrilla}, E. and {van Leeuwen}, M. and {van Reeven}, W. and {Abbas}, U. and {Abreu Aramburu}, A. and {Accart}, S. and {Aerts}, C. and {Aguado}, J.~J. and {Ajaj}, M. and {Altavilla}, G. and {{\'A}lvarez}, M.~A. and {{\'A}lvarez Cid-Fuentes}, J. and {Alves}, J. and {Anderson}, R.~I. and {Anglada Varela}, E. and {Antoja}, T. and {Audard}, M. and {Baines}, D. and {Baker}, S.~G. and {Balaguer-N{\'u}{\~n}ez}, L. and {Balbinot}, E. and {Balog}, Z. and {Barache}, C. and {Barbato}, D. and {Barros}, M. and {Barstow}, M.~A. and {Bartolom{\'e}}, S. and {Bassilana}, J.-L. and {Bauchet}, N. and {Baudesson-Stella}, A. and {Becciani}, U. and {Bellazzini}, M. and {Bernet}, M. and {Bertone}, S. and {Bianchi}, L. and {Blanco-Cuaresma}, S. and {Boch}, T. and {Bombrun}, A. and {Bossini}, D. and {Bouquillon}, S. and {Bragaglia}, A. and {Bramante}, L. and {Breedt}, E. and {Bressan}, A. and {Brouillet}, N. and {Bucciarelli}, B. and {Burlacu}, A. and {Busonero}, D. and {Butkevich}, A.~G. and {Buzzi}, R. and {Caffau}, E. and {Cancelliere}, R. and {C{\'a}novas}, H. and {Cantat-Gaudin}, T. and {Carballo}, R. and {Carlucci}, T. and {Carnerero}, M.~I. and {Carrasco}, J.~M. and {Casamiquela}, L. and {Castellani}, M. and {Castro-Ginard}, A. and {Castro Sampol}, P. and {Chaoul}, L. and {Charlot}, P. and {Chemin}, L. and {Chiavassa}, A. and {Cioni}, M.-R.~L. and {Comoretto}, G. and {Cooper}, W.~J. and {Cornez}, T. and {Cowell}, S. and {Crifo}, F. and {Crosta}, M. and {Crowley}, C. and {Dafonte}, C. and {Dapergolas}, A. and {David}, M. and {David}, P.},
        title = "{Gaia Early Data Release 3. Summary of the contents and survey properties}",
      journal = {\aap},
         year = 2021,
        month = may,
       volume = {649},
          eid = {A1},
        pages = {A1},
          doi = {10.1051/0004-6361/202039657},
archivePrefix = {arXiv},
       eprint = {2012.01533},
 primaryClass = {astro-ph.GA},
       adsurl = {https://ui.adsabs.harvard.edu/abs/2021A&A...649A...1G}
}

@ARTICLE{2023A&A...674A...1G,
       author = {{Gaia Collaboration} and {Vallenari}, A. and {Brown}, A.~G.~A. and {Prusti}, T. and {de Bruijne}, J.~H.~J. and {Arenou}, F. and {Babusiaux}, C. and {Biermann}, M. and {Creevey}, O.~L. and {Ducourant}, C. and {Evans}, D.~W. and {Eyer}, L. and {Guerra}, R. and {Hutton}, A. and {Jordi}, C. and {Klioner}, S.~A. and {Lammers}, U.~L. and {Lindegren}, L. and {Luri}, X. and {Mignard}, F. and {Panem}, C. and {Pourbaix}, D. and {Randich}, S. and {Sartoretti}, P. and {Soubiran}, C. and {Tanga}, P. and {Walton}, N.~A. and {Bailer-Jones}, C.~A.~L. and {Bastian}, U. and {Drimmel}, R. and {Jansen}, F. and {Katz}, D. and {Lattanzi}, M.~G. and {van Leeuwen}, F. and {Bakker}, J. and {Cacciari}, C. and {Casta{\~n}eda}, J. and {De Angeli}, F. and {Fabricius}, C. and {Fouesneau}, M. and {Fr{\'e}mat}, Y. and {Galluccio}, L. and {Guerrier}, A. and {Heiter}, U. and {Masana}, E. and {Messineo}, R. and {Mowlavi}, N. and {Nicolas}, C. and {Nienartowicz}, K. and {Pailler}, F. and {Panuzzo}, P. and {Riclet}, F. and {Roux}, W. and {Seabroke}, G.~M. and {Sordo}, R. and {Th{\'e}venin}, F. and {Gracia-Abril}, G. and {Portell}, J. and {Teyssier}, D. and {Altmann}, M. and {Andrae}, R. and {Audard}, M. and {Bellas-Velidis}, I. and {Benson}, K. and {Berthier}, J. and {Blomme}, R. and {Burgess}, P.~W. and {Busonero}, D. and {Busso}, G. and {C{\'a}novas}, H. and {Carry}, B. and {Cellino}, A. and {Cheek}, N. and {Clementini}, G. and {Damerdji}, Y. and {Davidson}, M. and {de Teodoro}, P. and {Nu{\~n}ez Campos}, M. and {Delchambre}, L. and {Dell'Oro}, A. and {Esquej}, P. and {Fern{\'a}ndez-Hern{\'a}ndez}, J. and {Fraile}, E. and {Garabato}, D. and {Garc{\'\i}a-Lario}, P. and {Gosset}, E. and {Haigron}, R. and {Halbwachs}, J.-L. and {Hambly}, N.~C. and {Harrison}, D.~L. and {Hern{\'a}ndez}, J. and {Hestroffer}, D. and {Hodgkin}, S.~T. and {Holl}, B. and {Jan{\ss}en}, K. and {Jevardat de Fombelle}, G. and {Jordan}, S. and {Krone-Martins}, A. and {Lanzafame}, A.~C. and {L{\"o}ffler}, W. and {Marchal}, O. and {Marrese}, P.~M. and {Moitinho}, A. and {Muinonen}, K. and {Osborne}, P. and {Pancino}, E. and {Pauwels}, T. and {Recio-Blanco}, A. and {Reyl{\'e}}, C. and {Riello}, M. and {Rimoldini}, L. and {Roegiers}, T. and {Rybizki}, J. and {Sarro}, L.~M. and {Siopis}, C. and {Smith}, M. and {Sozzetti}, A. and {Utrilla}, E. and {van Leeuwen}, M. and {Abbas}, U. and {{\'A}brah{\'a}m}, P. and {Abreu Aramburu}, A. and {Aerts}, C. and {Aguado}, J.~J. and {Ajaj}, M. and {Aldea-Montero}, F. and {Altavilla}, G. and {{\'A}lvarez}, M.~A. and {Alves}, J. and {Anders}, F. and {Anderson}, R.~I. and {Anglada Varela}, E. and {Antoja}, T. and {Baines}, D. and {Baker}, S.~G. and {Balaguer-N{\'u}{\~n}ez}, L. and {Balbinot}, E. and {Balog}, Z. and {Barache}, C. and {Barbato}, D. and {Barros}, M. and {Barstow}, M.~A. and {Bartolom{\'e}}, S. and {Bassilana}, J.-L. and {Bauchet}, N. and {Becciani}, U. and {Bellazzini}, M. and {Berihuete}, A. and {Bernet}, M. and {Bertone}, S. and {Bianchi}, L. and {Binnenfeld}, A. and {Blanco-Cuaresma}, S. and {Blazere}, A. and {Boch}, T. and {Bombrun}, A. and {Bossini}, D. and {Bouquillon}, S. and {Bragaglia}, A. and {Bramante}, L. and {Breedt}, E. and {Bressan}, A. and {Brouillet}, N. and {Brugaletta}, E. and {Bucciarelli}, B. and {Burlacu}, A. and {Butkevich}, A.~G. and {Buzzi}, R. and {Caffau}, E. and {Cancelliere}, R. and {Cantat-Gaudin}, T. and {Carballo}, R. and {Carlucci}, T. and {Carnerero}, M.~I. and {Carrasco}, J.~M. and {Casamiquela}, L. and {Castellani}, M. and {Castro-Ginard}, A. and {Chaoul}, L. and {Charlot}, P. and {Chemin}, L. and {Chiaramida}, V. and {Chiavassa}, A. and {Chornay}, N. and {Comoretto}, G. and {Contursi}, G. and {Cooper}, W.~J. and {Cornez}, T. and {Cowell}, S. and {Crifo}, F. and {Cropper}, M. and {Crosta}, M. and {Crowley}, C. and {Dafonte}, C. and {Dapergolas}, A. and {David}, M. and {David}, P. and {de Laverny}, P. and {De Luise}, F. and {De March}, R.},
        title = "{Gaia Data Release 3. Summary of the content and survey properties}",
      journal = {\aap},
         year = 2023,
        month = jun,
       volume = {674},
          eid = {A1},
        pages = {A1},
          doi = {10.1051/0004-6361/202243940},
archivePrefix = {arXiv},
       eprint = {2208.00211},
 primaryClass = {astro-ph.GA},
       adsurl = {https://ui.adsabs.harvard.edu/abs/2023A&A...674A...1G}
}

@ARTICLE{2026SSRv..222...32S,
       author = {{Scaringi}, Simone and {Knigge}, Christian and {de Martino}, Domitilla},
        title = "{Accreting White Dwarfs: An Unreview}",
      journal = {\ssr},
         year = 2026,
        month = mar,
       volume = {222},
       number = {3},
          eid = {32},
        pages = {32},
          doi = {10.1007/s11214-026-01290-x},
archivePrefix = {arXiv},
       eprint = {2603.10150},
 primaryClass = {astro-ph.HE},
       adsurl = {https://ui.adsabs.harvard.edu/abs/2026SSRv..222...32S}
}

@BOOK{2003cvs..book.....W,
       author = {{Warner}, Brian},
        title = "{Cataclysmic Variable Stars}",
         year = 2003,
          doi = {10.1017/CBO9780511586491},
       adsurl = {https://ui.adsabs.harvard.edu/abs/2003cvs..book.....W}
}

@ARTICLE{2009MNRAS.397.2170G,
       author = {{G{\"a}nsicke}, B.~T. and {Dillon}, M. and {Southworth}, J. and {Thorstensen}, J.~R. and {Rodr{\'\i}guez-Gil}, P. and {Aungwerojwit}, A. and {Marsh}, T.~R. and {Szkody}, P. and {Barros}, S.~C.~C. and {Casares}, J. and {de Martino}, D. and {Groot}, P.~J. and {Hakala}, P. and {Kolb}, U. and {Littlefair}, S.~P. and {Mart{\'\i}nez-Pais}, I.~G. and {Nelemans}, G. and {Schreiber}, M.~R.},
        title = "{SDSS unveils a population of intrinsically faint cataclysmic variables at the minimum orbital period}",
      journal = {\mnras},
         year = 2009,
        month = aug,
       volume = {397},
       number = {4},
        pages = {2170-2188},
          doi = {10.1111/j.1365-2966.2009.15126.x},
archivePrefix = {arXiv},
       eprint = {0905.3476},
 primaryClass = {astro-ph.SR},
       adsurl = {https://ui.adsabs.harvard.edu/abs/2009MNRAS.397.2170G}
}

@ARTICLE{2020AJ....159..198S,
       author = {{Szkody}, Paula and {Dicenzo}, Brooke and {Ho}, Anna Y.~Q. and {Hillenbrand}, Lynne A. and {van Roestel}, Jan and {Ridder}, Margaret and {DeJesus Lima}, Isabel and {Graham}, Melissa L. and {Bellm}, Eric C. and {Burdge}, Kevin and {Kupfer}, Thomas and {Prince}, Thomas A. and {Masci}, Frank J. and {Mr{\'o}z}, Przemyslaw J. and {Golkhou}, V. Zach and {Coughlin}, Michael and {Cunningham}, Virginia A. and {Dekany}, Richard and {Graham}, Matthew J. and {Hale}, David and {Kaplan}, David and {Kasliwal}, Mansi M. and {Miller}, Adam A. and {Neill}, James D. and {Patterson}, Maria T. and {Riddle}, Reed and {Smith}, Roger and {Soumagnac}, Maayane T.},
        title = "{Cataclysmic Variables in the First Year of the Zwicky Transient Facility}",
      journal = {\aj},
         year = 2020,
        month = may,
       volume = {159},
       number = {5},
          eid = {198},
        pages = {198},
          doi = {10.3847/1538-3881/ab7cce},
archivePrefix = {arXiv},
       eprint = {2002.08447},
 primaryClass = {astro-ph.SR},
       adsurl = {https://ui.adsabs.harvard.edu/abs/2020AJ....159..198S}
}

@ARTICLE{1996A&A...307..459M,
       author = {{Motch}, C. and {Haberl}, F. and {Guillout}, P. and {Pakull}, M. and {Reinsch}, K. and {Krautter}, J.},
        title = "{New cataclysmic variables from the ROSAT All-Sky Survey.}",
      journal = {\aap},
         year = 1996,
        month = mar,
       volume = {307},
        pages = {459-469},
       adsurl = {https://ui.adsabs.harvard.edu/abs/1996A&A...307..459M}
}

@ARTICLE{2002AJ....123..430S,
       author = {{Szkody}, Paula and {Anderson}, Scott F. and {Ag{\"u}eros}, Marcel and {Covarrubias}, Ricardo and {Bentz}, Misty and {Hawley}, Suzanne and {Margon}, Bruce and {Voges}, Wolfgang and {Henden}, Arne and {Knapp}, Gillian R. and {Vanden Berk}, Daniel E. and {Rest}, Armin and {Miknaitis}, Gajus and {Magnier}, Eugene and {Brinkmann}, J. and {Csabai}, I. and {Harvanek}, M. and {Hindsley}, R. and {Hennessy}, G. and {Ivezic}, Z. and {Kleinman}, S.~J. and {Lamb}, D.~Q. and {Long}, D. and {Newman}, P.~R. and {Neilsen}, E.~H. and {Nichol}, R.~C. and {Nitta}, A. and {Schneider}, D.~P. and {Snedden}, S.~A. and {York}, D.~G.},
        title = "{Cataclysmic Variables from The Sloan Digital Sky Survey. I. The First Results}",
      journal = {\aj},
         year = 2002,
        month = jan,
       volume = {123},
       number = {1},
        pages = {430-442},
          doi = {10.1086/324734},
archivePrefix = {arXiv},
       eprint = {astro-ph/0110291},
 primaryClass = {astro-ph},
       adsurl = {https://ui.adsabs.harvard.edu/abs/2002AJ....123..430S}
}

@ARTICLE{2024PASP..136e4201R,
       author = {{Rodriguez}, Antonio C.},
        title = "{From Active Stars to Black Holes: A Discovery Tool for Galactic X-Ray Sources}",
      journal = {\pasp},
         year = 2024,
        month = may,
       volume = {136},
       number = {5},
          eid = {054201},
        pages = {054201},
          doi = {10.1088/1538-3873/ad357c},
archivePrefix = {arXiv},
       eprint = {2401.09537},
 primaryClass = {astro-ph.HE},
       adsurl = {https://ui.adsabs.harvard.edu/abs/2024PASP..136e4201R}
}

@ARTICLE{2024A&A...690A.374G,
       author = {{Galiullin}, Ilkham and {Rodriguez}, Antonio C. and {El-Badry}, Kareem and {Szkody}, Paula and {Anand}, Abhijeet and {van Roestel}, Jan and {Sibgatullin}, Askar and {Dodon}, Vladislav and {Tyrin}, Nikita and {Caiazzo}, Ilaria and {Graham}, Matthew J. and {Laher}, Russ R. and {Kulkarni}, Shrinivas R. and {Prince}, Thomas A. and {Riddle}, Reed and {Vanderbosch}, Zachary P. and {Wold}, Avery},
        title = "{Searching for new cataclysmic variables in the Chandra Source Catalog}",
      journal = {\aap},
         year = 2024,
        month = oct,
       volume = {690},
          eid = {A374},
        pages = {A374},
          doi = {10.1051/0004-6361/202450734},
archivePrefix = {arXiv},
       eprint = {2408.00078},
 primaryClass = {astro-ph.HE},
       adsurl = {https://ui.adsabs.harvard.edu/abs/2024A&A...690A.374G}
}

@ARTICLE{2025PASP..137a4201R,
       author = {{Rodriguez}, Antonio C. and {El-Badry}, Kareem and {Suleimanov}, Valery and {Pala}, Anna F. and {Kulkarni}, Shrinivas R. and {Gaensicke}, Boris and {Mori}, Kaya and {Rich}, R. Michael and {Sarkar}, Arnab and {Bao}, Tong and {Lopes de Oliveira}, Raimundo and {Ramsay}, Gavin and {Szkody}, Paula and {Graham}, Matthew and {Prince}, Thomas A. and {Caiazzo}, Ilaria and {Vanderbosch}, Zachary P. and {van Roestel}, Jan and {Das}, Kaustav K. and {Qin}, Yu-Jing and {Kasliwal}, Mansi M. and {Wold}, Avery and {Groom}, Steven L. and {Reiley}, Daniel and {Riddle}, Reed},
        title = "{Cataclysmic Variables and AM CVn Binaries in SRG/eROSITA + Gaia: Volume Limited Samples, X-Ray Luminosity Functions, and Space Densities}",
      journal = {\pasp},
         year = 2025,
        month = jan,
       volume = {137},
       number = {1},
          eid = {014201},
        pages = {014201},
          doi = {10.1088/1538-3873/ada185},
archivePrefix = {arXiv},
       eprint = {2408.16053},
 primaryClass = {astro-ph.HE},
       adsurl = {https://ui.adsabs.harvard.edu/abs/2025PASP..137a4201R}
}

@ARTICLE{2025ApJ...991..125L,
       author = {{Liu}, Hao-Bin and {Gu}, Wei-Min and {Lu}, Yongqi and {Liu}, Teng and {Liu}, Jin-Zhong},
        title = "{Searching for Accreting Compact Object Binaries in SRG/eROSITA eRASS1}",
      journal = {\apj},
         year = 2025,
        month = oct,
       volume = {991},
       number = {2},
          eid = {125},
        pages = {125},
          doi = {10.3847/1538-4357/adfecd},
archivePrefix = {arXiv},
       eprint = {2505.10478},
 primaryClass = {astro-ph.HE},
       adsurl = {https://ui.adsabs.harvard.edu/abs/2025ApJ...991..125L}
}

@dataset{2026yCat.5162....0L,
       author = {{Luo}, A.-L. and {Zhao}, Y.-H. and {Zhao}, G. and {et al.}},
        title = "{VizieR Online Data Catalog: LAMOST DR11 catalogs (Luo+, 2026)}",
 howpublished = {VizieR On-line Data Catalog: V/162.  Originally published in: 2026RAA..in.prep..L},
         year = 2026,
        month = jan,
          eid = {V/162},
       adsurl = {https://ui.adsabs.harvard.edu/abs/2026yCat.5162....0L}
}

@ARTICLE{2020MNRAS.494.3799P,
       author = {{Pala}, A.~F. and {G{\"a}nsicke}, B.~T. and {Breedt}, E. and {Knigge}, C. and {Hermes}, J.~J. and {Gentile Fusillo}, N.~P. and {Hollands}, M.~A. and {Naylor}, T. and {Pelisoli}, I. and {Schreiber}, M.~R. and {Toonen}, S. and {Aungwerojwit}, A. and {Cukanovaite}, E. and {Dennihy}, E. and {Manser}, C.~J. and {Pretorius}, M.~L. and {Scaringi}, S. and {Toloza}, O.},
        title = "{A Volume-limited Sample of Cataclysmic Variables from Gaia DR2: Space Density and Population Properties}",
      journal = {\mnras},
         year = 2020,
        month = may,
       volume = {494},
       number = {3},
        pages = {3799-3827},
          doi = {10.1093/mnras/staa764},
archivePrefix = {arXiv},
       eprint = {1907.13152},
 primaryClass = {astro-ph.SR},
       adsurl = {https://ui.adsabs.harvard.edu/abs/2020MNRAS.494.3799P}
}

@ARTICLE{2021MNRAS.504.2420I,
       author = {{Inight}, K. and {G{\"a}nsicke}, Boris T. and {Breedt}, E. and {Marsh}, T.~R. and {Pala}, A.~F. and {Raddi}, R.},
        title = "{Towards a volumetric census of close white dwarf binaries - I. Reference samples}",
      journal = {\mnras},
         year = 2021,
        month = jun,
       volume = {504},
       number = {2},
        pages = {2420-2442},
          doi = {10.1093/mnras/stab753},
archivePrefix = {arXiv},
       eprint = {2103.06892},
 primaryClass = {astro-ph.SR},
       adsurl = {https://ui.adsabs.harvard.edu/abs/2021MNRAS.504.2420I}
}

@ARTICLE{2024A&A...682A...5V,
       author = {{Vincent}, O. and {Barstow}, M.~A. and {Jordan}, S. and {Mander}, C. and {Bergeron}, P. and {Dufour}, P.},
        title = "{Classification and parameterization of a large Gaia sample of white dwarfs using XP spectra}",
      journal = {\aap},
         year = 2024,
        month = feb,
       volume = {682},
          eid = {A5},
        pages = {A5},
          doi = {10.1051/0004-6361/202347694},
archivePrefix = {arXiv},
       eprint = {2308.05572},
 primaryClass = {astro-ph.SR},
       adsurl = {https://ui.adsabs.harvard.edu/abs/2024A&A...682A...5V}
}

\begin{appendix}

\section{ H\texorpdfstring{$\alpha$}{alpha}  line identification}
\label{app:lines}

\begin{figure*}
    \centering
    $\vcenter{\hbox{\includegraphics[width = 0.4\linewidth]{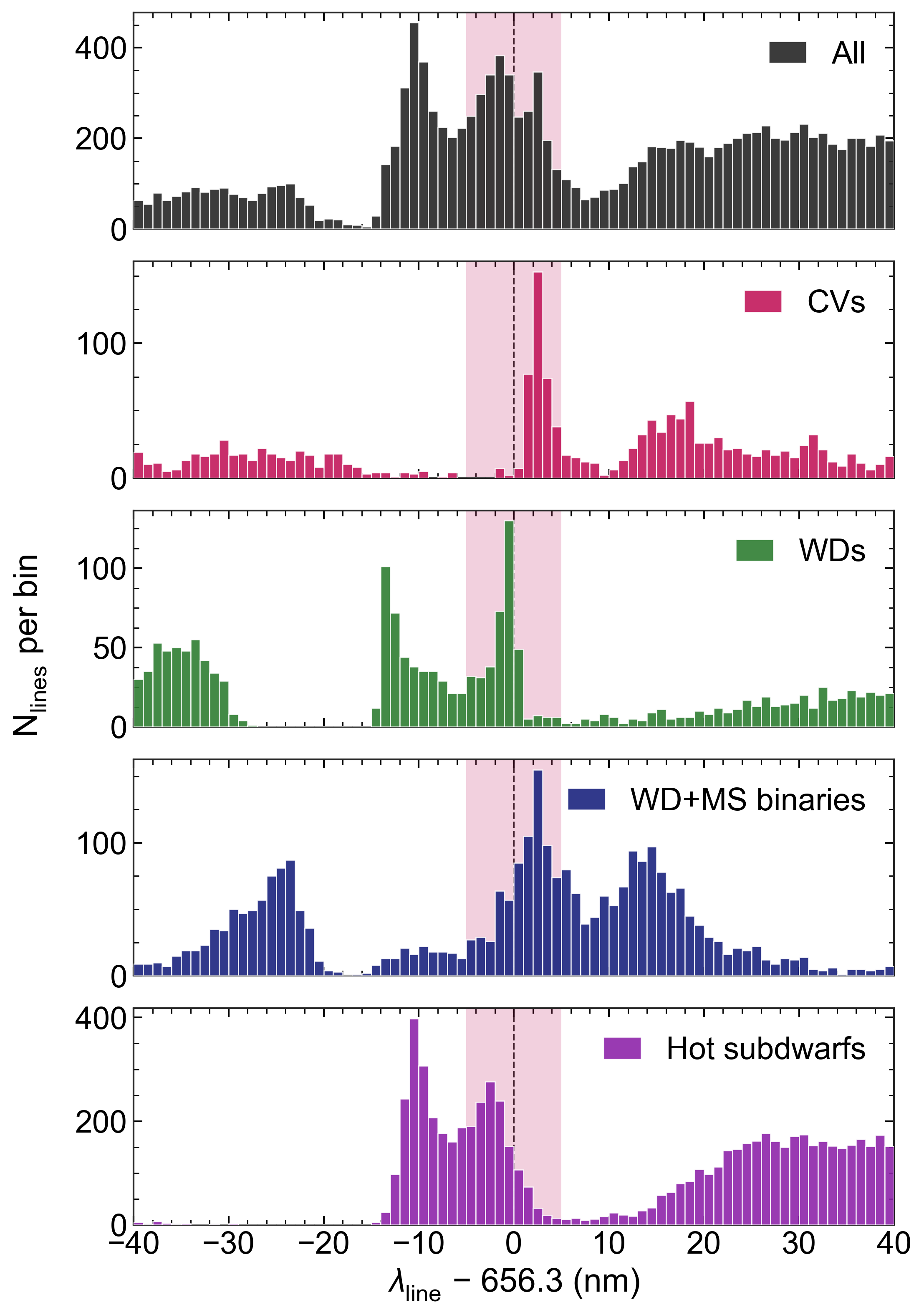}}}$
    $\vcenter{\hbox{\includegraphics[width = 0.4\linewidth]{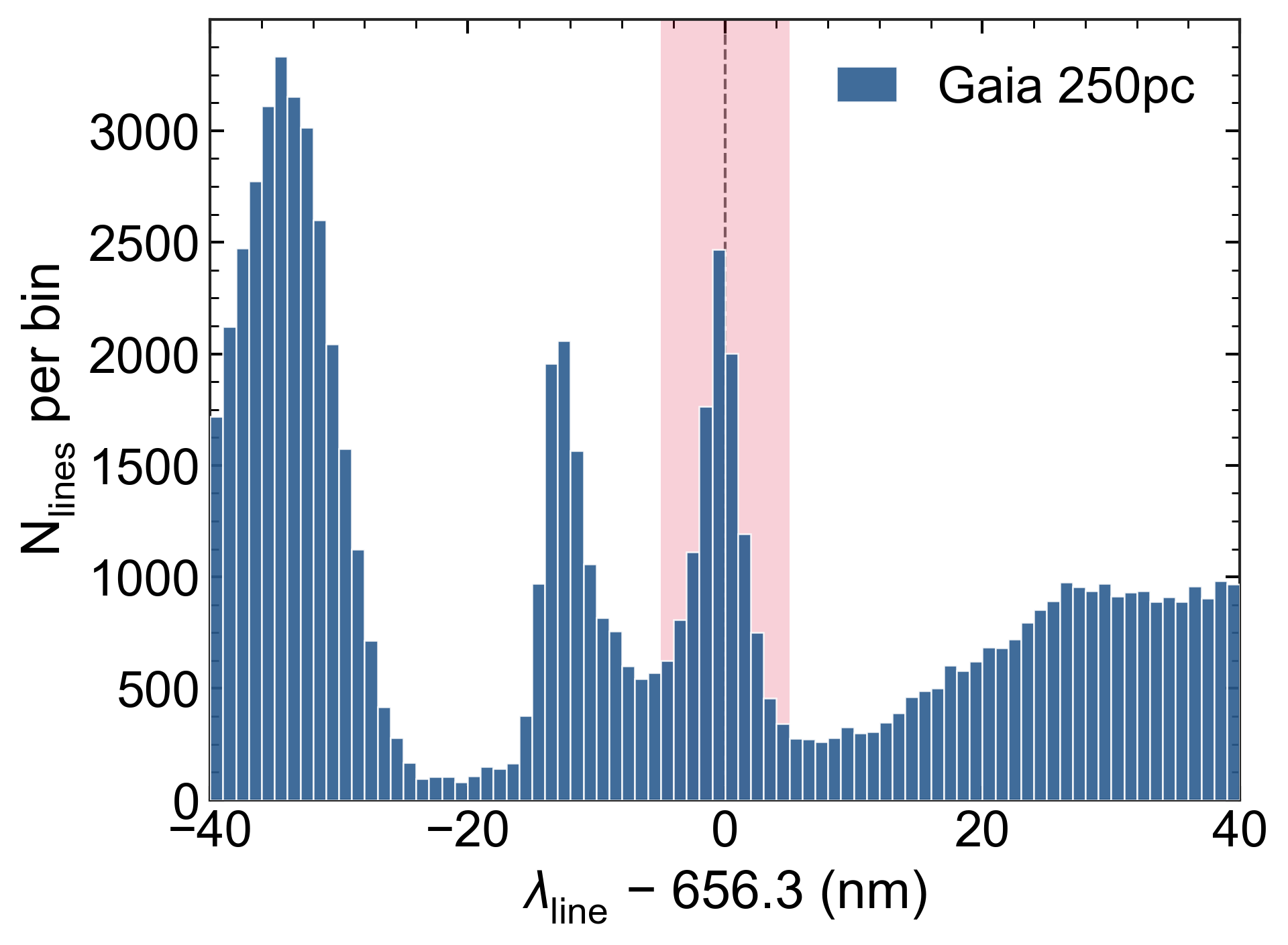}}}$
    \caption{Distribution of the detected extrema wavelengths ($\lambda_{\rm line} - 656.3$\,nm) for each stellar population (left) and the volume-limited {\it Gaia} sample within 250~pc located below the main sequence on the {\it Gaia} HR diagram (right). The empirical 5\,nm search window for the H$\alpha$ line is marked by the shaded pink area. }
    \label{fig:lines}
\end{figure*}

\begin{figure*}
    \centering
    $\vcenter{\hbox{\includegraphics[width = 0.4\linewidth]{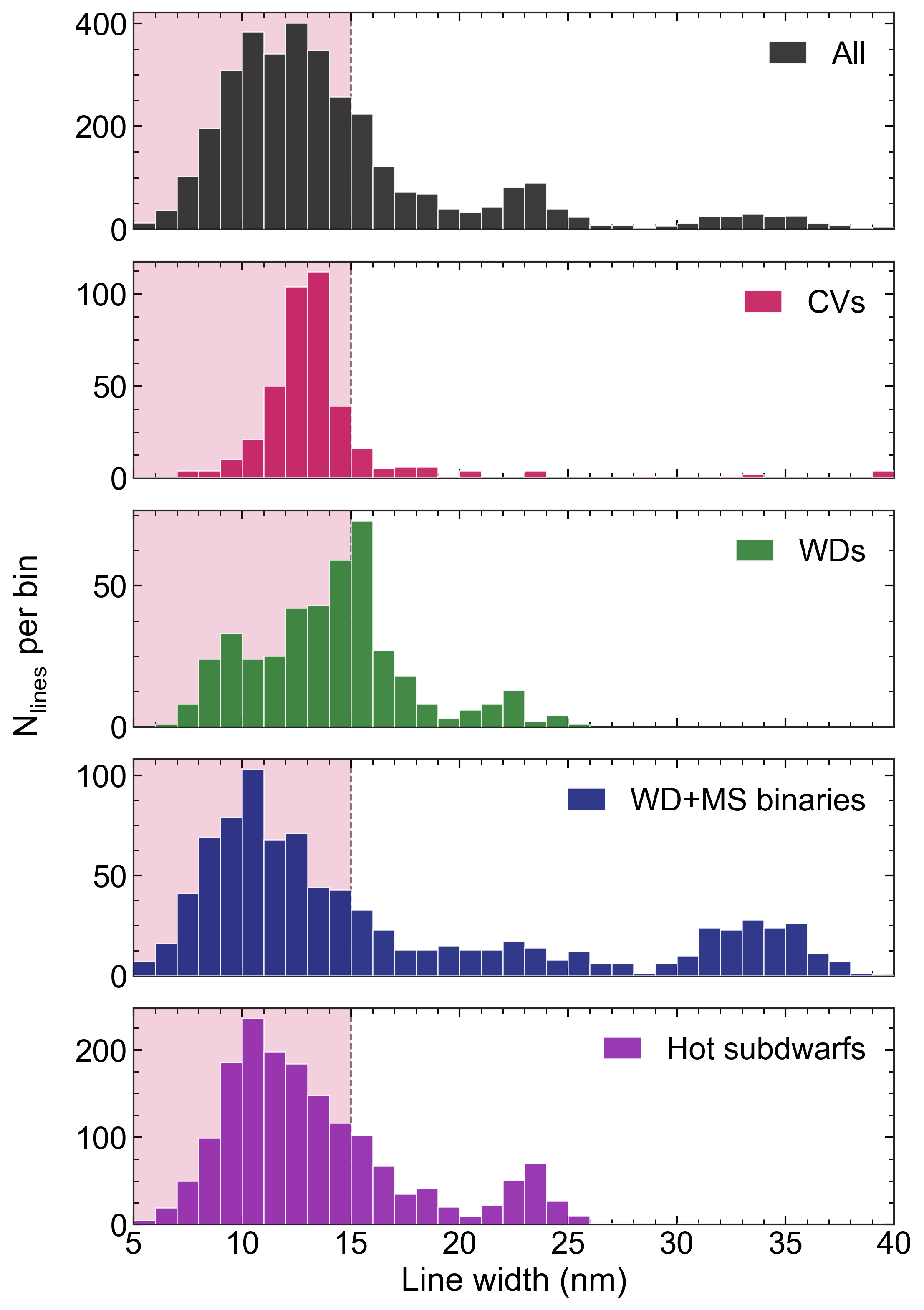}}}$
    $\vcenter{\hbox{\includegraphics[width = 0.4\linewidth]{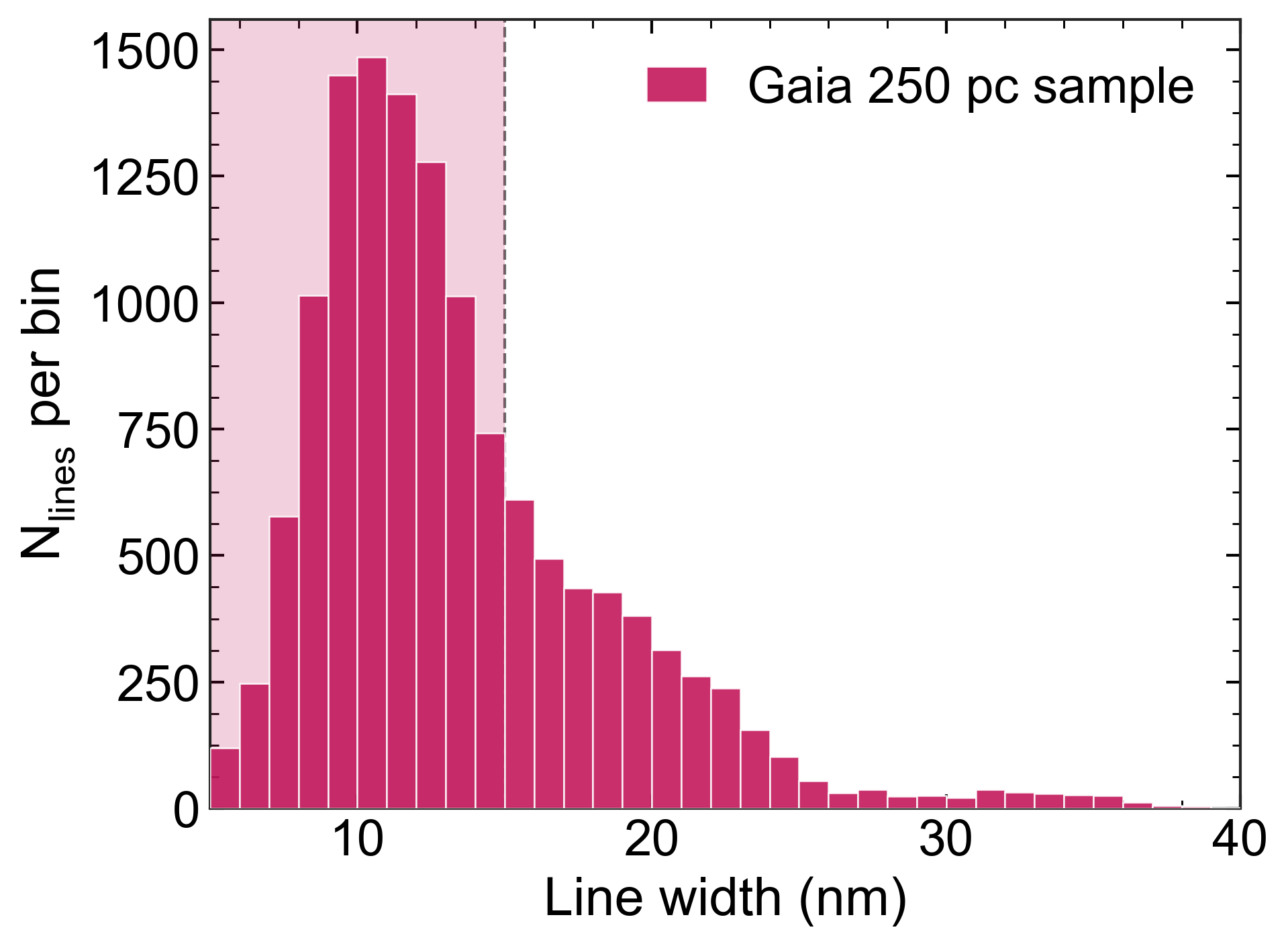}}}$
    \caption{Distribution of the detected line widths (in\,nm) within the 5\,nm search window for each stellar population (left) and the volume-limited {\it Gaia} sample within 250~pc located below the main sequence on the {\it Gaia} HR diagram (right). The empirical selection threshold (${\tt width} \le 15$\,nm) is marked by the shaded pink area.}
    \label{fig:widths}
\end{figure*}

In this Appendix, we discuss the empirical thresholds applied to our H$\alpha$ line identification pipeline using the output list from the {\tt find\_extrema} tool. 

Left panel of Figure~\ref{fig:lines} shows the distribution of all detected extrema wavelengths ($\lambda_{\rm line} - 656.3$~nm) across different stellar populations: CVs, WDs, WD+MS binaries, and hot subdwarfs. No filtering based on the {\tt sig\_pwl} parameter was applied to these detected extrema. The wavelength distribution for CVs is narrow and drops off significantly beyond $\approx 4$~nm, with a similar shape observed for WDs. In contrast, the WD+MS binaries exhibit a wider distribution. This broadening is primarily driven by the molecular absorption bands of M-dwarf companions, which are smoothed out by the low-resolution nature of the {\it Gaia}~XP spectra. The distribution for hot subdwarfs is also wide, extending past $\approx 6$~nm. Right panel of Figure~\ref{fig:lines} shows the distribution of all detected extrema wavelengths for the {\it Gaia} sample within 250~pc located below the main sequence on the {\it Gaia}~HR diagram. The 5~nm search window isolates the true H$\alpha$ line, specifically for CVs. In contrast, a wider window would include artifact peaks and pseudo-lines, introducing false-positive sources to the GEM diagram. While individual stellar subclasses in the left panel of Figure~\ref{fig:lines} show asymmetric or distinct profile variations, the collective distribution for the bulk of the {\it Gaia} 250~pc sample is remarkably symmetric. The empirical 5~nm window thus serves as a clean, symmetric boundary for H$\alpha$ line identification.

Figure~\ref{fig:widths} shows the distribution of the detected line widths within the 5~nm search window for each stellar population (left panel) and for the {\it Gaia} 250~pc sample (right panel). The line width distribution for CVs primarily extends up to 15~nm, with a few outliers above this value. In contrast, the distributions for WDs, hot subdwarfs, and WD+MS binaries extend well beyond 15~nm, reaching widths up to 40~nm. The ${\tt width} \le 15$~nm cut excludes broad, smoothed continuum features or fluctuations that do not correspond to true H$\alpha$ lines.

Based on the empirical distributions shown in Figures~\ref{fig:lines} and \ref{fig:widths}, we adopt a 5~nm search window and a line width threshold of ${\tt width} \le 15$~nm for the H$\alpha$ line identification in this work. All sources presented in the GEM diagram are selected using these exact criteria. We note that while alternative thresholds could be chosen to prioritize either the completeness or the purity of the sample, our current selections are optimized specifically for the GEM diagram and are applied uniformly to all objects.

\section{Gaia 250 pc sample}
\label{app:identification}

In this Appendix, we describe in more detail the identification of known sources from the {\it Gaia} 250~pc sample located above our empirical selection line on the GEM diagram. We present the individual properties of our three newly identified CV candidates.

\subsection{Known source identification}

\begin{figure*}
    \centering
    \resizebox{\linewidth}{!}{%
        \includegraphics[height=10cm]{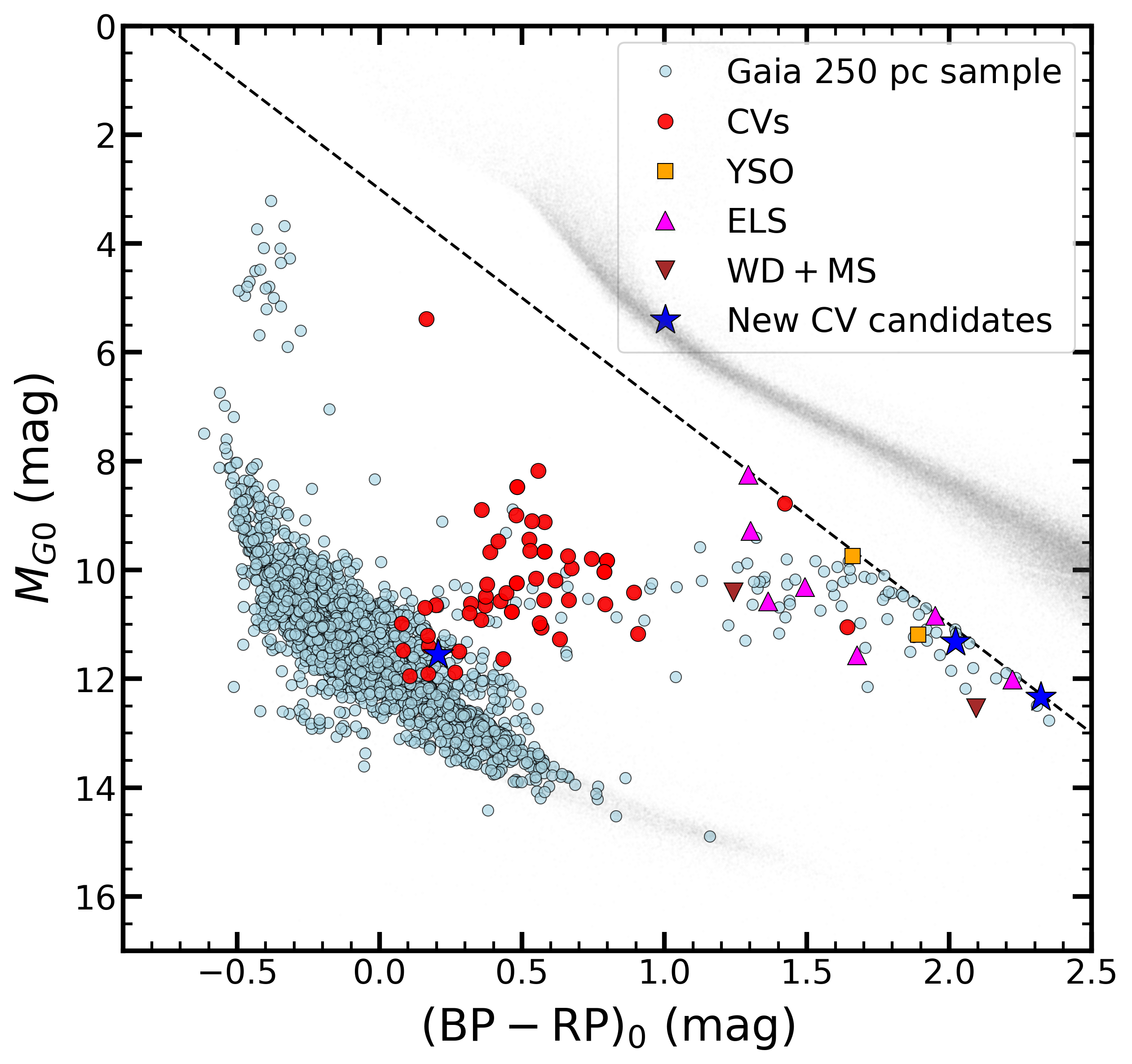}%
        \includegraphics[height=10cm]{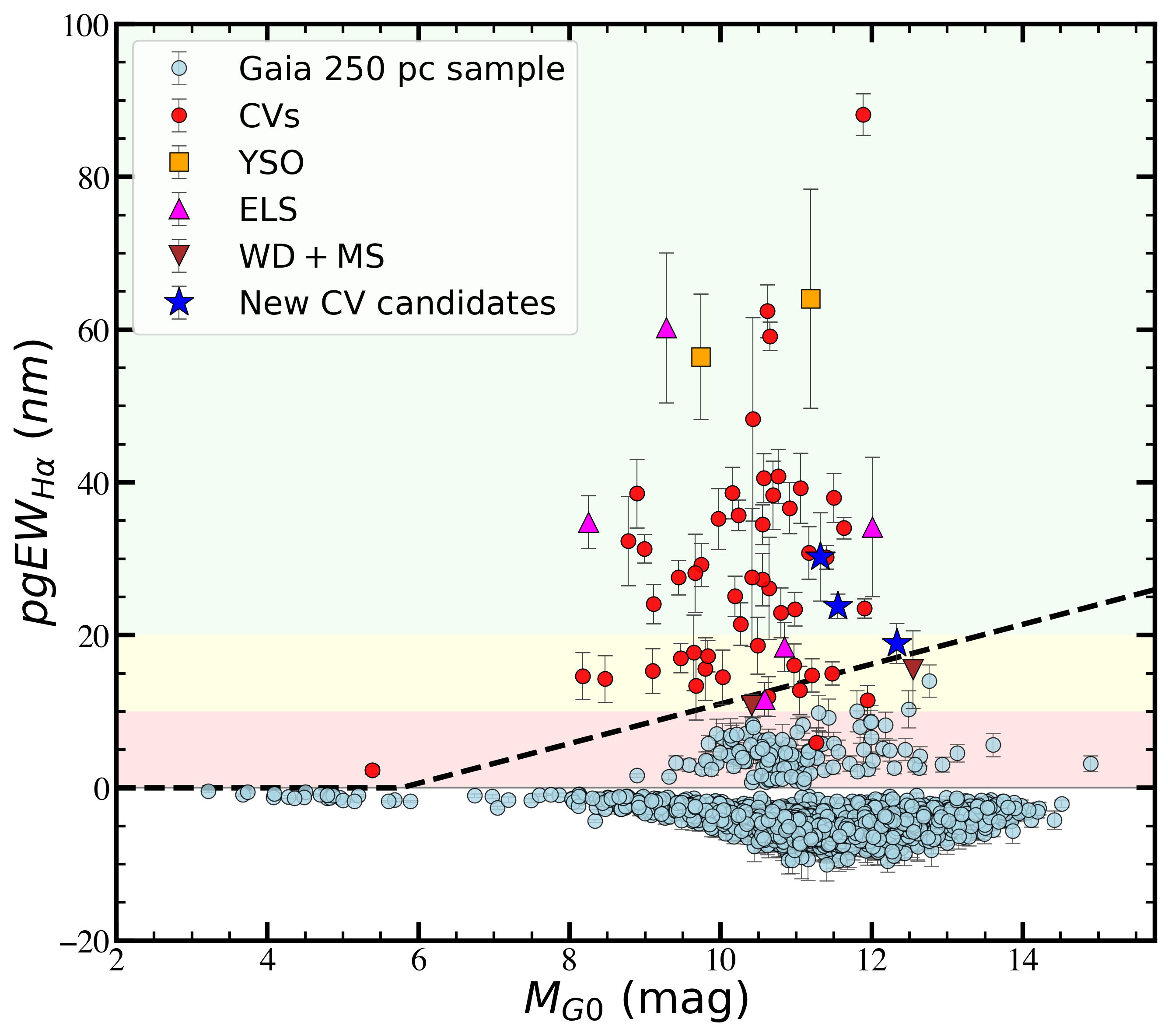}%
        }
    \caption{Same description as in  Figure~\ref{fig:250pc}, but focusing on the 62 objects located above our empirical selection cut. Symbols correspond to different object types identified via the SIMBAD database. New CV candidates are marked with blue stars.}
    \label{fig:250pciden}
\end{figure*}

\begin{figure*}
    \centering
    \resizebox{\linewidth}{!}{%
        \includegraphics[height=8cm]{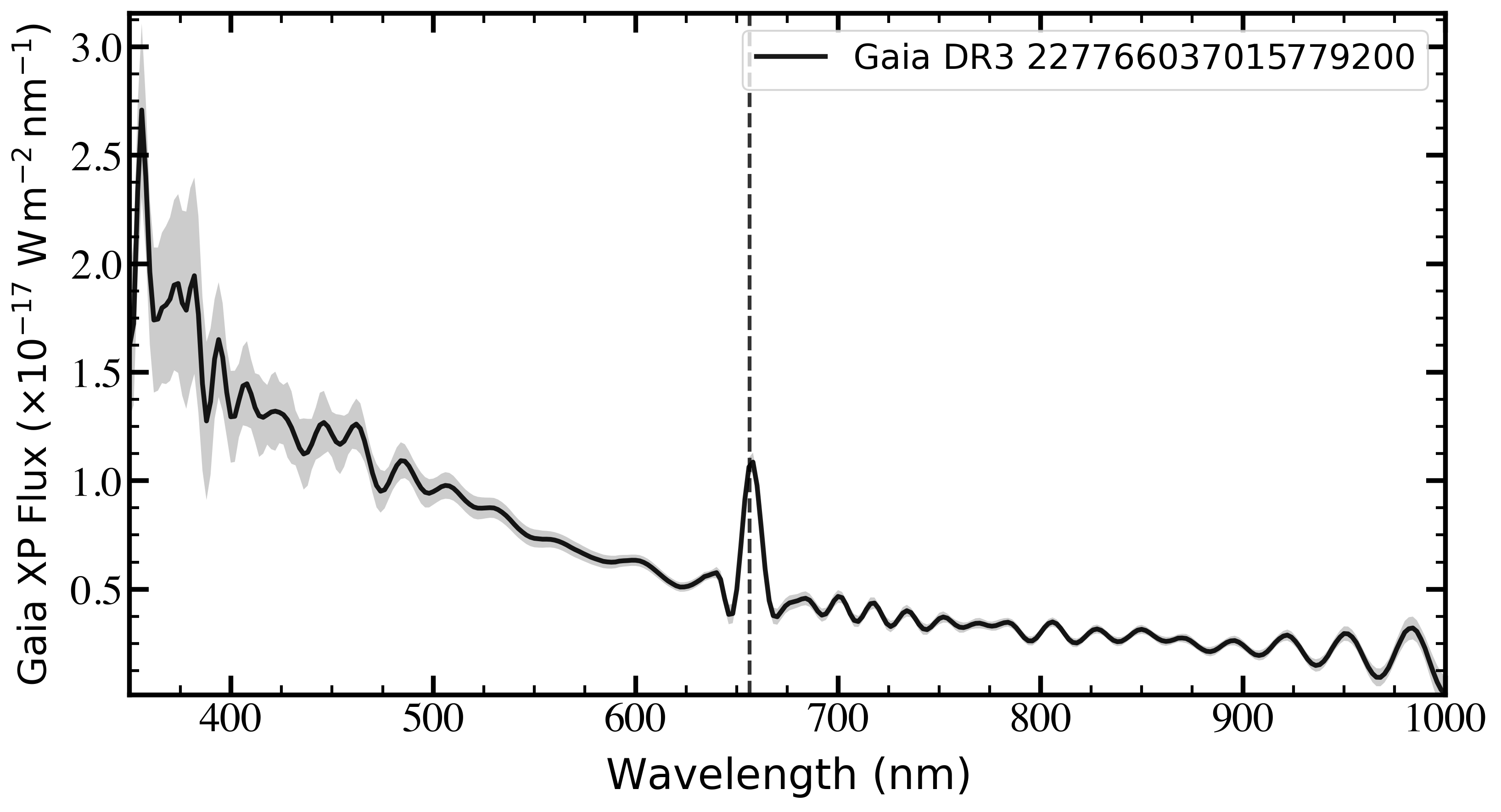}
        \includegraphics[height=8cm]{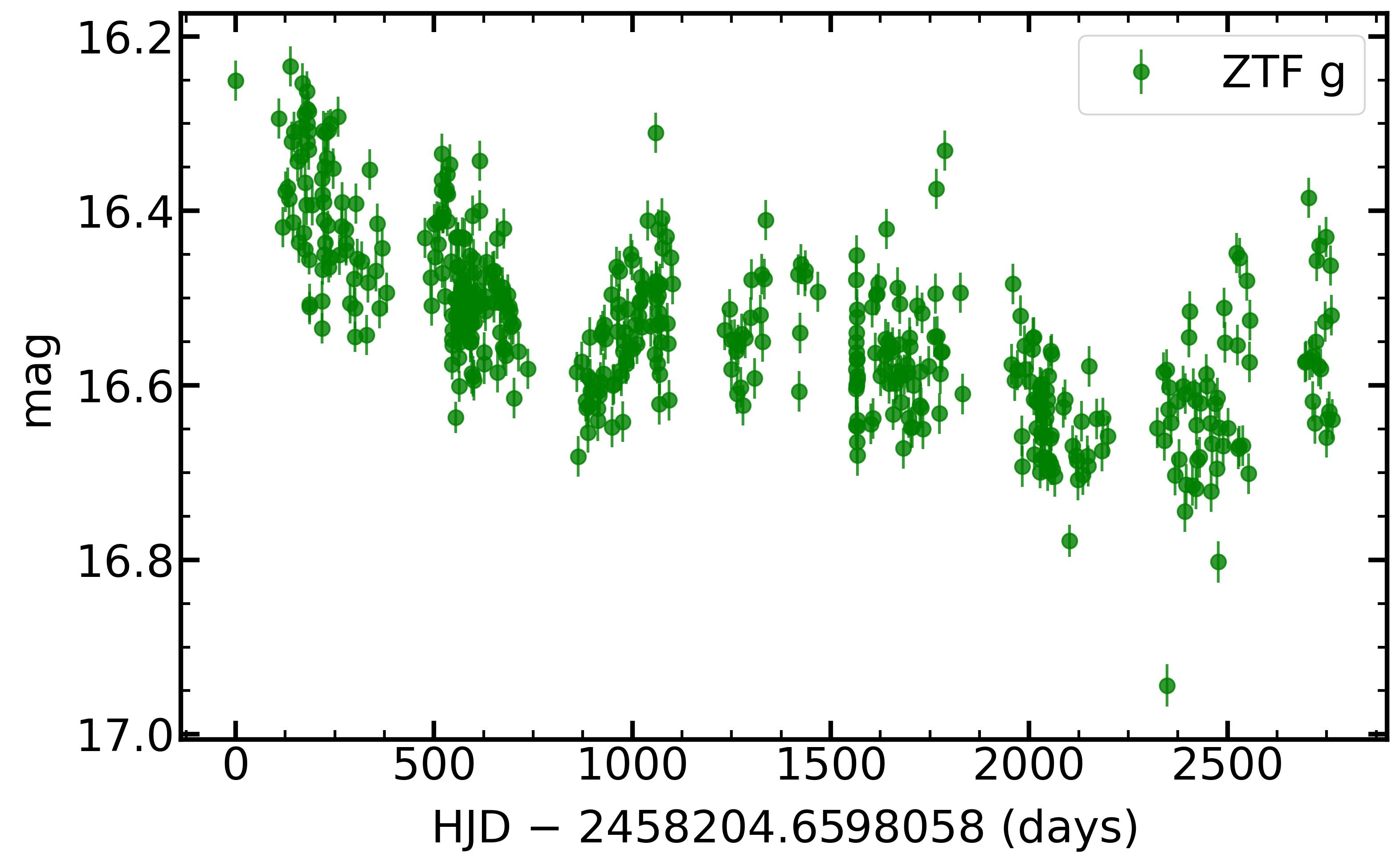}
        }
    \resizebox{\linewidth}{!}{%
        \includegraphics[height=8cm]{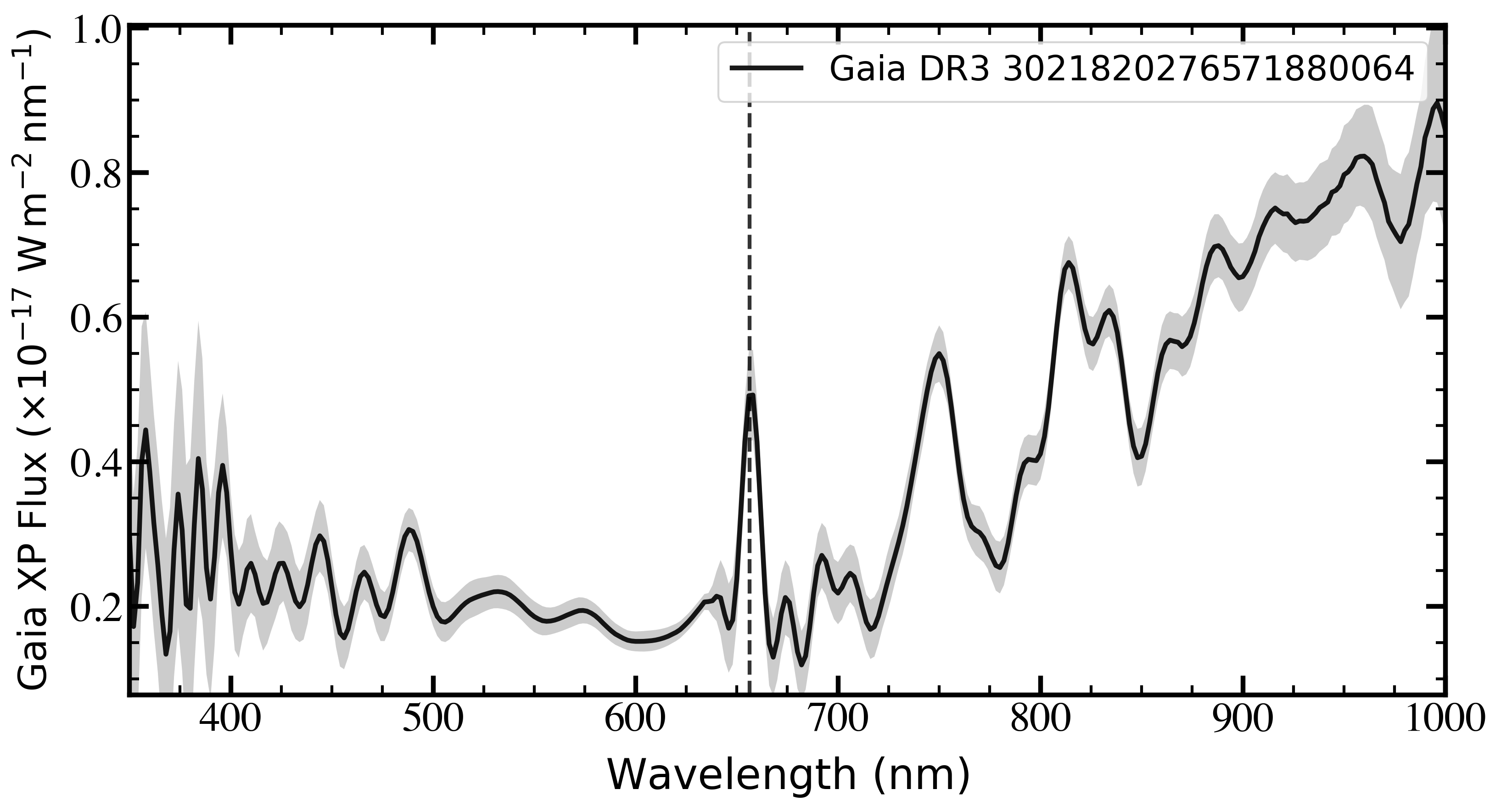}
        \includegraphics[height=8cm]{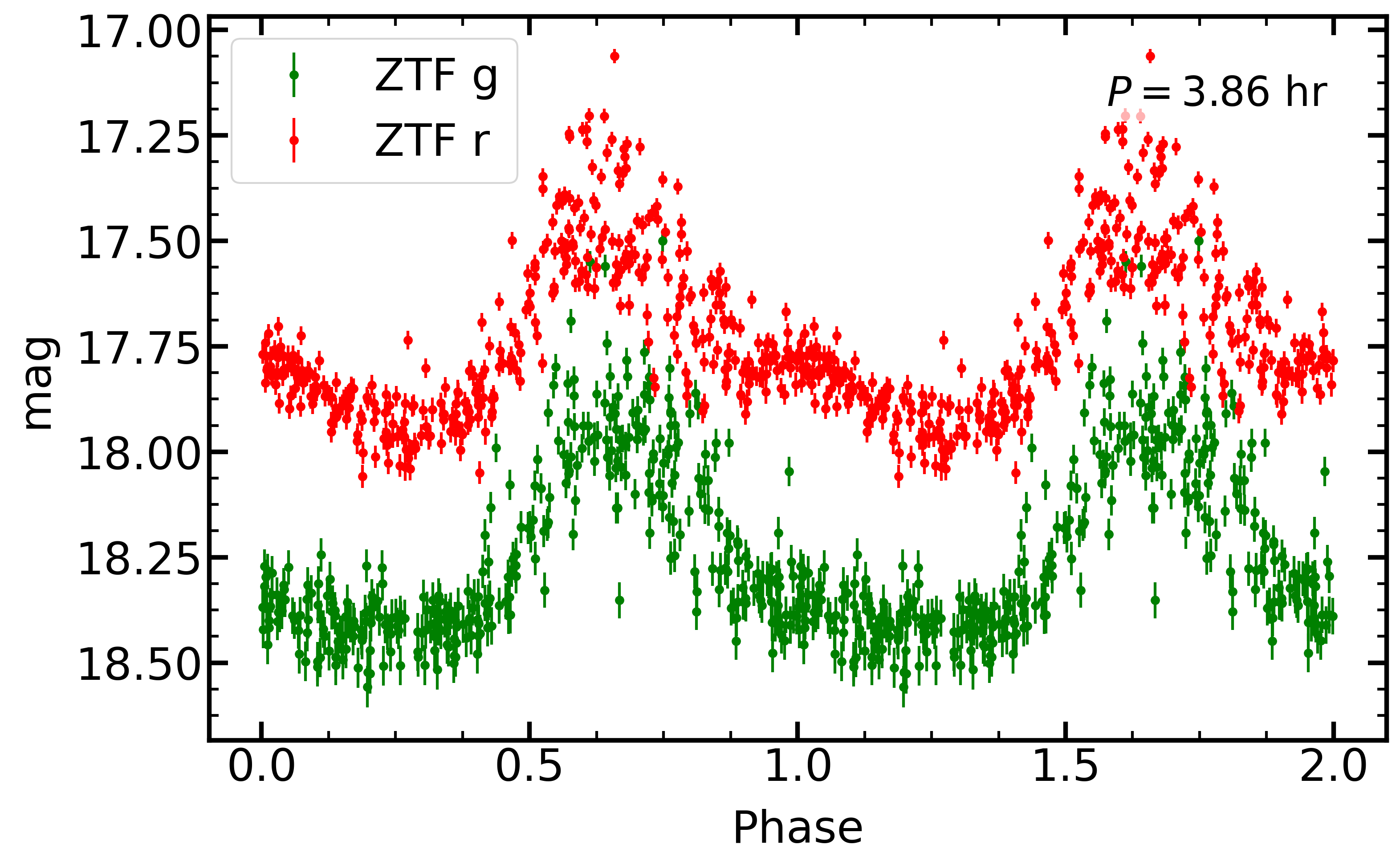}
        }
    \resizebox{\linewidth}{!}{%
        \includegraphics[height=8cm]{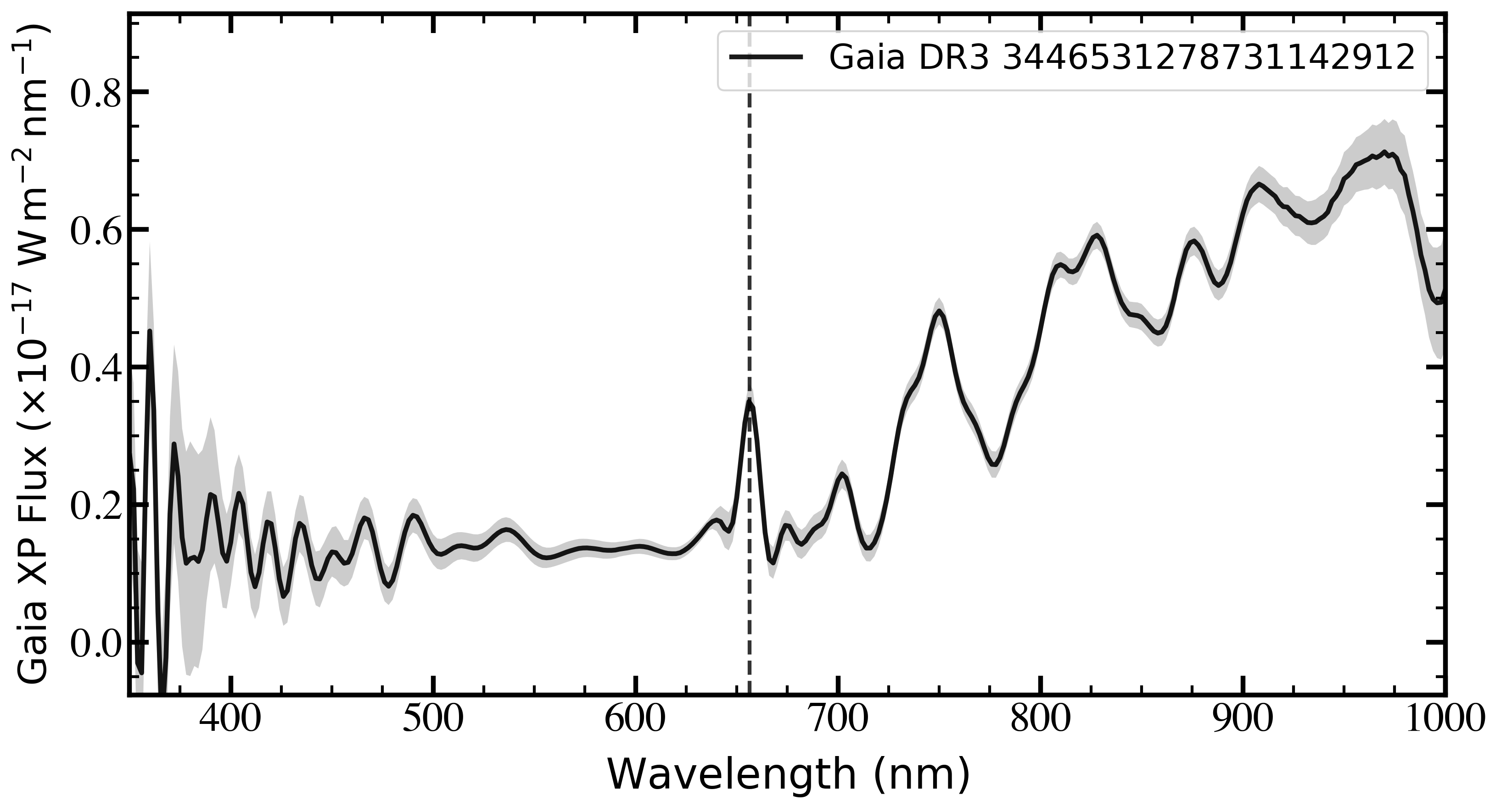}
        \includegraphics[height=8cm]{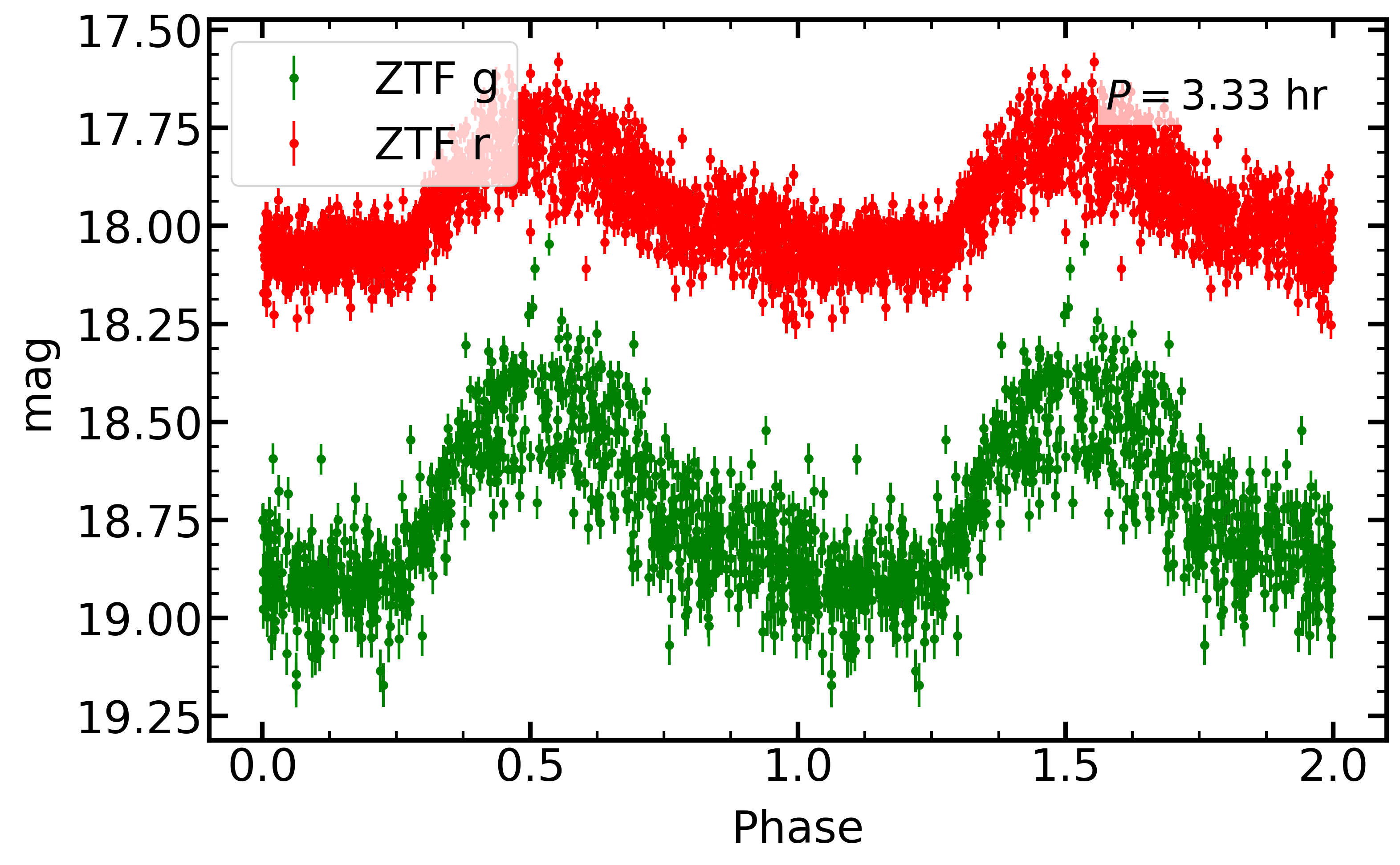}
        }
    \caption{{\it Gaia} XP spectra (left) and ZTF light curves (right) of the new CV candidates identified using the GEM diagram. The H$\alpha$ line is marked by a dashed line on the spectra. The ZTF light-curve panels show the long-term data (top) and the phase-folded data (middle and bottom). }
    \label{fig:cvcandidates}
\end{figure*}

Based on our empirical selection line in Equation~\ref{eq:selection}, we selected 62 objects as CV candidates satisfying the criterion $(\text{pgEW}_{\rm H\alpha} + \text{pgEW}_{\rm err}) \ge \text{pgEW}_{\rm H\alpha}^{\rm cut}$. We cross-matched this sample with the catalogs of known CVs described in Section~\ref{sec:data} using their {\it Gaia} source IDs. We also queried the SIMBAD database using a $2\,\arcsec$ search radius to find remaining known CVs. Out of these 62 objects, 43 objects were classified as known CVs. The remaining 19 sources had different SIMBAD classifications:

\begin{itemize}
 \setlength{\itemsep}{0.04em} 
    \item Emission-line stars (8 objects)
    \item White dwarfs (2 objects)
    \item Young stellar objects (2 objects)
    \item Planetary nebula candidate (1 object)
    \item Proper motion stars (1 object)
    \item Star (1 object)
    \item Eclipsing binaries (1 object)
    \item Possible CV candidates (1 object)
    \item Unclassified sources (2 objects)
\end{itemize}

We manually investigated these 19 sources by cross-matching them with the VizieR database. Out of these 19 objects, we identified 5 additional known CVs, which were originally classified in SIMBAD as a possible CV candidate (1 object), a star (1 object), a WD (1 object), or unclassified (2 objects). The remaining sources were grouped into four main categories based on their SIMBAD types and properties:
\begin{enumerate}
 \setlength{\itemsep}{0.5em} 
    \item \textbf{Emission-line stars (ELS)}: This group includes 7 objects originally classified as emission-line stars (6 objects) or a planetary nebula candidate (1 object) in the SIMBAD database.
    \item \textbf{Young stellar objects (YSO)}: This group includes 2 objects with young stellar object classifications from the SIMBAD database.
    \item \textbf{WD+MS binaries (WD+MS)}: This group includes 2 objects (originally classified as an eclipsing binary and a proper motion star) where visual inspection of their {\it Gaia} XP spectra showed clear M-dwarf type signatures alongside a hot compact component.
    \item \textbf{New CV candidates}: This group includes 3 objects originally classified as emission-line stars (2 objects) or a WD (1 object).
\end{enumerate}

Finally, out of our 62 selected CV candidates, 48 are known CVs, 7 are emission-line stars, 2 are young stellar objects, 2 are WD+MS binaries, and 3 are new CV candidates. Figure~\ref{fig:250pciden} shows the distribution of the 62 sources on the {\it Gaia} HR diagram (left) and GEM diagram (right) along with their identified groups. The young stellar objects, emission-line stars, and WD+MS binaries groups are distributed close to the main sequence on the {\it Gaia} HR diagram. While these stellar populations (except WD+MS binaries) are normally expected to lie on or above the main sequence, peculiar properties can cause them to shift downward and locate them within the CV selection zone on the GEM diagram. A detailed individual investigation of all selection outliers is beyond the scope of this paper.

Table~\ref{tab:objects} presents the full list of the 62 CV candidates located above the empirical selection line on the GEM diagram. For each object, we provide its celestial coordinates (R.A. and Dec.), {\it Gaia} apparent ($G$) and absolute magnitudes ($M_{G0}$), pseudo-Gaussian equivalent widths for the H$\alpha$ line ($\text{pgEW}_{\rm H\alpha}$), {\it Gaia} variability flags, and its primary SIMBAD classification. Individual notes for each object are given in the final column of the table. Two emission-line stars with $\text{pgEW}_{\rm H\alpha} > 200$~nm were removed from Figure~\ref{fig:gem} and Figure~\ref{fig:250pciden} for illustrative purposes.

To estimate how many known CVs might have been missed by our analysis, we cross-matched the entire {\it Gaia} 250~pc sample distributed below the main sequence (12247 objects; see Section~\ref{sec:250pc}) with the SIMBAD database and our reference CV catalogs (see Section~\ref{sec:data}) using a $2\,\arcsec$ search radius. This query revealed 74 known CVs in total within these 12247 objects\footnote{One object, Gaia~DR3~1111086122258605952, is classified as a CV by the SIMBAD database but showed an H$\alpha$ absorption line in the Gaia XP spectrum along with $\text{pgEW}_{\rm H\alpha} < 0$. Recent classification shows that this is a WD \citep{2024A&A...682A...5V}, so we removed it from the sample.}. Out of these 74 systems, 48 CVs were isolated by our empirical selection cut on the GEM diagram, indicating a completeness of $\sim$65\% for the {\it Gaia} 250~pc sample distributed below the main sequence. The remaining known CVs were missed because no H$\alpha$ emission line was detected in their {\it Gaia} XP spectra by the line-finding tool. Only two known CVs passed the H$\alpha$ line detection step but were located below the selection line on the GEM diagram, demonstrating that the majority of the known CVs are isolated by the empirical selection line with a high efficiency of $\sim$96\%.  We included these two sources in Figures~\ref{fig:250pc} and \ref{fig:250pciden} for completeness. We note that a detailed selection function and completeness analysis is beyond the scope of this work, as  {\it Gaia} 250~pc sample serves as a demonstration of the GEM diagram concept.

\subsection{New CV candidates}

We analyzed the X-ray and optical properties of the three new CV candidates identified with our empirical selection cut on the GEM diagram. We manually searched for available data in the VizieR database using a 2$\arcsec$ search radius. We used the X-ray fluxes of our targets to estimate their X-ray luminosities and X-ray-to-optical flux ratios ($F_\mathrm{X}/F_\mathrm{opt}$). Optical fluxes were derived from the {\it Gaia} $G$-band magnitudes, assuming the AB magnitude system and a central wavelength of $6000$~\AA. All X-ray fluxes were converted to the $0.5-7$~keV energy band using the WebPIMMS tool\footnote{\url{https://heasarc.gsfc.nasa.gov/cgi-bin/Tools/w3pimms/w3pimms.pl}}, assuming a power-law model with a photon index of 1.7. We investigated the presence of accretion activity using the X-ray main sequence diagram \citep{2024PASP..136e4201R}, where selection cuts from \cite{2024PASP..136e4201R} and the stricter criteria from \cite{2024A&A...690A.374G} were applied to independently verify our targets as CV candidates. We analyzed the optical variability of the CV candidates using ZTF light curves. We searched for periodicities in the range of 0.5 to 12 hours using the Lomb-Scargle periodogram, identifying the best-fit period based on the peak value of the power spectrum. The individual properties of the three new CV candidates are discussed below.

\begin{enumerate}
 \setlength{\itemsep}{1em} 
    \item \textbf{Gaia DR3 227766037015779200}: This source is classified as a WD in the SIMBAD database. The source has an apparent {\it Gaia} $G$-band magnitude of $16.6$~mag, a $\rm (BP-RP)$ color of $0.21$~mag, and a distance of $102$~pc. It is located close to the WD sequence on the {\it Gaia} HR diagram (see Figure~\ref{fig:250pciden}, left panel). The target was detected in the X-rays by both the ROSAT \citep{2016A&A...588A.103B,2022A&A...664A.105F} and Swift-XRT \citep{2023MNRAS.518..174E} observatories, showing a mean X-ray flux of $F_\mathrm{X} \approx 8.3 \times 10^{-13}$~erg~s$^{-1}$~cm$^{-2}$. This corresponds to an X-ray luminosity of $L_\mathrm{X} \approx 10^{30}$~erg~s$^{-1}$. Its X-ray-to-optical flux ratio is $F_\mathrm{X}/F_\mathrm{opt} \approx 0.2$, which significantly exceeds the selection thresholds for CV candidates proposed by both \cite{2024PASP..136e4201R} ($F_\mathrm{X}/F_\mathrm{opt} \gtrsim 5\times10^{-4}$) and \cite{2024A&A...690A.374G} ($F_\mathrm{X}/F_\mathrm{opt} \gtrsim 0.04$). The {\it Gaia} XP spectrum displays a blue continuum with an $\mathrm{H}\alpha$ emission line, a characteristic signature of CVs (see Figure~\ref{fig:cvcandidates}, left panel). Analysis of the ZTF light curve reveals only long-term variability (see Figure~\ref{fig:cvcandidates}, right panel). No significant periodicity was detected in the ZTF data, with the only notable peaks in the periodogram corresponding to 1-day daily aliases. Its position on the X-ray main sequence diagram, combined with the features in the {\it Gaia} XP spectrum, strongly indicates that this object is a highly probable CV candidate.
    
    \item \textbf{Gaia DR3 3021820276571880064}: This source is classified as an emission-line star in the SIMBAD database. The object has an apparent {\it Gaia} $G$-band magnitude of $17.2$~mag, a $\rm (BP-RP)$ color of $2.02$~mag, and a distance of $151$~pc. It is located close to the main sequence on the {\it Gaia} HR diagram (see Figure~\ref{fig:250pciden}, left panel). Notably, this object was an outlier in our original WD+MS sample used to define the GEM diagram (see Figure~\ref{fig:gem}). The target was detected in the X-rays by the SRG/eROSITA observatory \citep{2024A&A...684A.121F}, exhibiting an X-ray flux of $F_\mathrm{X} \approx 8.4 \times 10^{-14}$~erg~s$^{-1}$~cm$^{-2}$, which corresponds to an X-ray luminosity of $L_\mathrm{X} \approx 2 \times 10^{29}$~erg~s$^{-1}$. Its X-ray-to-optical flux ratio is $F_\mathrm{X}/F_\mathrm{opt} \approx 0.04$, which exceeds the baseline selection threshold for CV candidates proposed by \cite{2024PASP..136e4201R} ($F_\mathrm{X}/F_\mathrm{opt} \gtrsim 0.03$) but falls below the stricter criterion of \cite{2024A&A...690A.374G} ($F_\mathrm{X}/F_\mathrm{opt} \gtrsim 0.82$). The {\it Gaia} XP spectrum shows an $\mathrm{H}\alpha$ emission line along with M-dwarf features, such as TiO molecular bands (see Figure~\ref{fig:cvcandidates}, left panel). Analysis of the ZTF light curve reveals a period of $3.86$~hr (see Figure~\ref{fig:cvcandidates}, right panel), consistent with values from \cite{2020ApJS..249...18C}. Its position on the X-ray main sequence along with the GEM diagram suggests it is a potential CV candidate. However, the strong $\mathrm{H}\alpha$ emission could also be caused by an active M-dwarf companion, or be a blend with molecular bands at the {\it Gaia} low spectral resolution, rather than an accretion disk in a CV. The $3.86$~hr period is consistent with CVs, but could be a factor-of-two subharmonic or harmonic of the estimated period. Therefore, this object is a good CV candidate, but it could also be an active binary star.
    
    \item \textbf{Gaia DR3 3446531278731142912}: This source is classified as an emission-line star in the SIMBAD database. The object has an apparent {\it Gaia} $G$-band magnitude of $17.4$~mag, a $\rm (BP-RP)$ color of $2.32$~mag, and a distance of $103$~pc. It is located close to the main sequence on the {\it Gaia} HR diagram (see Figure~\ref{fig:250pciden}, left panel). The target was detected in X-rays by the ROSAT observatory \citep{2016A&A...588A.103B,2022A&A...664A.105F}, exhibiting an X-ray flux of $F_\mathrm{X} \approx 2 \times 10^{-13}$~erg~s$^{-1}$~cm$^{-2}$, which corresponds to an X-ray luminosity of $L_\mathrm{X} \approx 3 \times 10^{29}$~erg~s$^{-1}$. Its X-ray-to-optical flux ratio is $F_\mathrm{X}/F_\mathrm{opt} \approx 0.10$, which exceeds the baseline selection threshold for CV candidates proposed by \cite{2024PASP..136e4201R} ($F_\mathrm{X}/F_\mathrm{opt} \gtrsim 0.07$) but falls below the stricter criterion of \cite{2024A&A...690A.374G} ($F_\mathrm{X}/F_\mathrm{opt} \gtrsim 1.3$). Similar to the previous object,  the {\it Gaia} XP spectrum shows an $\mathrm{H}\alpha$ emission line along with M-dwarf features, such as TiO molecular bands (see Figure~\ref{fig:cvcandidates}, left panel). Analysis of the ZTF light curve reveals a period of $3.33$~hr (see Figure~\ref{fig:cvcandidates}, right panel), consistent with values from \cite{2020ApJS..249...18C}. Its position on the X-ray main sequence along with the GEM diagram suggests it is a potential CV candidate.  As discussed for the previous objects, $\mathrm{H}\alpha$ emission can be caused by an active M-dwarf companion or be an artifact of blending with molecular bands at the {\it Gaia} low spectral resolution. The $3.33$ hr period is consistent with CVs, but could be a factor-of-two subharmonic or harmonic of the estimated period. Therefore, this object is a good CV candidate, but it could also be an active binary star.

\end{enumerate}


\begin{table*}
\caption{Sample of 62 CV candidates located above the empirical selection line on the GEM diagram. }
\label{tab:objects}
\centering
\footnotesize
\renewcommand{\arraystretch}{0.96}
\setlength{\tabcolsep}{3.7pt}
\begin{tabular}{@{}llllllllll@{}}
\hline\hline
\noalign{\smallskip}
\# & \textit{Gaia} DR3 & R.A. & Dec. & $G$ & $M_G$ &
$\mathrm{pgEW}_{\mathrm{H}\alpha}$ & \textit{Gaia} &
SIMBAD & Note \\
& Source ID & (deg) & (deg) & (mag) & (mag) & (nm) &
Var. Flag & Class & \\
\noalign{\smallskip}
\hline
\noalign{\smallskip}

1 & 1030279027003254784 & 129.1779 & +53.4772 & 16.6 & 10.6 & 40.6 & VAR & CV (DN) & SW UMa \\
2 & 1037411284055821568 & 133.4343 & +57.8112 & 16.1 & 10.2 & 38.6 & VAR & CV (DN) & BZ UMa \\
3 & 1074721664954807040 & 175.9106 & +71.6891 & 15.2 & 8.8 & 32.3 & VAR & CV (IP) & DO Dra \\
4 & 1289860214647954816 & 225.6704 & +33.5732 & 17.3 & 10.9 & 36.6 & VAR & CV (DN) & NZ Boo \\
5 & 1558322303741820928 & 204.9212 & +48.7910 & 17.4 & 11.5 & 15.0 & VAR & CV (DN) & V355 UMa \\
6 & 1597621426298361216 & 238.3794 & +55.2707 & 17.3 & 11.0 & 12.8 & VAR & CV (P) & MQ Dra \\
7 & 176429285061830144 & 66.53892 & +35.6957 & 16.1 & 9.8 & 15.6 & VAR & CV (DN) & 1RXS J042608.9+354151 \\
8 & 1800384942558699008 & 324.5280 & +26.3325 & 15.6 & 10.7 & 59.2 & -- & CV (DN) & V627 Peg \\
9 & 1809844934461976832 & 301.9024 & +17.7040 & 15.2 & 11.9 & 23.5 & VAR & CV (DN) & WZ Sge \\
10 & 1920126431748251776 & 353.5057 & +39.3607 & 16.0 & 11.6 & 34.0 & VAR & CV (DN) & V455 And \\
11 & 2208124536065383424 & 342.6677 & +63.4771 & 16.4 & 11.9 & 88.2 & VAR & CV (DN) & GD 552 \\
12 & 2307289214897332480 & 358.2541 & --38.8632 & 16.3 & 11.4 & 30.2 & VAR & CV (DN) & BW Scl \\
13 & 2488974302977323008 & 38.61593 & --4.90868 & 15.2 & 10.6 & 26.2 & VAR & CV (DN) & ASASSN-14dx \\
14 & 2754909740118313344 & 6.295982 & +12.2866 & 17.5 & 11.5 & 38.0 & -- & CV (DN) & FL Psc \\
15 & 286497000368243200 & 83.14130 & +62.7979 & 15.5 & 9.4 & 27.6 & VAR & CV (DN) & V391 Cam \\
16 & 2923643719394227328 & 100.1986 & --24.3871 & 15.2 & 9.1 & 24.1 & VAR & CV (DN) & PU CMa \\
17 & 3071240270519385856 & 123.3269 & --1.05758 & 16.0 & 9.6 & 17.7 & VAR & CV (DN) & ASASSN-14ag \\
18 & 3116059834803771904 & 106.7040 & +3.41298 & 17.1 & 10.6 & 27.3 & VAR & CV (P) & PM J07068+0324 \\
19 & 3445477328117272576 & 82.13635 & +28.6436 & 16.4 & 9.7 & 13.4 & VAR & CV (P) & RX J0528+2838 \\
20 & 3955313418148878080 & 189.8833 & +21.1351 & 16.7 & 10.0 & 35.2 & -- & CV (DN) & IR Com \\
21 & 4107290939047483008 & 258.6777 & --29.7293 & 17.2 & 11.0 & 23.4 & -- & CV (DN) & OGLE BLG-DN-7 \\
22 & 4111991385628196224 & 257.0794 & --25.8091 & 15.4 & 10.2 & 25.1 & -- & CV (DN) & V2051 Oph \\
23 & 426306363477869696 & 17.55491 & +60.0765 & 16.3 & 10.6 & 34.5 & VAR & CV (DN) & HT Cas \\
24 & 4476137370261520000 & 270.1480 & +8.17071 & 16.6 & 11.2 & 30.8 & VAR & CV (P) & V2301 Oph \\
25 & 4485786821746343680 & 265.1911 & +6.06403 & 15.8 & 9.0 & 31.3 & VAR & CV (P) & SDSS J1740+0603 \\
26 & 4706297108508032128 & 1.640169 & --69.0094 & 15.1 & 8.2 & 14.7 & VAR & CV (DN) & HV 8001 \\
27 & 5078976609103251456 & 43.90873 & --22.7842 & 17.2 & 10.6 & 62.4 & VAR & CV (DN) & IQ Eri \\
28 & 5084805635638179584 & 53.86947 & --25.7394 & 16.6 & 9.7 & 29.2 & VAR & CV (P) & UZ For \\
29 & 5091939988633728640 & 66.17152 & --20.1200 & 11.8 & 5.4 & 2.3 & VAR & CV (NL) & IM Eri \\
30 & 5242787486412627072 & 151.5915 & --70.2346 & 15.6 & 10.8 & 22.9 & VAR & CV (DN) & OY Car \\
31 & 5397102019219938944 & 168.4998 & --37.6799 & 16.4 & 10.4 & 48.3 & VAR & CV (DN) & V436 Cen \\
32 & 5524430207364715520 & 134.6379 & --41.7979 & 16.6 & 10.6 & 12.0 & -- & CV (DN) & CU Vel \\
33 & 5665509162794290944 & 146.4624 & --19.7338 & 16.8 & 10.5 & 18.6 & VAR & CV (DN) & NSV 4618 \\
34 & 6039131391540808832 & 243.8122 & --28.6259 & 14.6 & 9.1 & 15.3 & VAR & CV (DN) & V893 Sco \\
35 & 6153373329716524800 & 189.5674 & --38.7127 & 17.2 & 10.7 & 38.3 & VAR & CV (IP) & V1025 Cen \\
36 & 6185040879503491584 & 193.1003 & --29.2488 & 13.2 & 9.5 & 17.0 & VAR & CV (IP) & EX Hya \\
37 & 6226943645600487552 & 229.9803 & --25.0067 & 16.5 & 11.2 & 14.8 & VAR & CV (DN) & GW Lib \\
38 & 6453536224527716224 & 318.9213 & --58.6818 & 17.3 & 10.4 & 27.6 & VAR & CV (P) & CD Ind \\
39 & 6557154200328277120 & 348.8825 & --30.8135 & 16.8 & 10.3 & 21.5 & VAR & CV (IP) & CC Scl \\
40 & 667150159287327360 & 119.7209 & +16.2792 & 15.5 & 8.9 & 38.5 & VAR & CV (IP) & DW Cnc \\
41 & 6896767366186700416 & 318.0387 & --8.82707 & 16.8 & 11.1 & 39.3 & VAR & CV (DN) & VY Aqr \\
42 & 6911950900211768704 & 316.9922 & --5.29488 & 16.4 & 10.0 & 14.5 & VAR & CV (P) & HU Aqr \\
43 & 855167540988615296 & 160.9859 & +58.1255 & 17.3 & 11.0 & 16.1 & VAR & CV (DN) & IY UMa \\

44 & 3021820276571880064 & 93.35673 & --3.04419 & 17.2 & 11.3 & 30.3 & -- & EM & New CV candidate$^{\dagger}$ \\
45 & 3446531278731142912 & 81.01973 & +29.6338 & 17.4 & 12.3 & 18.9 & -- & EM & New CV candidate$^{\dagger}$ \\
46 & 4835425609500365184 & 59.19386 & --44.8056 & 16.7 & 10.6 & 11.6 & VAR & EM & ELS group$^{\dagger}$ \\
47 & 5093945085525624448 & 57.73343 & --20.8044 & 14.8 & 9.3 & 60.3 & -- & EM & ELS group$^{\dagger}$   \\
48 & 5105520847021315712 & 51.79829 & --17.3162 & 17.1 & 11.6 & 244.9 & VAR & EM & ELS group$^{\dagger}$ \\
49 & 5788603818854455680 & 180.4302 & --78.5965 & 17.1 & 12.0 & 34.2 & -- & EM & ELS group$^{\dagger}$  \\
50 & 6064536604140907136 & 205.9949 & --55.3288 & 13.9 & 8.3 & 34.8 & -- & EM & ELS group$^{\dagger}$ \\
51 & 6097584212806297984 & 214.2756 & --42.9363 & 17.4 & 10.8 & 18.4 & -- & EM & ELS group$^{\dagger}$ \\
52 & 5409115210248936832 & 144.5049 & --50.1616 & 15.8 & 10.3 & 214.4 & -- & PN? & ELS group$^{\dagger}$ \\
53 & 1951666884868218112 & 322.2563 & +36.5179 & 17.3 & 10.8 & 40.8 & VAR & WD & CV (DN) \citep{2021MNRAS.504.2420I} \\
54 & 227766037015779200 & 64.16321 & +40.0426 & 16.6 & 11.6 & 23.8 & VAR & WD & New CV candidate$^{\dagger}$ \\
55 & 6086028036361965952 & 195.4860 & --47.0349 & 15.3 & 9.7 & 56.5 & -- & YSO & YSO group$^{\dagger}$  \\
56 & 6243210920137332096 & 239.8717 & --22.6050 & 16.9 & 11.2 & 64.1 & -- & YSO & YSO group$^{\dagger}$ \\
57 & 5362212384974013184 & 165.8220 & --48.5844 & 16.3 & 10.4 & 10.8 & VAR & EB & WD + MS group$^{\dagger}$ \\
58 & 4929395130249009024 & 23.13838 & --49.5267 & 16.1 & 12.5 & 15.5 & -- & PM & WD + MS group$^{\dagger}$\\
59 & 285957277597658240 & 79.76650 & +63.0610 & 15.0 & 9.7 & 28.1 & VAR & STAR & CV (DN) \citep{2020MNRAS.494.3799P} \\
60 & 4120637734561313664 & 266.4073 & --17.9397 & 16.7 & 9.8 & 17.3 & -- & CV? & CV \citep{2023AJ....165..163C} \\
61 & 5343601913741261312 & 178.8603 & --56.6976 & 14.0 & 8.5 & 14.3 & -- & -- & CV, V1040 Cen \citep{2021MNRAS.504.2420I} \\
62 & 5825198967486003072 & 228.5004 & --65.0935 & 16.1 & 10.2 & 35.7 & VAR & -- & CV, EK Tra \citep{2025MNRAS.536.1057I} \\

\noalign{\smallskip}
\hline
\end{tabular}
The {\it Gaia} variability status (``VAR'' denotes {\tt VARIABLE}, while ``--'' denotes {\tt NOT\_AVAILABLE}). The adopted abbreviations for CV subtypes are DN (dwarf nova), P (polar), IP (intermediate polar), and NL (nova-like variable). Object types from SIMBAD database: EM (emission-line object), WD (white dwarf), YSO (young stellar object), EB (eclipsing binary), PN (planetary nebula), and PM (high-proper-motion object). ($\dagger$) - See Appendix~\ref{app:identification} for details on the different groups.
\end{table*}

\end{appendix}

\end{document}